%% file: ThetaMetric.tex
\documentclass[a4paper]{article}
\usepackage{typearea}
\usepackage{graphicx}  
\usepackage{hyperref}
\usepackage{makeidx}
\usepackage{amsmath}

\usepackage{mathtools}
\usepackage{amssymb}
\usepackage{amsthm}
\usepackage{mathrsfs}
\usepackage{color}
\usepackage{bm}
\usepackage{booktabs}
\usepackage{multirow}
\usepackage{multicol}
\usepackage{ytableau}
\usepackage{cite} 
\usepackage{pifont}
\DeclareSymbolFont{extraup}{U}{zavm}{m}{n}
\DeclareMathSymbol{\varheart}{\mathalpha}{extraup}{86}
\DeclareMathSymbol{\vardiamond}{\mathalpha}{extraup}{87}

\input{./setting}

\numberwithin{equation}{subsection}
\begin{document}
\typearea{16}
\title{Yang--Mills--Utiyama Theory
and
Graviweak Correspondence
}
\author{Yoshimasa Kurihara\footnote{yoshimasa.kurihara@kek.jp}
\\
{\small 
The High Energy Accelerator Organization (KEK), 
Tsukuba, Ibaraki 305-0801, Japan}
}
\date{\today}

\maketitle
\begin{abstract} 
This report provides a geometrical Yang--Mills theory, including gravity.
The theory treats the space-time symmetry of the local Lorentz group in the same manner as the internal gauge symmetry.
We extend the theory into a more expansive space, which includes both Euclidean and Lorentzian metrics at its boundary.
In the extended space, we can transport topological achievements from the Euclidean Yang--Mills theory into the Lorentzian Yang--Mills theory.
This extension also provides a relation between space-time and internal symmetry. 
The perspective provided by the extended theory suggests the novel relation between gravity and the weak force, leading us to the graviweak correspondence.
\end{abstract}

  \tableofcontents


\input{Sec1}

\input{Sec2}
\input{Sec-3-YMU}

\input{Sec3}

\input{Sec4}

\input{Sec5}

%
%
\section*{Acknowledgements}
I would like to thank Dr Y$.$ Sugiyama and Prof$.$ J$.$ Fujimoto for their continuous encouragement and fruitful discussions.
~\vspace{1cm}

\appendix
\noindent
\textbf{\Large Appendix}
~\vspace{-5mm}
\numberwithin{equation}{section}
\input{Appendix}

\numberwithin{equation}{subsection}
\input{amphoYMB}

\bibliographystyle{unsrt}
\bibliography{common-ref}      
\end{document}

%% file: setting.tex
\newtheorem{definition}{Definition}[section]

\newtheorem{remark}[definition]{Remark}

\theoremstyle{definition}

\newcommand{\QED}{\nobreak \ifvmode \relax \else
      \ifdim\lastskip<1.5em \hskip-\lastskip
      \hskip1.5em plus0em minus0.5em \fi \nobreak
      \vrule height0.75em width0.5em depth0.25em\fi}
	\def\slash#1{\not\!#1}

	\def\sslash#1{\not\!{\hspace{-.3em}\not\!#1}}

\def\CRed#1{{\color{red}#1}}

\def\hlf{{1\hspace{-.1em}/\hspace{-.05em}2}}

\def\VL{{\textsc{v}\hspace{-.06em}\textsc{l}}}
\def\Sl{{\textsc{s}\hspace{-.06em}\textsc{l}}}
\def\VE{{\textsc{v}\hspace{-.06em}\textsc{e}}}
\def\SE{{\textsc{s}\hspace{-.06em}\textsc{e}}}

\def\ncdot{{\!\cdot}}

\def\YMU{Y\hspace{-.15em}M\hspace{-.1em}U\hspace{.2em}}

\def\cod{{\hat{d}}}

\def\wwwQ{\www_{\hspace{-.1em}\theta}}

\def\bphi{\bm{\phi}}
\def\phiU{{\phi_{\textsc{u}}^{~}}}
\def\phiD{{\phi_{\textsc{d}}^{~}}}
\def\bphiU{{\bm\phi_\textsc{~}^\textsc{u}}}
\def\bphiD{{\bm\phi_\textsc{~}^\textsc{d}}}

\def\bphiw{{\bm\phi_{\textsc{w}}^{~}}}

\def\bphiUw{{\bm\phi_\textsc{w}^\textsc{u}}}
\def\bphiDw{{\bm\phi_\textsc{w}^\textsc{d}}}
\def\hbphiUw{{\hat{\bm\phi}_\textsc{w}^\textsc{u}}}
\def\hbphiDw{{\hat{\bm\phi}_\textsc{w}^\textsc{d}}}

\def\bphih{{\bm\phi_{\textsc{h}}^{~}}}
\def\phiUh{{\phi_{\textsc{h}}^{\textsc{u}}}}
\def\phiDh{{\phi_{\textsc{h}}^{\textsc{d}}}}
\def\bphiUh{{\bm\phi_\textsc{h}^\textsc{u}}}
\def\bphiDh{{\bm\phi_\textsc{h}^\textsc{d}}}
\def\hbphiUh{{\hat{\bm\phi}_\textsc{h}^\textsc{u}}}
\def\hbphiDh{{\hat{\bm\phi}_\textsc{h}^\textsc{d}}}

\def\Ca{{C\hspace{-.1em}aus}}
\def\LCa{{\CCC\aaa\uuu\sss}}

\def\<{\langle}
\def\>{\rangle}
\def\la{\left\langle}
\def\ra{\right\rangle}

\def\bpsi{\bm{\xi}}

\def\sl{{S\hspace{-.1em}L}}
\def\SL2C{{S\hspace{-.1em}L(2,\C)}}
\def\GLTC{{G\hspace{-.1em}L(2,\C)}}

\def\SUW{SU^{~}_{\hspace{-.15em}\textsc{w}}\hspace{-.1em}(2)}

\def\GcP{{G^{~}_{\hspace{-.1em}\cP}}}

\def\pij#1{\pi_{\hspace{-.1em}#1}}
\def\dpij#1{\dot{\pi}_{\hspace{-.1em}#1}}

\def\VVs{{\VV^{~}_{\hspace{-.2em}s}}}
\def\Vs#1{{\VV^{#1}_{\hspace{-.2em}s}}}

\def\PH#1{{P^{#1}_{\hspace{-.15em}\hat{\textsc{h}}}}}

\def\VH#1{{\VV^{#1}_{\hspace{-.15em}\hat{\textsc{h}}}}}

\def\({\left(}
\def\){\right)}

\def\Tr{{\mathrm{Tr}\hspace{-.05em}}}

\def\stars{{\star\star}}

\def\bcdot{{\circ}}
\def\bcdots{{\circ\circ}}

\def\bullets{{\bullet\bullet}}

\def\TM{{T\hspace{-.2em}\M}}
\def\TsM{{T^*\hspace{-.2em}\M}}

\def\TsMB{{T^*\hspace{-.2em}\M_\bullet}}
\def\TME{{T\hspace{-.2em}\ME}}
\def\TsME{{T^*\hspace{-.2em}\ME}}
\def\TMM{{T\hspace{-.2em}\MM}}
\def\TsMM{{T^*\hspace{-.2em}\MM}}

\def\TMM{{T\hspace{-.2em}\MM}}
\def\TsMM{{T^*\hspace{-.2em}\MM}}

\def\Gso{{{G}_{\hspace{-.2em}S\hspace{-.1em}O}}}
\def\Gsu{{{G}_{\hspace{-.2em}S\hspace{-.1em}U}}}

\def\aaa{{\mathfrak a}}
\def\Aa{{\mathscr{A}}}

\def\AAA{{\mathfrak A}}

\def\AAAL{{\mathfrak A}_\textsc{l}}
\def\AAAQ{{\mathfrak A}_\theta}
\def\AAAO{{\mathfrak A}_0}

\def\bbb{\mathfrak{b}}

\def\ooo{\mathfrak{o}}

\def\C{\mathbb{C}}
\def\Z{\mathbb{Z}}

\def\Ff{\mathbb{F}}

\def\uuu{\mathfrak{u}}

\def\TT{{\mathcal T}}
\def\TTT{{\mathfrak T}}

\def\tt{\tilde{t}}

\def\ds{{\slash{d}}}
\def\cds{{\slash{\delta}}}

\def\v{{\bm{v}}}

\def\vvv{\mathfrak{v}}
\def\VVV{\mathfrak{V}}
\def\VV{{\mathbb{V}}}

\def\u{{\bm{u}}}

\def\f{\mathscr{F}}
\def\FF{{\mathcal F}}

\def\FFF{\mathfrak{F}}
\def\FFFE{\mathfrak{F}_\textsc{e}}
\def\FFFL{\mathfrak{F}_\textsc{l}}
\def\FFFQ{\mathfrak{F}_\theta}
\def\FFFO{\mathfrak{F}_0}

\def\SOt{{r}}
\def\g{\bm{g}}
\def\gB#1{\bm{g}^{~}_{\hspace{-.1em}#1}}
\def\gBa#1{\bm{g}^{a}_{#1}}
\def\gBs#1{\bm{g}_{\hspace{-.1em}#1}}
\def\GGG{\mathfrak{G}}
\def\ggg{\mathfrak{g}}

\def\sss{\mathfrak{s}}

\def\SSS{{\mathfrak S}}

\def\Ss{{S}}

\def\LL{{\cal L}}

\def\LLL{{\mathfrak L}}

\def\lll{\mathfrak{l}}

\def\JJJ{{\mathfrak J}}

\def\HQ{\mathbb{H}}
\def\H{\mathscr{H}}

\def\bbb{{\mathfrak b}}

\def\C{\mathbb{C}}
\def\CCC{{\mathfrak C}}

\def\Varepsilon{{\mathcal E}}

\def\EEE{\mathfrak{E}}
\def\eee{{\mathfrak e}}
\def\eeeE{\eee_\textsc{e}}
\def\eeeL{\eee_\textsc{l}}
\def\eeeQ{\eee_\theta}

\def\I{\mathscr{I}}

\def\M{{\cal M}}
\def\MM{{\mathscr{M}^{~}_{\hspace{-.1em}\textsc{g}}}}

\def\N{\mathbb{N}}

\def\R{\mathbb{R}}

\def\Re{\mathrm{Re}}
\def\Im{\mathrm{Im}}
\def\Ri{R}

\def\RRR{\mathfrak{R}}

\def\W{{\cal W}}

\def\www{{\mathfrak w}}
\def\WWW{{\mathfrak W}}

\def\kE{\kappa^{~}_{\hspace{-.1em}\textsc{e}}}

\def\={{\hspace{-.1em}=\hspace{-.1em}}}

\def\LLambda{\bm{\Lambda}}
\def\Wedge{{\Omega}}

\def\OmegaL{\Omega_\textsc{l}}
\def\OmegaQ{\Omega_\theta}

\def\SO{{\hspace{-.1em}S\hspace{-.1em}O\hspace{-.1em}}}
\def\SU{{\hspace{-.1em}S\hspace{-.1em}U\hspace{-.1em}}}

\def\Sp{{\hspace{-.1em}s\hspace{-.1em}p\hspace{.1em}}}
\def\SG{{\hspace{-.2em}s\hspace{-.1em}g}}

\def\Cl{{\mathcal C\hspace{-.1em}l}}

\def\sss{{\mathfrak s}}
\def\uuu{{\mathfrak u}}

\def\cP{{\hspace{-.1em}\mathrm{c}\hspace{-.05em}\textsc{p}}}

\def\cov{{\hspace{-.1em}cov}}

\def\cG{{c^{~}_{\hspace{-.1em}g\hspace{-.1em}r}}}

\def\dSp{{\slash{d}^{~}_{\hspace{-.1em}\Sp}}}

\def\gdg{\bm{g}^{-1}_\SU\hspace{.1em}d\bm{g}_\SU^{~}}

\def\HD{{\widehat{\mathrm{H}}}}

\def\HDB{{\widehat{\mathrm{H}}_\bullet^{~}}}
\def\HDBi{{\widehat{\mathrm{H}}_\bullet^{-1}}}

\def\HDE{{\widehat{\mathrm{H}}_\textsc{e}^{~}}}
\def\HDL{{\widehat{\mathrm{H}}_\textsc{l}^{~}}}
\def\HDQ{{\widehat{\mathrm{H}}_\theta^{~}}}

\def\H{{\cal{H}}}

\def\TxT{{2\hspace{-.2em}\times\hspace{-.2em}2}}
\def\FxF{{4\hspace{-.2em}\times\hspace{-.2em}4}}

\def\ME{{{\M^{~}_{\hspace{-.1em}\textsc{e}}}}}
\def\ML{{{\M^{~}_{\hspace{-.1em}\textsc{l}}}}}
\def\MB{{{\M^{~}_{\hspace{-.1em}\bullet}}}}
\def\MQ{{{\M^{~}_{\hspace{-.1em}\theta}}}}

\def\bsigE{{{\bm\sigma^{~}_{\hspace{-.1em}\textsc{e}}}}}

\def\etaE{{{\eta^{~}_{\textsc{e}}}}}

\def\etaL{{{\eta^{~}_{\textsc{l}}}}}

\def\etaQ{{\eta^{~}_\theta}}

\def\betaE{{{\bm\eta^{~}_{\textsc{e}}}}}

\def\betaL{{{\bm\eta^{~}_{\textsc{l}}}}}
\def\betaB{{{\bm\eta^{~}_{\bullet}}}}
\def\betathe{{\bm\eta^{~}_\theta}}
\def\betaQ{{\bm\eta^{~}_\theta}}

\def\TML{{\TM_{\hspace{-.1em}\textsc{l}}}}
\def\TsML{{\TsM_{\hspace{-.1em}\textsc{l}}}}

\def\MQ{{\M^{~}_{\hspace{-.1em}\theta}}}
\def\TMQ{{\TM^{~}_{\hspace{-.1em}\theta}}}
\def\TsMQ{{\TsM^{~}_{\hspace{-.1em}\theta}}}

%% file: Sec1.tex
\section{Introduction}
The standard theory of particle physics was finally established in 2011 owing to the experimental discovery of the Higgs boson\footnote{For the standard theory of particle physics, see, e.g., ref.\cite{weinberg1995quantum}, and for its experimental support, see ref.\cite{PhysRevD.110.030001}}.
Since then, the Standard theory has been supported by various precise experimental data.
The standard theory coherently treats three fundamental forces, electric, weak, and strong, and describes the behaviour of elementary particles with these interactions.
The standard theory consists of three layers. The first layer is the classical Yang--Mills theory\cite{Yang:1954ek}, which includes $U\!(1)$, $SU\!(2)$, and $SU\!(3)$ gauge symmetries respectively corresponding to electromagnetic, weak and strong interactions.
The second layer is the quantum field theory with its perturbative expansion, which includes a renormalisation.
Though the original gauge symmetry is broken due to the gauge fixing during quantisation, a new kind of symmetry, the BRST symmetry\cite{BECCHI1976287,Tyutin:1975qk}, appears and ensures the theory's unitarity. 
The perturbative quantisation of the Yang--Mills theory with the compact Lie group gauge symmetry has been established, and the renormalisability has been proven for all orders of perturbation.
The third layer is the spontaneous $U\!(1){\otimes}SU\!(2)$ symmetry breaking, providing the mass to the gauge bosons and fermions, so-called the Higgs mechanism.
The spontaneous symmetry breaking has the broken symmetry of its vacuum, and the gauge boson obtains its effective mass while maintaining the gauge symmetry at the Lagrangian level.
On the other hand, \textit{ad hoc} Yukawa coupling with the Higgs boson provides its mass for a fermion.

While the standard theory of particle physics considers three forces except the gravitational force among the four fundamental forces in nature, general relativity treats the gravitational force classically.
General relativity is supported by various experiments, including recent observations of the gravitational wave\cite{PhysRevLett.116.061102}.
General relativity is also the Yang--Mills type gauge theory with the local $SO(1,3)$ gauge symmetry, as first Utiyama has clarified\cite{PhysRev.101.1597}.
A relation between the Yang--Mills theory and gravity is discussed by Cofano, Fu and Krasnov\cite{PhysRevD.92.065012} and by Denfiz\cite{Dengiz:2016eoo}.
Unfortunately, a perturbative renormalisation theory of gravity has not yet been established, in contrast to the Yang--Mills theory with the compact Lie group.
In other words, a theory of gravity is still in the first layer of the standard theory of particle physics.

After the original work of Yang and Mills in 1954, the Yang--Mills theory was refined geometrically and inspired mathematicians as a tool for investigating three- and four-dimensional Euclidean manifolds.
On the other hand, the Yang--Mills theory, including gravitation with the Lorentzian metric, which is indeed the target of physicists, is not at that level. 
The primary objective of this study is to present the Yang--Mills theory with symmetries $SU\!(2)$ and the local $SO(1,3)$ using the language of geometry, then to accomplish the first layer of the standard theory, including gravity.
We refer to this theory as the \emph{Yang--Mills--Utiyama} (\YMU\hspace{-.2em}) theory in this report.
This objective has two goals: One is to give new insight into constructing quantum gravity by describing general relativity using the same mathematical language as the Yang--Mills theory: it may bridge the first layer to the second layer of the standard theory, including gravity. 
The other is to allow the importing of modern mathematical harvests into physics.
The secondary objective of this study is to provide novel tools to make it possible.
The current author has developed a mathematical tool\cite{Kurihara:2022sso} to connect the Lorentzian metric to the Euclidean one topologically\footnote{This report refers to the tool as the \textit{amphometric}, named the $\theta$-metric in ref.\cite{Kurihara:2022sso}. The parametrisation of the metric in this report differs from that in the reference.}.
It allows topological results obtained in the Euclidean metric space to be imported into the Lorentzian metric space. 
This study refines and adapts that method to the Yang--Mills theory.

In the third layer of the standard theory, the Yukawa coupling between the Higgs boson and the fermion gives the fermion a mass.
The theory insists that a fermion coupled to the $SU\!(2)$ vacuum through the Higgs boson obtains the \textit{pole mass}, which is a position of the divergence of the fermion propagator.
Consider the relationship between a gravitational mass and the quantum field theoretic mass in the nucleon-effective theory here.
Even though a proton is not an elementary particle with an internal structure consisting of quarks and gluons, the effective theory of nucleons treats it as a simple particle with a definite mass, charge, spin, and others.
In the effective theory, a proton \textit{pole mass} is equivalent to the \textit{gravitational mass} of 1mol of hydrogen gas divided by the Avogadro constant, which is $1.672621898(21)\times10^{-27}$kg\cite{PhysRevD.110.030001}.
The theory with this mass well describes a proton-electron scattering\cite{RevModPhys.94.015002}.
We can recognise this as evidence of the equivalence theorem in the quantum field theoretic context: ``\textit{The pole in the fermion propagator is equivalent to the gravitational mass}.''
Suppose we recognise the equivalence principle as a true statement for the quantum field theory.
This fact insists that a source of fermionic matter's gravitational mass is the vacuum's weak $SU\!(2)$ charge.
This is one of the variants of the  \textit{graviweak correspondence}.
Several authors\cite{Nesti:2007ka, Alexander:2007mt, Alexander:2012ge} have tried unifying the weak and gravitational forces.

This study is also inspired by recent works by Woit\cite{Woit:2021bmb,Woit:2023idu}, who has proposed a novel approach to possible chirality asymmetric method to treat both Minkowski space-time and the Yang--Mills theory. 
The Lie-algebra homomorphism $\sss\ooo(4)=\sss\uuu(2)\oplus\sss\uuu(2)$ and $\sss\ooo(1,3)\otimes\C=\sss\lll(2,\C)\oplus\overline{\sss\lll(2,\C)}$ gives us a hint about the graviweak correspondence.
This Lie-algebra homomorphism between rotational and unitary groups is accidental, appearing only in low-dimensional spaces such as $\sss\ooo(3)=\sss\uuu(2)$; it is not valid for higher dimensional rotation.
An angular momentum is an eigenvector of the  $SO(3)$ generator, and a spinor is that of $SU\!(2)$ one. 
Eigenvalues of $SU\!(2)$ spinor can be treated as a half-integer valued angular momentum in physics owing to this accidental Lie-algebra homomorphism.
In the higher-dimensional spaces, a pair $SO(N)$ and $SU\!(N\!-\!1)$ does not share their Lie-algebra; thus, the $SU\!(N\!-\!1)$ spinor is not interpreted as the $SO(N)$ angular momentum when $N\geq5$.
This observation tells us that accidental homomorphism in four-dimensional space-time plays an essential role in fixing the properties of fermions.

Only left-handed fermions have a weak charge.
On the other hand, we can describe a gravitational interaction using only half the space-time degree of freedom, say the right-handed part.
It suggests the origin of the $\SU(2)$ gauge symmetry is owing to another half of the space-time degree of freedom.
This is one aspect of the graviweak correspondence.
If it is true, we can understand the universe being weakly charged.
This observation tells us that accidental homomorphism in four-dimensional space-time plays an essential role in fixing the properties of fermions.
This study's tertiary and last objective is to pursue the graviweak correspondence using the tool to connect Euclidean and Lorentzian metrics developed for achieving the secondary objective.

The structure of this report is as follows:
After this introductory section, \textbf{section 2} provides mathematical preliminaries utilised in this study.
In this section, we also introduce mathematical notations used throughout this report. 
\textbf{Section 3} is devoted to accomplishing the primary objective, establishing the general relativistic Yang--Mills theory using geometrical language.
After clearly defining principal bundles where the theory is defined, we write down the Lagrangian and the equation of motion of the Yang--Mills theory with symmetries $SU\!(2)$ and the local $SO(1,3)$, namely the \YMU theory, using the geometrical language.
\textbf{Section 4} introduces a novel mathematical tool to connect Euclidean and Minkowskian spaces topologically and analytically.
 Mathematical tools introduced in \textbf{section 1} and the \YMU Lagrangian are redefined in the broader space, including Euclidean and Minkowskian spaces.
This section achieves the secondary objective of this study. 
The perspective provided in \textbf{section 4} suggests the graviweak correspondence discussed in \textbf{section 5}.
We present the possible resolution of why the Higgs-fermion Yukawa coupling provides a fermion's gravitational mass.
Two supplementary sections present a list of induced representations of $SL(2)$ group (\textbf{Appendix A}) and a detailed discussion of the instanton in Euclidean and Minkowskian spaces as an example of the validity of the method introduced in \textbf{section 3} (\textbf{Appendix B}). 

This report uses the following physical units:
a speed of light is set to unity $c=1$, and a gravitational constant $\kE=8\pi G_{\hspace{-.1em}\text{N}}/c^2$ ($G_{\hspace{-.1em}\text{N}}$ is the Newtonian constant of gravitation) and reduced Plank-constant $\hbar=h/2\pi$ are written explicitly.
In these units, physical dimensions of fundamental constants are $[\hbar\hspace{.1em}\kE]=L^2=T^2$ and $[\hbar/\kE]=E^2=M^2$, where $L$, $T$, $E$ and $M$ are, respectively, length, time, energy and mass dimensions.

%% file: Sec2.tex
\section{Mathematical preliminaries}\label{Mathematicalpreliminaries}
The standard theory of particle physics is based on the quantum gauge theory with $U\!(1){\otimes}SU\!(2)$ and $SU\!(3)$ groups, which are called the Yang-Mills theory\cite{Yang:1954ek}.
The former describes electroweak interaction, and the latter represents the strong interaction in nature.
In addition to the gauge symmetry, the Yang--Mills theory has the Poinca\'{r}e symmetry owing to the \emph{special principle of relativity}.
General relativity is also the gauge theory\cite{PhysRev.101.1597} with the $GL(4)$ group owing to the \emph{general principle of relativity} and describing a gravitational interaction.
In this report, we call the general relativistic Yang--Mills theory, including the gravitational force as the gauge interaction, as Yang--Mills--Utiyama (\YMU\hspace{-.2em}) theory.

At the classical level, the \YMU theory utilises mathematics of differential geometry, algebraic topology, and group representation theory.
This section prepares mathematical preliminaries used in the  \YMU theory.
%
%
\input{Sec2-2}

%
%
\input{Sec2-4}

%
%
\input{Sec2-5}

%
%
\input{Sec2-Clifford}

%
%
\input{Sec2-ChiranSpinor}

%
%
\input{Sec2-3}

%
%
\input{Sec2-GRfields}

%
%
\input{Sec2-6}

%% file: Sec2-2.tex
%
%
\subsection{General relativity}
General relativity\cite{1915SPAW.......844E}, developed by Einstein, is a geometrical theory that describes the gravitational force as an intrinsic property of space-time.
Einstein constructed his gravitational theory under two fundamental principles: the \emph{general relativistic principle} and the \emph{equivalent principle}.
The former requires that space-time is the smooth manifold constructing the principal bundle with the $GL(4)$ structure group, and physical fields are given as sections in the bundle.
The latter insists that at any point in the manifold, we can find a local smooth manifold with the vanishing Levi-Civita connection, which means the locally vanishing gravity.
Physical matter fields and the forces among them correspond to sections and curvatures of the principal bundle, respectively.
This section introduces mathematical set-ups to treat general relativity.

\subsubsection{Section, connection, and curvature}\label{app1}
The fundamental force field corresponds to a curvature in a four-dimensional smooth manifold, with the compact Lie group as the structure group, and the physical matter field corresponds to the section of the bundle.
This section introduces a principal bundle and its \textit{section}, \textit{connection}, and \textit{curvature}.
We denote the principal bundle by the tuple $(E,\pi,M,G)$, where the smooth manifolds $E$ and $M$ are total and base spaces, respectively, a diffeomorphism $\pi$ is a bundle (projection) map and $G$ is a principal group.

We introduce the vector space $V$ defined in total space $E$ as the section whose space is denoted as $\Gamma(M,V(E))$.  
When we emphasise the vector $\v{\hspace{.1em}\in\hspace{.1em}}V$  has symmetry owing to the structure group $G$, we explicitly write as $\Gamma(M,V(E),G)$.
The group operator $g^{~}_\textsc{g}{\hspace{.1em}\in\hspace{.1em}}G$ acts on section $s\in\Gamma(M,V(E),G)$ from the left.
A connection is a Lie algebra valued one-form $\AAA_\textsc{g}\in\Wedge^1(T^*\hspace{-.2em}M)\otimes Ad(\ggg^{~}_G)$, where $\Wedge^p_{~}(T^*\hspace{-.2em}M)$ is a space of $p$-forms in the cotangent bundle $T^*\hspace{-.1em}M$ and $\ggg^{~}_G$ is a Lie algebra of the structure group $G$. 
The differential operator, covariant under the structural group, namely the covariant differential, is provided using the connection as
\begin{align}
d_\textsc{g}:\Omega^p\rightarrow\Omega^{p+1}:
\aaa\hspace{.2em}{\mapsto}\hspace{.2em}d^{~}_\textsc{g}\aaa:=d\aaa-ic^{~}_\textsc{g}\hspace{.1em}[\AAA^{~}_\textsc{g},\aaa]_\wedge,
~~\text{where}~~
[\AAA^{~}_\textsc{g},\aaa]_\wedge:=\AAA^{~}_\textsc{g}\wedge\aaa-(-1)^{p}\hspace{.1em}\aaa\wedge\AAA^{~}_\textsc{g}.\label{connec}
\end{align}
Corresponding curvature two-form $\FFF_\textsc{g}\in\Wedge^2(T^*\hspace{-.2em}M)\otimes\ggg^{~}_G$ is defined owing to the structure equation as
\begin{align}
\FFF^{~}_\textsc{g}&:=d\AAA^{~}_\textsc{g}-ic^{~}_\textsc{g}\hspace{.2em}\AAA^{~}_\textsc{g}\wedge\AAA^{~}_\textsc{g}=\sum_I\(
d\AAA^{\hspace{.2em}I}_\textsc{g}+\frac{c^{~}_\textsc{g}}{2}\sum_{J,K}f^I_{\hspace{.1em}J\hspace{-.1em}K}\hspace{.1em}\AAA^{\hspace{.2em}J}_\textsc{g}\wedge\AAA^{\hspace{.2em}K}_\textsc{g}
\)\tau^{~}_I,\label{adjointG}
\intertext{with}
\left[\tau^{~}_J,\tau^{~}_K\right]&=\tau^{~}_J\tau^{~}_K-\tau^{~}_K\tau^{~}_J
=:i\sum_If^I_{\hspace{.1em}J\hspace{-.1em}K}\tau^{~}_I,\label{strconst}
\end{align}
where $\tau^{~}_I$ and $f^I_{\hspace{.1em}JK}$ are the generator and the structure constant of the structure group.
The structure constant $f^I_{\hspace{.1em}JK}$ is defined by (\ref {strconst}).
Here, we introduce a coupling constant $c^{~}_\textsc{g}\in\C$ in the covariant differential following the physics convention, though it does not appear in the standard mathematics textbook.
When a single bundle is considered, the coupling constant can be absorbed in defining a connection and a curvature scaled as $-ic^{~}_\textsc{g}\hspace{.1em}\AAA_\textsc{g}\mapsto\AAA_\textsc{g}$ and $-ic^{~}_\textsc{g}\hspace{.1em}\FFF_\textsc{g}\mapsto\FFF_\textsc{g}$. 
Nevertheless, we explicitly write a coupling constant in this study because when multiple bundles are twisted over the same base manifold, each bundle's coupling constant cannot be absorbed by rescaling and provides relative strength of interactions in physics.

The covariant differential of section $\bm\phi=\sum_I\phi^I\tau_I\in\Gamma(M,V(E),G)$ is given using (\ref{connec}) for the zero-form object $\bm\phi$ as
\begin{align}
d_\textsc{g}\bm\phi&=d\bm\phi-ic^{~}_\textsc{g}\left[\AAA^{~}_\textsc{g},\bm\phi\right]_\wedge=
\sum_I\(d\phi^I+{c^{~}_\textsc{g}}\sum_{J,K}f^I_{\hspace{.1em}JK}\hspace{.1em}\AAA^{\hspace{.2em}J}_\textsc{g}\hspace{.1em}\phi^K\)\tau_I\in\Gamma(M,V(E),G).\label{cdiffc}
\end{align}

\subsubsection{Inertial bundle}\label{STManifold}
We introduce the inertial bundle in general relativity. 
A base manifold of the inertial bundle is the pseud-Riemannian manifold $(\MM,\bm{g})$, namely the global space-time manifold, as a model of the universe, where $\MM$ is a smooth and oriented four-dimensional manifold, and $\bm{g}$ is a metric tensor in $\MM$ with a negative sign such that det$[\bm{g}]<0$.
In an open  neighbourhood $U_p\subset\MM$, orthonormal bases in $T_{\hspace{-.1em}p}\MM$ and $T^*_{\hspace{-.1em}p}\MM$ are respectively introduced as $\partial/\partial x^\mu$ and $dx^\mu$.
We use the abbreviation $\partial_\mu:=\partial/\partial x^\mu$ throughout this report.
Two trivial vector-bundles $\TMM:=\bigcup_p T_{\hspace{-.1em}p}\MM$ and  $\TsMM:=\bigcup_p T^*_{\hspace{-.1em}p}\MM$ are referred to as  a tangent and cotangent bundles in $\MM$, respectively.
General relativity requires the universe to be a four-dimensional smooth manifold equipping the $GL(4)$ symmetry as the structure group.

In the global space-time manifold, the Levi-Civita (affine) connection is defined owing to the metric tensor as
\begin{align*}
\Gamma^\lambda_{~\mu\nu}(x\!\in\!\MM):=\frac{1}{2}\hspace{.1em}{\sum_{\sigma=0}^3}\hspace{.1em}g^{\lambda\sigma}\left(
\partial_\mu g_{\nu\sigma}(x)+\partial_\nu g_{\mu\sigma}(x)-\partial_\sigma g_{\mu\nu}(x)
\right),
\end{align*}
where $g_{\mu\nu}:=[\bm{g}]_{\mu\nu}$.
An inertial system, in which the Levi-Civita connection vanishes  $\Gamma^\lambda_{~\mu\nu}(p)=0$, exists at any point $p{\hspace{.1em}\in\hspace{.1em}}\MM$. 
A local inertial manifold at $p$ denoted as $\ML(p)$, where suffix ``$\textsc{l}$'' stands for the principal \emph{Lorentzian group} ($\SO(1,3)$ group).

An inertial bundle is the principal bundle such that:
\begin{align}
\(\MM\otimes\ML,\pi^{~}_\textsc{i},\MM,SO(1,3)\),\label{LorentzBndl}
\end{align}
where $\ML:=\bigcup_p\ML(p)$.
A projection map acts on the total space as
\begin{align*}
\pi^{~}_\textsc{i}:\MM\otimes\ML\rightarrow\MM:\ML(p)\mapsto{p}.
\end{align*}

The existence of the local inertial manifold at any point in $\MM$ is nothing other than ``Einstein's equivalence principle'' in physics.
We denote an orthonormal basis on $\TML$ as $\partial_a$.
As for suffixes of vectors in $\ML$, Roman letters are used for components of the standard basis throughout this study; while Greek letters are used for them in $\MM$.
The metric tensor in $\ML$ is denoted as $\betaL:=\mathrm{diag}[1,-1,-1,-1]$ using the standard basis.
The metric tensor and the Levi-Civita tensor (complete anti-symmetric tensor) $\bm\epsilon$, whose component is $[\bm\epsilon]^{~}_{0123}=\epsilon^{~}_{0123}=+1$, are constant tensors in $\ML$.

A pull-back\footnote{A pull-back (push-forward) of a map $\bullet$ is denoted as $\bullet^\sharp$ ($\bullet_\sharp$) instead of $\bullet^*$ ($\bullet_*$) in this study, avoiding confusion with the complex conjugate.} of bundle map $\pi^{~}_\textsc{i}$ induces one-form object $\eee\in\Gamma\(\ML,\TsML,SO(1,3)\)$ represented using the standard basis as
\begin{align*}
\pi_{\textsc{i}}^\sharp:
\Omega^1(\TsMM)\rightarrow\Gamma\(\ML,\TsML,SO(1,3)\):
dx^\mu\mapsto \eee^a:=\Varepsilon^a_\mu(p) dx^\mu,
\end{align*}
where $\Varepsilon^a_\mu\in{C^\infty(\MM)}$ is a smooth function defined in the open neighbour $U\subset\MM$, namely a vierbein, which is determined using the bi-local function $\bm\xi(y,x)$ for $x,y{\hspace{.1em}\in\hspace{.1em}}U{\subset}\MM$ as 
\begin{align*}
\xi^a(x,x)=0{\quad\text{and}\quad}\Varepsilon^a_\mu(x):=\left.\frac{\partial\xi^a(y,x)}{\partial y^\mu}\right|^{~}_{y=x}.
\end{align*}
The Einstein convention for repeated indices (one in up and one in down) is exploited throughout this study.
The vierbein maps a vector in $\MM$ to that in $\ML$ and \textit{vice versa}.
We use a Fraktur letter to represent differential forms\footnote{A Fraktur letter is also used showing Lie algebra.} defined in the cotangent bundle in this report.
Vierbain inverse $[\Varepsilon^{-1}]_a^\mu=\Varepsilon_a^\mu$ is also called the vierbein.
One-form object $\eee$ is referred to as the vierbein form.
The vierbein form provides an orthonormal basis in $\TsML$; it is a dual basis of $\partial_a$ such as $\eee^a\partial_b=\Varepsilon^a_\mu\Varepsilon^\nu_b dx^\mu\partial_\nu=\delta^a_b$ and is a rank-one tensor (vector) in $\TML$.
The four-dimensional volume form is represented using vierbein forms as 
\begin{align}
\vvv&:=\frac{1}{4!}\epsilon_{\bcdots\bcdots}\hspace{.1em}\eee^\bcdot\wedge\eee^\bcdot\wedge\eee^\bcdot\wedge\eee^\bcdot
=\mathrm{det}[\bm\Varepsilon]\hspace{.2em}dx^0\wedge dx^1\wedge dx^2\wedge dx^3
,~~\mathrm{det}[\bm\Varepsilon]>0.\label{VolForm}
\end{align}
Dummy \textbf{Roman} indices are often abbreviated to a small circle $\bcdot$ (or $\star$) when the dummy-index pair of the Einstein convention is obvious as above.
When multiple circles appear in an expression, the pairing must be on a left-to-right order at upper and lower indices.
Metric tensors  $\bm{g}$ and $\betaL$ are related each other like
\begin{align*}
g_{\mu\nu}&=\left[\bm{\Varepsilon}^t\betaL
\hspace{.1em}\bm{\Varepsilon}\right]_{\mu\nu}
=\etaL_{\hspace{.05em}\bcdots}\hspace{.2em}\Varepsilon^\bcdot_\mu\Varepsilon^\bcdot_\nu,
~~\text{yielding}~~
\textrm{det}[\bm{g}]=\textrm{det}[\betaL]\textrm{det}[\bm{\Varepsilon}]^2=-\textrm{det}[\bm{\Varepsilon}]^2<0.
\end{align*}
Thus, we obtain from (\ref{VolForm}) that
\begin{align}
\vvv&=\sqrt{-\mathrm{det}[\bm{g}]}\hspace{.2em}dx^0{\wedge}dx^1{\wedge}dx^2{\wedge}dx^3\!.\label{vvvmg}
\end{align}
Similarly, the two-dimensional surface and three-dimensional volume forms are defined as
\begin{align}
\SSS_{ab}:=\frac{1}{2}\epsilon_{ab\bcdots}\hspace{.1em}\eee^\bcdot\wedge\eee^\bcdot~~\text{and}~~
\VVV_{a}:=\frac{1}{3!}\epsilon_{a\bcdots\bcdot}\hspace{.1em}\eee^\bcdot\wedge\eee^\bcdot\wedge\eee^\bcdot\!,
\end{align}
respectively.

We introduce a connection one-form concerning the local $\SO(1,3)$ group, 
\begin{align*}
\www^{ab}=\omega^{~ab}_{\mu}\hspace{.1em}dx^\mu{\hspace{.2em}\in\hspace{.2em}}
V^2(\TML)\otimes\Omega^1(\TsML){\otimes}Ad(\ggg^{~}_{\SO(1,3)}),
\end{align*}
namely the \emph{spin connection form}.
Owing to (\ref{connec}), the $\SO(1,3)$-covariant differential operator $d_\www$ is defined owing to the spin-connection form as
\begin{align}
d_\www\aaa^a:=d\aaa^a+\cG\hspace{.1em}\eta^{~}_{\textsc{l}\hspace{.1em}\bcdots}\hspace{.1em}\www^{a\bcdot}\wedge\aaa^\bcdot~~\text{for}~~
\aaa{\hspace{.1em}\in\hspace{.1em}}\Omega^1(\TsML).\label{dwww}
\end{align}
Two-form object 
\begin{align}
\TTT^a:=d_\www\eee^a\in V^1(\TML)\otimes\Omega^2(\TsML)\label{torsionFM}
\end{align}
is referred to as a \emph{torsion form}.

The ($p$-tensor)-($q$-form) object $\aaa{\in}V^p(\TML)\otimes\Omega^q(\TsML)$ has a component representation owing to the standard basis such that
\begin{align}
\aaa=a^{i_1\cdots i_p}_{\hspace{2em}j_1\cdots j_q}
(\partial_{i_1}\otimes\cdots\otimes\partial_{i_p})
(\eee^{j_1}\wedge\cdots\wedge\eee^{j_q}),
\end{align}
Raising and lowering Roman indices for both tensor- and form-indices are done using the metric tensor $\betaL$ as in (\ref{dwww}).
Since we represent the tensor basis in $\TML$ with lower indices and the form bases in $\TsML$ with upper indices, the tensor coefficient has upper indices and the form coefficient has lower indices, which is called the \emph{intrinsic position} of indices in this report.

From the gauge theoretical point of view, choosing the inertial frame among general coordinate frames breaks the local $G\!L(4)$ gauge symmetry as the gauge fixing does.
After the gauge fixing by choosing one of the inertial frames, the inertial manifold still keeps a local $\SO(1,3)$ symmetry, and its spin connection behaves as a massless gauge boson, a graviton.  

In this study, a vector ${\bm u}{\hspace{.1em}\in\hspace{.1em}}V(\TML)$ is defined as a column vector, and its component is represented with a superscript as the intrinsic position. 
Components of its transpose ${\bm u}^t$ are represented as a row vector with a subscript as the intrinsic position. 
A local $\SO(1,3)$ invariant bilinear form of two vectors ${\bm u},{\bm v}{\hspace{.1em}\in\hspace{.1em}}V(\TML)$ are defined utilizing the metric tensor as
\begin{align}
\<\u,\v\>^{~}_\textsc{l}=\u^t\ncdot\betaL\ncdot\v=\etaL_\bcdots\hspace{.1em}u^\bcdot v^\bcdot=\<\v,\u\>^{~}_\textsc{l}\in\R,\label{TMinner}
\end{align}
yielding the pseudo-norm as $\|\u\|^{~}_\textsc{l}:={\<\u,\u\>^{\hlf}_\textsc{l}}^{~}$\hspace{-.4em}.
In this report, operator ``$\cdot$'' shows the standard multiplication between two matrices or a matrix and a vector.
This bilinear form degenerates in the inertial manifold with the Lorentzian metric and can have a negative value.
Local $\SO(1,3)$ group action $\Gso{\in}End(\TsML)$ is known as the Lorentz transformation.
We denote the ($\FxF$)-matrix representation of the $\SO(1,3)$ group as $\bm{g}^{~}_\textsc{l}$, which acts on a vector from the left as
$
\u'=\bm{g}^{~}_\textsc{l}\cdot\u
$.
The Lorentz transformation of the vierbein form and the spin-connection form are provided as
\begin{align*}
\Gso:\hspace{.25em}\eee\hspace{.15em}\mapsto\Gso(\hspace{.15em}\eee\hspace{.15em})&=\hspace{.15em}\eee'\hspace{.15em}=\bm{g}^{~}_\textsc{l}\ncdot\eee,\\
\Gso:\www\mapsto\Gso(\www)&=\www'=
\bm{g}^{~}_\textsc{l}\ncdot\www\cdot\bm{g}_\textsc{l}^{-1}+{\cG}^{\hspace{-.5em}-1}\bm{g}^{~}_\textsc{l}\ncdot d\hspace{-.1em}\bm{g}_\textsc{l}^{\hspace{-.1em}-1},
\intertext{yielding the Lorentz transformation of the covariant differential (\ref{dwww}) as}
\Gso(d_\www\eee)&=d_{\www'}\eee'.
\end{align*}

The $O(1,3)$ group is not simply connected but separated into two subgroups according to the sign of  $[\bm{g}^{~}_\textsc{l}]^0_{\hspace{.3em}0}$.
We define the \emph{orthochronous} and \emph{anti-orthochronous} transformations of $\SO(1,3)$ as
\begin{align*}
SO^{\uparrow}(1,3)&:=\{\bm{g}^{~}_\textsc{l}{\hspace{.1em}\in\hspace{.1em}}SO(1,3)\big|[\bm{g}^{~}_\textsc{l}]^0_{\hspace{.3em}0}>0\},~~\text{and}~~
SO^{\downarrow}_{~}(1,3):=\{\bm{g}^{~}_\textsc{l}{\hspace{.1em}\in\hspace{.1em}}SO(1,3)\big|[\bm{g}^{~}_\textsc{l}]^0_{\hspace{.3em}0}<0\},
\end{align*}
respectively.
Elements of  $\SO^{\uparrow}(1,3)$ are called the proper Lorentz transformation in physics.
There are two essential elements of subgroup $O(1,3){\setminus}SO(1,3)$ such that:  
\begin{align*}
\bm{O}^\textsc{t}:=\text{diag}[-1,1,1,1]~~\text{and}~~
\bm{O}^\textsc{p}:=\text{diag}[1,-1,-1,-1],
\end{align*}
where $\bm{O}^\textsc{t}$ and $\bm{O}^\textsc{p}$ are called time-reversal and parity-reversal operators in physics, respectively.

A \emph{space-time curvature} is defined owing to the structure equation as
\begin{align}
\RRR^{ab}&:=d\www^{ab}+\cG\www^a_{~\bcdot}\wedge\www^{\bcdot b}
\in V^2(\TML)\otimes\Omega^2(\TsML){\otimes}Ad(\ggg^{~}_{\SO(1,3)}),\label{RRR}
\end{align}
which is a two-form valued rank-$2$ tensor  represented using the standard basis as
\begin{align*}
\RRR^{ab}&=
\sum_{c<d}R^{ab}_{\hspace{.7em}cd}\hspace{.1em}\eee^{c}\wedge\eee^{d}=
\frac{1}{2}R^{ab}_{\hspace{.7em}\bcdots}\hspace{.1em}\eee^\bcdot\wedge\eee^\bcdot.
\end{align*}
Tensor coefficient $R^{ab}_{\hspace{.7em}cd}$ is referred to as the \emph{Riemann curvature} tensor.
\emph{Ricci curvature} tensor and \emph{scalar curvature} are defined, respectively, owing to the Riemann curvature tensor as
\begin{align}
R_{ab}:=\eta_{\textsc{l}\hspace{.1em}a\star}\hspace{.1em}\Ri^{\bcdot\star}_{\hspace{.7em}\bcdot b}
~~\textrm{and }~~
R:=\Ri^{\bcdot\star}_{\hspace{.7em}\bcdot\star}.\label{RicciR}
\end{align}
The first and second Bianchi identities are
\begin{align}
d_\www(d_\www\eee^a)=\cG\hspace{.2em}
\eta^{~}_{\textsc{l}\hspace{.1em}\bcdots}\hspace{.1em}\RRR^{a\bcdot}\wedge\eee^{\bcdot} ~~\mathrm{and}~~
d_\www\hspace{.1em}\RRR^{ab}=0.\label{RBianchi}
\end{align}

We also introduce a Riemannian manifold with Euclidean metric $\betaE:=\text{diag[}1,1,1,1]$, denoted as $\ME$.
For objects belonging to the Euclidean space, we put the subscript $\bullet^{~}_\textsc{e}$ instead of $\bullet^{~}_\textsc{l}$. E.g., the bilinear form in the Euclidean manifold is 
\begin{align*}
\v,\u{\hspace{.1em}\in\hspace{.1em}}V(\TME),~~\<\u,\v\>^{~}_\textsc{e}=\u^t\ncdot{\betaE}\ncdot\v
=\etaE_\bcdots\hspace{.1em}u^\bcdot v^\bcdot=\<\v,\u\>^{~}_\textsc{e}\in\R^+,
\end{align*}
which is positive definite for non-zero vectors and non-degenerate, such as $\|\u\|^{2}_\textsc{e}:=\<\u,\u\>^{~}_\textsc{e}=0\iff\u=0$.
The volume form of the Euclidean space is provided as
\begin{align*}
(\ref{VolForm})\big|_{\textsc{l}\rightarrow\textsc{e}}\implies\vvv_\textsc{e}&
{:=}\mathrm{det}[\bm\Varepsilon_\textsc{e}]\hspace{.2em}dx^0\wedge dx^1\wedge dx^2\wedge dx^3
~~\text{with}~~
\textrm{det}[\bm{g}^{~}_\textsc{e}]=\textrm{det}[\betaE]\textrm{det}[\bm{\Varepsilon}]^2=\textrm{det}[\bm{\Varepsilon}]^2>0;
\intertext{thus,}
\vvv_\textsc{e}&=\sqrt{\mathrm{det}[\bm{g}^{~}_\textsc{e}]}\hspace{.2em}dx^0{\wedge}dx^1{\wedge}dx^2{\wedge}dx^3.
\end{align*}

%
%
\subsubsection{co-Poincar\'{e} bundle}
The Yang--Mills theory in the Lorentzian metric space has the Poincar\'{e} symmetry in addition to the gauge symmetry.
On the other hand, general relativity is not invariant under the four-dimensional translation.
The current author has introduced a modified Poincar\'{e} symmetry, keeping general relativity invariant, namely the co-Poincar\'{e} symmetry\cite{doi:10.1063/1.4990708,Kurihara_2020}.
This section introduces a principal bundle with the co-Poincar\'{e} group as the structure group.

The Poincar\'{e} group is the semidirect product of Lorentz group and the four-dimensional translation group as
\begin{align*}
I\hspace{-.15em}S\hspace{-.1em}O(1,3)=SO(1,3)\ltimes T^4,
\end{align*}
whose Lie algebra is given as 
\begin{align}
\left[P_{a},P_{b}\right]&=0,\notag\\
\left[J_{ab},P_{c}\right]&=-\eta_{ac}\hspace{.1em}P_{b}+\eta_{bc}\hspace{.1em}P_{a},\label{JJ}\\
\left[J_{ab},J_{cd}\right]&=
-\eta_{ac}\hspace{.1em}J_{bd}+\eta_{bc}\hspace{.1em}J_{ad}
-\eta_{bd}\hspace{.1em}J_{ac}+\eta_{ad}\hspace{.1em}J_{bc},\notag
\end{align}
where $P_a$ and $J_{ab}$ are generators of the $T^4$ group and the $\SO(1,3)$ group, respectively.
Although the Yang--Mills Lagrangian is invariant under the Poincar\'{e} group, the Einstein--Hilbert gravitational Lagrangian does not respect the symmetry\cite{0264-9381-29-13-133001}, obstructing quantisation of general relativity.
The current author has discovered a novel symmetry named the co-Poincar\'{e} symmetry\cite{doi:10.1063/1.4990708} which allows the construction of general relativity in four-dimensional space-time and defined the invariant quadratic in four-dimensional space-time owing to the co-Poincar\'{e} group\cite{doi:10.1140/epjp/s13360-021-01463-3}.

The co-Poincar\'{e} group, denoted as $\GcP$, is the Poincar\'{e} group that replaced the translation operator with the co-translation operator.
A principal co-Poincar\'{e} bundle is defined as a tuple 
\begin{align}
\left(\MM\otimes\ML,\pi^{~}_{\textsc{i}},\MM,\GcP\right),
\end{align}
which is the same as the inertial bundle except the structure group $\GcP$.
A generator of the co-translation is defined as $P_{ab}:=P_a\iota_b$, where $\iota_b$ is a contraction concerning trivial frame field $\partial_b$.
The generator of the co-Poincar\'{e} group can be represented by two ($\FxF$)-matrices as
\begin{subequations}
\begin{align}
\left[\Theta_I\right]_{ab}:=
\begin{cases}
P_{ab},&I=1\\
J_{ab},&I=2
\end{cases}\!,
\end{align}
whose the Lie algebra are provided as
\begin{align}
\left[P_{ab},P_{cd}\right]&=0,~~~
\left[J_{ab},P_{cd}\right]=-\eta_{ac}\hspace{.1em}P_{bd}+\eta_{bc}\hspace{.1em}P_{ad}
~~\text{and}~~
 \left[J_{ab},J_{cd}\right]=\text{(\ref{JJ})}.\label{coPoincare1}
\end{align}
\end{subequations}
We denote the structure constants of the co-Poincar\'{e} group obtained from the above Lie algebra as
\begin{subequations}
\begin{align}
&[\Theta_I,\Theta_J]:=\FF_{~IJ}^{K}\hspace{.1em}\Theta_K,\label{coPoincare2}
\intertext{yielding}
&\left\{
\begin{array}{l}
\FF_{~11}^{1}=\FF_{~11}^{2}=\FF_{~12}^{2}=
\FF_{~21}^{2}=\FF_{~22}^{1}=0,\\
\left[\FF_{~12}^{1}\right]_{ab;cd}^{ef}=
-\left[\FF_{~21}^{1}\right]_{cd;ab}^{ef}=
\eta_{ac}\hspace{.1em}\delta^e_b\delta^f_d-\eta_{bc}\hspace{.1em}\delta^e_a\delta^f_d,\\
\left[\FF_{~22}^{2}\right]_{ab;cd}^{ef}=
-\eta_{ac}\hspace{.1em}\delta_b^e\delta_d^f+\eta_{bc}\hspace{.1em}\delta_a^e\delta_d^f
-\eta_{bd}\hspace{.1em}\delta_a^e\delta_c^f+\eta_{ad}\hspace{.1em}\delta_b^e\delta_c^f.
\end{array}
\right.\label{FgrLie}
\end{align}
\end{subequations}

Co-Poincar\'{e} connection form $\AAA_\cP$ and curvature form $\FFF_\cP$ are, respectively, introduced as Lie-algebra valued one- and two-form objects concerning the co-Poincar\'{e} group, and they are expressed using the trivial basis as
\begin{subequations}
\begin{align}
\AAA^{I}_\cP&=
\begin{cases}
J_\bcdots\otimes\www^\bcdots,&\hspace{1.2em}I=1,\\
P_\bcdots\otimes\SSS^\bcdots,&\hspace{1.2em}I=2,
\end{cases}
\hspace{2em}\in\Wedge^1(\TsML)\otimes Ad(\ggg_\cP),\label{cPLieA1}\\
\FFF^{I}_\cP&=
\begin{cases}
J_\bcdots\otimes\RRR^\bcdots,&I=1,\\
P_\bcdots\otimes d_\www\SSS^\bcdots,&I=2,
\end{cases}
\hspace{2em}\in\Wedge^2(\TsML)\otimes Ad(\ggg_\cP),\label{cPLieA2}
\end{align}
where $\ggg_\cP$ is a Lie algebra of the co-Poincar\'{e} symmetry.
\end{subequations}
We note that $P^\bcdots\otimes d_\www\SSS_\bcdots\in\Omega^2(\TsML)$ due to $\iota_\bullet:\Omega^p\rightarrow\Omega^{p-1}$, e.g., $\iota_a\eee^b=\delta^b_a$.

%% file: Sec2-4.tex
%
%
\subsection{Hodge-dual operator}\label{HDO}
A Hodge-dual operator is an indispensable theoretical tool for describing the Yang--Mills theory.
One natural gauge-invariant equation of motion for the gauge field is, e.g., $d_G\FFF^{~}_G=\JJJ$, where $\JJJ$ is a three-form object owing to the source field.
However, the left-hand side is identically zero owing to the Bianchi identity.
We cannot exploit it as an equation of motion.
In reality, the gauge-invariant equation of motion is $\HD(d_G\HD(\FFF^{~}_G))=\JJJ$, where  $\HD$ is the Hodge-dual operator.
Together with the structure equation, the equation of motion governs the system's dynamics.
The Hodge-dual operator induces dynamics into the Yang--Mills theory while keeping the gauge invariance of the equation of motion.
 
 Another role of the Hodge-dual operator is introducing a $\Z_2$-grading structure for the gauge fields.
Since the Hodge-dual operator maps a $p$-form object to an $(n\hspace{-.15em}-\hspace{-.15em}p)$-form object in an $n$-dimensional smooth Riemannian manifold, the Hodge-dual operator is an endomorphism for $n$-form objects in $2n$-dimensional manifolds. 
It induces the $\Z_2$-grading vector space.
This section discusses the $\Z_2$-grading vector space owing to the Hodge-dual operator in the inertial manifold with the Euclidean and the Lorentzian metric tensors.

\subsubsection{Definition and properties}
Suppose $p$-form objects $\aaa,\bbb$ have coefficient representations such that: 
\begin{align*}
\aaa&:=\frac{1}{p!}{a}_{i_1\cdots i_p}\eee^{i_1}\wedge\cdots\wedge\eee^{i_p}
\hspace{.2em}\in\Wedge^p(\TsMB),~~
\bbb:=\frac{1}{p!}{b}_{i_1\cdots i_p}\eee^{i_1}\wedge\cdots\wedge\eee^{i_p}
\hspace{.2em}\in\Wedge^p(\TsMB),\quad
(0{\leq}i_j{\leq}n-1).
\end{align*}
The $SO(4)$ or $\SO(1,3)$ invariant bilinear form is defined using the metric tensor $\betaB$ $(\bullet\in\{\textsc{e},\textsc{l}\})$ as
\begin{align*}
\<\aaa,\bbb\>^{~}_\bullet:=\frac{1}{p!}\hspace{.1em}
\eta_\bullet^{\hspace{.2em}i_1j_1}\cdots\eta_\bullet^{\hspace{.2em}i_pj_p}\hspace{.1em}
a^{~}_{i_1{\cdots}i_p}\hspace{.1em}b^{~}_{j_{1}{\cdots}j_p}=\<\bbb,\aaa\>^{~}_\bullet\in\R,
\end{align*}
which is consistent with the inner product in $\TML$ given by (\ref{TMinner}).
A (pseudo-)norm squared of $\aaa$ is given as
\begin{align}
\|\aaa\|_\bullet^{2}&:=\<\aaa,\aaa\>^{~}_\bullet,\label{pnormsq}
\end{align}
which is non-degenerate and positive-definite for $\aaa\neq0$ in the Euclidean space. 
We introduce the Hodge-dual operator in this space:
\begin{definition}[Hodge-dual operator]\label{ctm}
Suppose $\MB$ is an $n$-dimensional oriented and smooth manifold and $\aaa,\bbb\in\Wedge^p(\TsMB)$ are $p$-form objects.
The Hodge-dual operator, denoted as $\HD$, is defined to give
\begin{align}
\aaa\wedge\HDB(\bbb):=
\mathrm{det}[\betaB]^\hlf\hspace{.1em}\<\aaa,\bbb\>^{~}_\bullet\hspace{.1em}\vvv^{~}_\bullet,~~\bullet\in\{\textsc{e},\textsc{l}\}.\label{HodgeD}
\end{align}
\end{definition}
\noindent
Our definition has the extra factor $\mathrm{det}[\betaB]^\hlf$ over the standard definition for later convenience.
This factor works to unify Euclidean and Lorentzian Hodge-dual operators in \textbf{Section \ref{AIndex}}.
Since $\betaB$ is the constant tensor, the Hodge-dual operator is invariant under the $GL(4)$ group.  

The Hodge-dual operator has a component representation as
\begin{align*}
\HDB:\Omega^p\rightarrow\Omega^{n-p}:\bbb\mapsto\hat{\bbb}:=\HDB(\bbb)
=\frac{\text{det}[\betaB]^\hlf}{p!(n-p)!}{b}_{i_1{\cdots}i_{p}}
\left[\bm\epsilon\right]^{i_1{\cdots}i_{p}}_{\hspace{2.4em}i_{p+1}{\cdots}i_{n}}
\eee^{i_{p+1}}\wedge\cdots\wedge\eee^{i_{n}},
\end{align*}
where 
\begin{align*}
\left[\bm\epsilon\right]^{i_1{\cdots}i_{p}}_{\hspace{2.4em}i_{p+1}{\cdots}i_{n}}
:=\frac{1}{p!}\eta_\bullet^{\hspace{.2em}i_1j_1}{\cdots}\eta_\bullet^{\hspace{.2em}i_pj_p}\hspace{.1em}
\epsilon_{j_1\cdots j_{p}i_{p+1}{\cdots}i_{n}},
\end{align*}
and $\bm\epsilon$ is a $n$-dimensional completely anti-symmetric tensor with $[\bm\epsilon]_{01\cdots n-1}=+1$. 
The operator acts on tensor coefficients of $\bbb$  as
\begin{align*}
\HDB:{b}_{i_1{\cdots}i_p}\mapsto&\hat{b}_{i_{p+1}{\cdots}i_{n}}
:=\frac{\mathrm{det}[\betaB]^\hlf}{p!}{b}_{i_{1}{\cdots}i_{p}}
\left[\bm\epsilon\right]^{i_{1}{\cdots}i_{p}}_{\hspace{2.4em}i_{p+1}{\cdots}i_{n}},
\intertext{with}
&\hat{\bbb}=\frac{1}{(n-p)!}\hat{b}_{i_1{\cdots}i_{n-p}}\hspace{.1em}
\eee^{i_1}\wedge\cdots\wedge\eee^{i_{n-p}}.
\end{align*}
We obtain the Hodge-dual inverse as
\begin{align}
\HDB\(\HDB(\aaa)\)&=\text{det}[\betaB](-1)^{p(n-p)}\aaa\implies
\HDBi(\aaa)=\text{det}[\betaB]^{-1}(-1)^{p(n-p)}{\HDB}(\aaa).\label{DHi}
\end{align}
The Hodge-dual operator relates with the volume form as
\begin{align*}
\HDB(1)&=\text{det}[\betaB]^\hlf\hspace{.1em}\eee^0\wedge\cdots\wedge\eee^{n-1}=
\text{det}[\betaB]^\hlf\hspace{.1em}\vvv=
\text{det}[\betaB]\mathrm{det}[\bm\Varepsilon]\hspace{.2em}dx^0\wedge dx^1\wedge dx^2\wedge dx^3.
\end{align*}

We define a co-differential operator ${\cod}$ for $p$-form object $\aaa$ in an $n$-dimensional oriented manifold as
\begin{align}
{\cod}:\Omega^p(\TsMB)\rightarrow\Omega^{p-1}(\TsMB):\aaa\mapsto{\cod}\aaa
:=&(-1)^{p}\hspace{.2em}\HDBi\hspace{-.2em}\(d\HDB(\aaa)\)=
\text{det}[\betaB]^{-1}(-1)^{p(n-p+1)}\hspace{.2em}\HDB\(d\HDB(\aaa)\),\label{codiff}
\end{align}
where we use (\ref{DHi}).
The co-differential is nilpotent like ${\cod}\hspace{.1em}{\cod}=0$ as well as the external differential.
We note that the Hodge-dual operator depends on the metric tensor; thus, the co-differential also does.
We omit subscript $d_\bullet$ ($\hat{d}_\bullet$) for simplicity.
The co-differential operator is a {weak}-adjoint of external differential in the compact manifold $\MB$ with $\aaa\in\Omega^{p-1}(\TsMB)$ and $\bbb\in\Omega^{p}(\TsMB)$ such as 
\begin{align*}
\int\<{d\aaa,\bbb}\>^{~}_\bullet\vvv^{~}_\bullet&=\text{det}[\betaB]^{-\hlf}\hspace{.1em}\int d\aaa\wedge\HDB(\bbb)=\text{det}[\betaB]^{-\hlf}\hspace{.1em}
(-1)^p\int\aaa{\wedge}d\HDB(\bbb),\notag\\&=
\text{det}[\betaB]^{-\hlf}\hspace{.1em}\int\aaa{\wedge}\HDB\(\(-1\)^p\HDBi\hspace{-.2em}
\(d\HDB\(\bbb\)\)\)
=\int\<{\aaa,{\cod}\bbb}\>^{~}_\bullet\vvv^{~}_\bullet.
\end{align*}
The Storks theorem and an assumption that $\MQ$ is compact are used here.

The Laplace--Beltrami operator is defined owing to the co-differential as
\begin{align}
\Delta^{~}_\bullet&:=d\hspace{.1em}{\cod}+{\cod}\hspace{.1em}d,\label{LBop}
\end{align}
which is a linear endomorphism on $\Omega^p(\TsMQ)$ and is commutable with the Hodge-dual operator.
We note that it is self {weak}-adjoint like 
\begin{align*}
\<\Delta^{~}_\bullet\aaa,\bbb\>^{~}_\bullet=\<d\cod\aaa,\bbb\>^{~}_\bullet+\<{\cod}d\aaa,\bbb\>^{~}_\bullet
&=\<\aaa,d{\cod}\bbb\>^{~}_\bullet+\<\aaa,{\cod}d\bbb\>^{~}_\bullet
=\<\aaa,\Delta^{~}_\bullet\bbb\>^{~}_\bullet.
\end{align*}

Suppose $n=2m$ ($m\in\Z$) is an even integer.  
The Hodge-dual operator is endomorphism on the $(m)$-form object since $\HDB(\aaa\in\Omega^{m})\in\Omega^{m}$.
For $\aaa{\in}\Omega^{m}(\TsM)$, the Hodge-dual operator fulfils
\begin{align*}
\text{(\ref{DHi})}\implies\HDB\(\HDB(\aaa)\)&=(-1)^{m^2}\text{det}[\betaB]\hspace{.1em}\aaa\in\Omega^p(\TsM);
\end{align*}
thus, $\HDB$ has two-eigenvalues of $\pm1$ or $\pm i$ depend on a value of $n$ and a metric.
An orthonormal basis of $m$-form objects is constructed using the standard basis as
\begin{align*}
\widetilde\SSS^I:=\eee^{i_{p(1}}\wedge\cdots\wedge\eee^{i_{m)}},~~\textrm{with}~~I=1,\cdots,l:=\frac{n!}{(m!)^2},
\end{align*}
where $p(1\cdots m)$ runs over all possible permutations.
We note that $l$ is even integer for any even $n$.
Owing to the definition of the Hodge-dual operator, we can always take $(l/2)$-pairs of bases $(\widetilde\SSS^I,\widetilde\SSS^J)$ fulfiling
\begin{align*}
\HDB(\widetilde\SSS^I)=(-1)^{m}\hspace{.1em}\widetilde\SSS^J~~\textrm{and}~~
\HDB(\widetilde\SSS^J)=\widetilde\SSS^I.
\end{align*}
Thus, we can construct new bases using pairs $(\widetilde\SSS^I,\widetilde\SSS^J)$ as
\begin{align}
\widetilde\SSS^\pm:=\widetilde\SSS^I\pm(-1)^{m}\hspace{.1em}\widetilde\SSS^J,\label{SISJ}
\end{align}
yielding 
\begin{align*}
\HDB\left(\widetilde\SSS^+\right)&=
(-1)^{m}\left(\widetilde\SSS^J+\widetilde\SSS^I\right)=
+(-1)^{m}\widetilde\SSS^+\in\VH+(\Omega^{m}),\\
\HDB\left(\widetilde\SSS^-\right)&=
(-1)^{m}\left(\widetilde\SSS^J-\widetilde\SSS^I\right)=
-(-1)^{m}\widetilde\SSS^-\in\VH-(\Omega^{m}).
\end{align*}
The $m$-form objects $\widetilde\SSS^\pm$ are linearly independent from one another, and any $m$-form objects can be represented using a linear combination of these forms.
Therefore, a space of $m$-form objects splits into two subspaces concerning eigenvalues of the Hodge-dual operator like 
\begin{align}
\VV(\Omega^{m})=\VV_{~}^+(\Omega^{m})\oplus\VV_{~}^-(\Omega^{m}).\label{superspace} 
\end{align}
Consequently, we confirm that the Hodge-dual operator induces the $\Z_2$-grading vector space in $\Omega^{m}$.


So far, in this section, we have discussed the Hodge-dual operator in the general $n$-dimensional manifold.
Hereafter, we restrict a space-time manifold to four-dimensional space equipping the Euclidean or the Lorentzian metric.
Projection operators owing to the Hodge-dual operator are defined as
\begin{align}
\PH{\pm}:=\frac{1}{2}\left(1\pm\HDB\right),\label{projHodge}
\end{align}
which fulfils the definition of the projection operator $\PH{\pm}\hspace{.1em}\PH{\pm}=\PH{\pm}$ and $\PH{\pm}\hspace{.1em}\PH{\mp}=0$.
This operator projects a two-form object onto one of two eigenspaces such that 
\begin{align}
\PH{\pm}\aaa&=\aaa^\pm
\implies
\HDB(\aaa^\pm)=\pm\aaa^\pm\in\VH{\pm}\(\Omega^{2}_{~}\(\TsMB\)\).\label{Hapm}
\end{align}
The two-form object belonging to $\VH{+}$ ($\VH{-}$) is referred to as the self-dual (SD) (the anti-self-dual (ASD)) forms, respectively.
We also call the sign of the eigenvalue as the \emph{chirality} owing to the analogy with the spinor space.
We assign chiral-right to SD-forms and chiral-left to ASD-forms.

\subsubsection{two-form bases of $\Z_2$-grading vector space}
\paragraph{Euclidean metric:}
For the standard one-form bases $\eeeE^a{\in}\Omega^1(\TsME)$ $(a=0,1,2,3)$, the Hodge-dual operator on two-form objects provides that
\begin{subequations}
\begin{align}
&\HDE(\eeeE^0\wedge\eeeE^1)=\eeeE^2\wedge\eeeE^3,~~~
\HDE(\eeeE^0\wedge\eeeE^2)=-\eeeE^1\wedge\eeeE^3,~~~
\HDE(\eeeE^0\wedge\eeeE^3)=\eeeE^1\wedge\eeeE^2,\notag\\
&\HDE(\eeeE^1\wedge\eeeE^2)=\eeeE^0\wedge\eeeE^3,~~~
\HDE(\eeeE^1\wedge\eeeE^3)=-\eeeE^0\wedge\eeeE^2,~~~
\HDE(\eeeE^2\wedge\eeeE^3)=\eeeE^0\wedge\eeeE^1.\notag
\intertext{Thus, we obtain the $\Z_2$-grading two-form bases as, e.g., }
&\SSS^{+\hspace{.1em}a}_\textsc{e}:=\left\{
\eeeE^0\wedge\eeeE^1+\eeeE^2\wedge\eeeE^3,\hspace{.2em}
\eeeE^0\wedge\eeeE^2-\eeeE^1\wedge\eeeE^3,\hspace{.2em}
\eeeE^0\wedge\eeeE^3+\eeeE^1\wedge\eeeE^2\right\},\label{SpE}\\
&\SSS^{-\hspace{.1em}a}_\textsc{e}:=\left\{
\eeeE^0\wedge\eeeE^1-\eeeE^2\wedge\eeeE^3,\hspace{.2em}
\eeeE^0\wedge\eeeE^2+\eeeE^1\wedge\eeeE^3,\hspace{.2em}
\eeeE^0\wedge\eeeE^3-\eeeE^1\wedge\eeeE^2\right\},\label{SmE}
\end{align}
\end{subequations}
yielding
\begin{align}
\HDE\(\SSS^{\pm}_\textsc{e}\)=\pm\SSS^{\pm}_\textsc{e}{\hspace{.1em}\in\hspace{.1em}}\VH{\pm},~~
\PH{\pm}\(\SSS^{\pm}_\textsc{e}\)=\SSS^{\pm}_\textsc{e},~~\text{and}~~
\PH{\pm}\(\SSS^{\mp}_\textsc{e}\)=0.\label{DHES}
\end{align}

\paragraph{Lorentzian metric:}
We can introduce the $\Z_2$-grading two-form bases in the Lorentzian metric space as the same as it in the Euclidean space. 
Using the orthonormal bases $\eeeL^a{\in}\Omega^1(\TsML)$ $(a=0,1,2,3)$ in the Lorentzian metric space, we provide the Hodge-dual operator on two-form objects as
\begin{align*}
&\HDL(\eeeL^0\wedge\eeeL^1)=\hspace{.9em}i\hspace{.1em}\eeeL^2\wedge\eeeL^3,~~~
\HDL(\eeeL^0\wedge\eeeL^2)=-i\hspace{.1em}\eeeL^1\wedge\eeeL^3,~~~
\HDL(\eeeL^0\wedge\eeeL^3)=\hspace{.9em}i\hspace{.1em}\eeeL^1\wedge\eeeL^2,\\
&\HDL(\eeeL^1\wedge\eeeL^2)=-i\hspace{.1em}\eeeL^0\wedge\eeeL^3,~~~
\HDL(\eeeL^1\wedge\eeeL^3)=\hspace{.9em}i\hspace{.1em}\eeeL^0\wedge\eeeL^2,~~~
\HDL(\eeeL^2\wedge\eeeL^3)=-i\hspace{.1em}\eeeL^0\wedge\eeeL^1,
\end{align*}
where imaginary unit ``$i$'' appears owing to factor det$[\betaL]$ in the definition.
The $\Z_2$-grading two-form bases are provided as, e.g.,
\begin{subequations}
\begin{align}
&\SSS^{+\hspace{.1em}a}_\textsc{l}:=\left\{
\eeeL^0\wedge\eeeL^1+i\hspace{.1em}\eeeL^2\wedge\eeeL^3,\hspace{.2em}
\eeeL^0\wedge\eeeL^2-i\hspace{.1em}\eeeL^1\wedge\eeeL^3,\hspace{.2em}
\eeeL^0\wedge\eeeL^3+i\hspace{.1em}\eeeL^1\wedge\eeeL^2\right\},\label{SpL}\\
&\SSS^{-\hspace{.1em}a}_\textsc{l}:=\left\{
\eeeL^0\wedge\eeeL^1-i\hspace{.1em}\eeeL^2\wedge\eeeL^3,\hspace{.2em}
\eeeL^0\wedge\eeeL^2+i\hspace{.1em}\eeeL^1\wedge\eeeL^3,\hspace{.2em}
\eeeL^0\wedge\eeeL^3-i\hspace{.1em}\eeeL^1\wedge\eeeL^2\right\},\label{SmL}
\end{align}
\end{subequations}
yielding $(\ref{DHES})$ with ``$\textsc{e}$''$\rightarrow$``$\textsc{l}$''.
The Hodge-dual operator induces the $\Z_2$-grading spaces for curvature two-form $\FFFL$ in the Lorentzian metric space, namely SD and ASD curvatures, as
\begin{align*}
\FFFL^\pm:=\PH{\pm}\hspace{.1em}\FFFL
\implies
\HDL(\FFFL^\pm)=\pm\FFFL^\pm
\in \VH{\pm}\left(\OmegaL^2\right),
\end{align*}
where $\FFFL^\pm$ is, respectively,  spanned by $\widetilde\SSS^\pm_\textsc{l}$.

%% file: Sec2-5.tex
%
%
\subsection{Group representations}\label{SRVS}
The special principle of relativity requires that representations of the space-time symmetry group present the classical field in the local inertial space.
The Yang-Mills theory involves the space-time symmetry group and the $\SU(\!N\!)$ gauge group.
These groups have homomorphism among various Lie groups.
E.g., four-dimensional vector spaces with Euclidean and Lorentzian metrics, respectively, have accidental Lie-algebra homomorphism as
\begin{subequations}
\begin{align}
\sss\ooo(4)&=\sss\uuu(2)\oplus\sss\uuu(2),\label{ismorSO4}\\
\sss\ooo(1,3)\otimes\C&=\sss\lll(2,\C)\oplus\overline{\sss\lll(2,\C)}.\label{ismorSO13}
\intertext{The subspaces of them have}
\sss\ooo(3)&=\sss\uuu(2),\label{ismor20}\\
\sss\ooo(3)\otimes\C&=\sss\uuu(2)\otimes\C=\sss\lll(2,\C).\label{ismor2}
\end{align}
\end{subequations}
According to these homomorphisms, four-dimensional vectors in the Euclidean and Lorentzian metric spaces have complex valued ($\TxT$)-matrix representations.

Spinor spaces  have following group homomorphisms:
\begin{subequations}
\begin{align}
Spin(4)&=SU(2)\times SU(2)~~\text{and}~~Spin(1,3)=\SL2C,\label{Spin4SU2SU2}
\end{align}
inducing corresponding Lie algebra homomorphisms.
The spin group doubly covers the rotation group as
\begin{align}
Spin(4)=SO(4)\times_{\Z_2}U(1)~~\text{and}~~
Spin(1,3)=SO(1,3)\times_{\Z_2}U(1).\label{Spin4SU2SU2-2}
\end{align}
\end{subequations}
This section introduces irreducible representations of ${SO(4)}$ and ${SO(1,3)}$, according to Refs.\cite{talagrand_2022,woit2017quantum}.
Owing to abovementioned homomorphisms, irreducible representations of ${SO(4)}$ and ${SO(1,3)}$ groups can be constructed using those of $SU(2)$ and $\SL2C$, respectively.
We introduce $SU(2)$ and $\SL2C$ irreducible representations below.

\subsubsection{Irreducible representations}\label{RPs}
\paragraph{a) Euclidean metric space:}
We introduce a space of two complex numbers on the unit three-sphere,
\begin{align}
S^3&:=\left\{z_1,z_2\in\C\hspace{.1em}\big|\hspace{.1em}|z_1|^2+|z_2|^2=1\right\}.\label{S3}
\end{align}
Each point on $S^3$ induces an $SU(2)$ matrix  
\begin{align*}
\bm{M}&:=\(
\begin{array}{rr}
z_1 & -z_2\\
z_2^* & z_1^*
\end{array}
\)
~~\text{yielding}~~
\text{det}[\bm{M}]=1,~~
\bm{M}^\dagger\ncdot\bm{M}=\bm{1}_2,
\end{align*}
where $\bm{1}_2$ is a unit ($\TxT$)-matrix.
We define a $SU(2)$-spinor as a polynomial of the form
\begin{align}
\Vs{\textsc{e}}\ni
\lambda_{s}^{~}(z_1,z_2)&=\sum_{0\le k\le 2s}c_k z_1^{k}z_2^{2s-k},~~\text{where}~~c_k\in\R,~~\{z_1,z_2\}\in S^3,\label{lambda}
\end{align}
where $\Vs{\textsc{e}}$ is a space of $\SU(2)$ spinors.
Owing to homomorphism (\ref{ismorSO4}), we call the $\SU(2)$ spinor the \emph{Euclidean} spinor and put the suffix ``\textsc{e}'' on one with the $\SU(2)$ symmetry.
A parameter $s$ takes a positive half-integer, namely a \emph{spin} of spinor $\lambda_{s}^{~}$.

For group action $\gB{\textsc{e}}{\hspace{.1em}\in\hspace{.1em}}SU(2)$, its representation $\pij{s}\(\gB{\textsc{e}}\)$ acts on the spinor $\lambda_{s}^{~}(z_1,z_2)$ as
\begin{align}
\(\pij{s}(\gB{\textsc{e}})\bcdot\lambda_{s}^{~}\)(z_1,z_2)=\lambda_{s}^{~}(z'_1,z'_2),~~~
\text{with}~~~
\(\!
     \begin{array}{c}
      z'_1\\
      z'_2
     \end{array}
\!\)=\g^{-1}_\textsc{e}
\(\!
     \begin{array}{c}
      z^{~}_1\\
      z^{~}_2
     \end{array}
\!\).\label{gEz1z2}
\end{align}
The complex column vector in (\ref{gEz1z2}) represents the Euclidean spinor.
The Lie-algebra representation is provided by the derivative of the group representation such as
\begin{align*}
\dpij{s}\(\ggg^{~}_\textsc{e}\)\bcdot\lambda_{s}^{~}&:=
\left.\frac{d}{dt}\lambda_{s}^{~}(e^{t\ggg}\hspace{.1em}z_1,e^{t\ggg}\hspace{.1em}z_2)\right|_{t=0},
\end{align*}
where $\ggg^{~}_\textsc{e}\in\sss\uuu(2)$ is any ($\TxT$) unitary-matrix.

We define an $\SU(2)$-invariant inner product for spinors as the Gaussian integral such that
\begin{align}
\<\lambda'_{s},\lambda_{s}^{~}\>&:=\frac{1}{\pi^2}\int_{\C^2}{\lambda'_{s}}^*\lambda_{s}^{~}
e^{-(|z_1|^2+|z_2|^2)}dx^{~}_1\hspace{.1em}dy^{~}_1\hspace{.1em}dx^{~}_2\hspace{.1em}dy^{~}_2,~~z^{~}_a:=x^{~}_a+iy^{~}_a<\infty,\label{inner}
\intertext{yielding the norm}
 \|\lambda_{s}^{~}\|&:=\<\lambda_{s}^{~},\lambda_{s}^{~}\>^\hlf.\label{Enorm}
\end{align}

Representation $\pij{s}$ is the irreducible, labelled by the half-integer $s$.
Each spinor $\lambda_{s}^{~}$ consists of $2s+1$ monomials corresponding to the spin projection on the specific axis in physics.
When we set monomial's coefficient $c^{~}_k$ for $\lambda_{s}^{~}$ in (\ref{lambda})  as 
\begin{align*}
c^{~}_k =\(k!(2s-k)!\)^{-\hlf},
\end{align*}
monomials $c^{~}_k z^{k}_1z^{2s-k}_2$ compose the orthonormal basis $\{\lambda_{s}^{~}\}$ by means of (\ref{inner}).

When $s=\hlf$, we have a matrix representation as follows:
The $SU(2)$-spinor at $\bm{z}:=(z_1,z_2)^t$ is represented using  the coefficient vector $\tilde{\bm{c}}:=(\tilde{c}_1,\tilde{c}_2)^t$ as
\begin{align*}
\lambda_{\hlf}^{~}(z_1,z_2)&=\tilde{\bm{c}}^t\ncdot\bm{z}=
(\tilde{c}_1,\tilde{c}_2)
\(\!
     \begin{array}{c}
      z_1\\
      z_2
     \end{array}
\!\)
=\tilde{c}_1z_1+\tilde{c}_2z_2,
\end{align*}
where $\tilde{c}_i$ is relabeling of $c_i$ as $(c_0,c_1)\rightarrow(\tilde{c}_2,\tilde{c}_1)$.
Accordingly, the group action is written as
\begin{align*}
\(\pij{\hlf}(\gB{\textsc{e}})\bcdot\lambda_{\hlf}^{~}\)(z_1,z_2)=
\lambda_{\hlf}^{~}(z'_1,z'_2)=
\tilde{\bm{c}}^t\ncdot\(\g^\dag_\textsc{e}\ncdot\bm{z}\)=
\(\gB{\textsc{e}}\ncdot\tilde{\bm{c}}\)^\dag\ncdot\bm{z}.
\end{align*}
Therefore, we obtain $\pij{\hlf}(\gB{\textsc{e}})=\gB{\textsc{e}}$; thus, $\dpij{\hlf}(\ggg^{~}_{E})=\ggg^{~}_{E}$.

The irreducible unitary representation of $Spin(4)\hspace{-.1em}=\hspace{-.1em}SU(2){\times}SU(2)$ consists of a tensor product 
\begin{align}
\pij{s^{~}_{\hspace{-.1em}1}\hspace{-.2em}{\otimes}s^{~}_{\hspace{-.1em}2}}^{\textsc{e}}&:=
\(
\pi_{\hspace{-.1em}s^{~}_{\hspace{-.1em}1}}\!(\gB{\textsc{e}})
\hspace{.1em}{\otimes}\hspace{.1em}
\pi_{\hspace{-.1em}s^{~}_{\hspace{-.1em}2}}\!(\gB{\textsc{e}}),
\VV^{\textsc{e}}_{\hspace{-.2em}s^{~}_{\hspace{-.1em}1}}{\otimes}\VV^{\textsc{e}}_{\hspace{-.2em}s^{~}_{\hspace{-.1em}2}}
\) ~~\text{with}~~s^{~}_{\hspace{-.1em}1},s^{~}_{\hspace{-.1em}2}\hspace{-.1em}=
\hspace{-.1em}0,\frac{1}{2},1,\frac{3}{2},2,\cdots, \label{piElr}
\end{align}
yielding the Lie-algebra representation of $\sss\ooo(4)$ as $\dpij{s^{~}_{\hspace{-.1em}1}\hspace{-.2em}{\otimes}s^{~}_{\hspace{-.1em}2}}^{\textsc{e}}:=\dot\pi_{s^{~}_{\hspace{-.1em}1}}{\otimes}\hspace{.2em}\dot\pi_{s^{~}_{\hspace{-.1em}2}}$, owing to homomorphism (\ref{ismorSO4}).

\paragraph{b) Lorentzian metric space:}
Similarly, we can label $\SL2C$ spinors as the complex $SU(2)$ representation owing to the Lie-algebra homomorphism of (\ref{ismor2}).
$\SL2C$ has two fundamental (non-unitary) complex-representations for $\gB{\textsc{l}}{\hspace{.1em}\in\hspace{.1em}}\SL2C$ as
\begin{align}
\pij{\hlf}(\gB{\textsc{l}})=\gB{\textsc{l}}=\(\gBs{\textsc{l}}^{t}\)^{-1}~~~\text{and}~~~
{\pi}_{\overline\hlf}^{~}(\gB{\textsc{l}})=\bm{g}_{\textsc{l}}^*=\(\gBs{\textsc{l}}^\dag\)^{-1}.\label{pipibar}
\end{align}
Owing to the homomorphism (\ref{ismorSO13}), we call the $\SL2C$ representations the \emph{Lorentzian} and use the suffix ``\textsc{l}'' on one having the $\SL2C$ symmetry.

In this case, $\gB{\textsc{l}}$ acts on $\{z^{~}_1,z^{~}_2\}{\hspace{.1em}\in\hspace{.1em}}\C^2$ as (\ref{gEz1z2}). 
Corresponding Lie algebra $\dpij{\hlf}(\ggg_\textsc{l})=\ggg_\textsc{l}$ is non-unitary, thus, $\ggg_\textsc{l}^\dag\neq\ggg_\textsc{l}^{-1}\implies\pij{\hlf}(\gB{\textsc{l}})\neq{\pi}_{\overline\hlf}^{~}(\gB{\textsc{l}})$.
Since $SO(1,3)$ is not simply connected, $\SL2C$, which is simply connected, cannot cover whole $SO(1,3)$ but covers $SO^{\uparrow}\hspace{-.1em}(1,3)$ entirely.

Owing to the Lie algebra homomorphism (\ref{ismorSO13}), we obtain the representation of $\sss\ooo(1,3)$ as a tensor product of these two $\SL2C$ representations as
\begin{align}
\dpij{s^{~}_{\hspace{-.1em}1}\hspace{-.2em}{\otimes}\bar{s}^{~}_{\hspace{-.1em}2}}^\textsc{l}&:=
\dpij{s^{~}_{\hspace{-.1em}1}}\!(\ggg^{~}_{\textsc{l}})\hspace{.1em}{\otimes}\hspace{.1em}
\dpij{\hspace{-.1em}\overline{s}^{~}_{\hspace{-.1em}2}}\!(\ggg^{~}_{\textsc{l}})
~~\implies~~
\pij{s^{~}_{\hspace{-.1em}1}\hspace{-.2em}{\otimes}\bar{s}^{~}_{\hspace{-.1em}2}}^\textsc{l}:=
\(
\pij{s^{~}_{\hspace{-.1em}1}}\!(\gB{\textsc{l}})
\hspace{.1em}{\otimes}\hspace{.1em}
\pij{\hspace{-.1em}\overline{s}^{~}_{\hspace{-.1em}2}}\!(\gB{\textsc{l}}),
\VV^{\textsc{l}}_{\hspace{-.2em}s^{~}_{\hspace{-.1em}1}}{\otimes}\VV^{\textsc{l}}_{\hspace{-.2em}\overline{s}^{~}_{\hspace{-.1em}2}}
\),\label{piLslsr}
\end{align}
where $s^{~}_{1}$ and $s^{~}_{2}$ take a value on half-integers as in (\ref{piElr}).
The vector spaces equipping Lie algebra $\sss\lll(2,\C)$ ($ \overline{\sss\lll(2,\C)}$) is denoted as $\VV^{\textsc{l}}_{\hspace{-.2em}s}$ ($\VV^{\textsc{l}}_{\hspace{-.2em}\overline{s}}$), respectively.
We also introduce Lorentzian bi-spinor spaces as
\begin{align}
\VV^{\textsc{l}}_{\hspace{-.2em}\overline\hlf,\hlf}:=\VV^{\textsc{l}}_{\hspace{-.2em}\overline\hlf}{\otimes}\VV^{\textsc{l}}_{\hspace{-.2em}\hlf}.\label{LbispinorS}
\end{align}

%% file: Sec2-Clifford.tex
\subsection{Clifford algebra}\label{Cliffordalgebra}
We introduce the Clifford algebra to define the spin group.
The Clifford algebra induces the $\Z_2$-grading vector space corresponding to left and right chiralities.
The chiral structure of the Yang-Mills theory is an essential ingredient in the standard theory of particle physics. 
While the Hodge-dual operator induces the $\Z_2$-grading vector space for two-form objects in the four-dimensional manifold, the Clifford algebra does it for spinors.
In physics, a spinor field corresponds to a fermion, and a curvature two-form does to a bosonic force field.
Thus, both fermions and bosons in particle physics are categorised owing to chirality, thanks to the Clifford algebra and the Hodge-dual operator. 

Suppose $V_{\hspace{-.1em}n}$ is an $n$-dimensional vector space defined in a manifold equipping metric tensor $\bm\eta$.
The Clifford algebra in $V_{\hspace{-.1em}n}$, denoted as $\Cl(V_{\hspace{-.1em}n})$, is an algebra fulfiling 
\begin{subequations}
\begin{align}
\zeta,\chi\hspace{.1em}{\hspace{.1em}\in\hspace{.1em}}\hspace{.1em}\Cl(V_{\hspace{-.1em}n}),~~~
\zeta\chi+\chi\zeta
=2\<\zeta,\chi\>_{\Cl}^{~},\label{ClQ}
\end{align}
where $\<\bullet,\bullet\>_{\Cl}^{~}$ is the bilinearr form defined owing to  the metric tensor as
\begin{align}
\<\zeta,\chi\>_{\Cl}^{~}&:={\bm\zeta}^{t}\cdot\bm\eta\cdot{\bm\chi}
=\eta^{~}_\bcdots\hspace{.1em}\zeta^\bcdot\chi^\bcdot\!.\label{Cl}
\end{align}
\end{subequations}
We represent the Clifford algebra primarily as a column vector and denote its component with a superscript; its transpose is a row vector denoted with a subscript.
Corresponding to each orthonormal base vector $e^a\in V_{\hspace{-.1em}n}$,  the Clifford algebra $\gamma^a$  exists as
\begin{align*}
\gamma:V_{\hspace{-.1em}n}\rightarrow\Cl(V_{\hspace{-.1em}n}):e^a\mapsto\gamma^a,~~\text{with}~~a=0,1,\cdots,n-1
\end{align*}
A space of $\Cl(V_{\hspace{-.1em}n})$ is spanned by the power set $\{(\gamma^0)^{i_0}(\gamma^1)^{i_1}\cdots(\gamma^{n-1})^{i_{n-1}}\}$ with $i_a=0$ or $1$; thus, $\Cl(V_{\hspace{-.1em}n})$ is a ($2^n$)-dimensional space.

Suppose $\Cl_0(V_{\hspace{-.1em}n})$ is a subgroup containing an even number of Clifford algebras.
The spinor group in $V_{\hspace{-.1em}n}$, denoted as $Spin(V_{\hspace{-.1em}n})$, is defined as
\begin{align*}
Spin(V_{\hspace{-.1em}n}):=\left\{\gamma{\hspace{.1em}\in\hspace{.1em}}\Cl_0(V_{\hspace{-.1em}n})\big|\overset{*}{\gamma}\gamma
=\pm\bm{1}_\Sp,\hspace{.2em}\overset{*}{\gamma}\v\gamma{\hspace{.1em}\in\hspace{.1em}}V_{\hspace{-.1em}n}~\text{for}~\forall\v{\hspace{.1em}\in\hspace{.1em}}V_{\hspace{-.1em}n} \right\}\!,
\end{align*}
where $\gamma=\gamma^{a^{~}_1}_{~}\cdots\gamma^{a^{~}_m}_{~}$ and $\overset{*}{\gamma}:=\gamma^{a^{~}_m}_{~}\cdots\gamma^{a^{~}_1}_{~}$.
It forms a group concerning the Clifford product.
We define operator $\tau(\gamma){\hspace{.1em}\in\hspace{.1em}}End(V_{\hspace{-.1em}n})$ acting on $\v{\in}V_{\hspace{-.1em}n}$ as 
\begin{align}
\tau(\gamma)(\v)&:=\gamma{\v}\overset{\hspace{.2em}*}{\gamma},\label{ggvgg}
\end{align}
which preserves the bilinear form of two vectors $\v,\u\in V_{\hspace{-.1em}n}$ as
\begin{align*}
\<\tau(\gamma)(\v),\tau(\gamma)(\u)\>=
\<\gamma{\v}\overset{\hspace{.2em}*}{\gamma},\gamma{\u}\overset{\hspace{.2em}*}{\gamma}\>
=\(\text{sign}[\gamma\overset{*}{\gamma}]\)^2\<{\v},{\u}\>=\<{\v},{\u}\>.
\end{align*}
When $n\geq3$, operator $\tau(\bullet)$ induces double covering of $O(n)$, such that $Spin(V_{\hspace{-.1em}n})/\pm1{\rightarrow}O(n)$. 

Hereafter, we discuss the Clifford algebra in four-dimensional spaces.
When $V_{\hspace{-.1em}n}$ has the $SO(4)$ or $SO(1,3)$ symmetry,  we denote $Spin(V_{\hspace{-.1em}n})$ as $Spin(4)$ or $Spin(1,3)$, respectively.
We define the chiral operator as 
\begin{align}
Spin(V_{\hspace{-.1em}n}){\ni}\Gamma^{~}_{\hspace{-.1em}s}:=\sqrt{\frac{1}{\text{det}[\bm\eta]}}\gamma^0\gamma^1\gamma^2\gamma^3,
~~&\text{where}~~
\text{det}[\bm\eta]=
\begin{cases}
+1&\bm\eta=\betaE\\
-1&\bm\eta=\betaL
\end{cases},\label{Gamma}
\end{align}
yielding $\Gamma^{2}_{\hspace{-.1em}s}=\bm{1}_\Sp$ owing to the algebra (\ref{ClQ}).
The chiral operator induces the $\Z_2$-grading structure in $Spin(V_{\hspace{-.1em}n})$ as follows:
Since $\Gamma^{2}_{\hspace{-.1em}s}=\bm{1}_\Sp$, the chiral operator has two eigenvectors with eigenvalues $\pm1$.
We denote a space of the eigenvectors of the chiral operator as $\VVs$, namely the spinor space.
Owing to the definition, we obtain that
\begin{align}
(\ref{Gamma})&\implies \Gamma^{~}_{\!s}\hspace{.1em}\gamma^a=-\gamma^a\hspace{.1em}\Gamma^{~}_{\!s},
~~\text{for}~~^\forall\!a\in\{0,1,2,3\}.\label{GggG}
\end{align}
We define two projection operators using the chiral operator as
\begin{align}
P^\pm_{\hspace{-.1em}s}&:=\frac{1}{2}\(\bm{1}_\Sp\pm\Gamma^{~}_{\hspace{-.1em}s}\)\hspace{.3em}{\hspace{.1em}\in\hspace{.1em}}\hspace{.3em}End(\VVs)
\implies
P^\pm_{\hspace{-.1em}s}P^\pm_{\hspace{-.1em}s}=P^\pm_{\hspace{-.1em}s}
~~\text{and}~~
P^\pm_{\hspace{-.1em}s}P^\mp_{\hspace{-.1em}s}=0.\label{proj2}
\end{align}
The spinor space splits into a disjoint pair, 
\begin{subequations}
\begin{align}
\VVs=\Vs{+}{\oplus}\Vs{-},\label{GammaxiB}
\end{align}
such as
\begin{align}
\xi^\pm&:=P^\pm_{\hspace{-.1em}s}\xi\hspace{.3em}{\hspace{.1em}\in\hspace{.1em}}
\hspace{.3em}\Vs{\pm},\label{GammaxiA}
~~\text{where}~~
\Vs{\pm}:=\text{Im}\(P^\pm_{\hspace{-.1em}s}\)=\text{Ker}\(P^\mp_{\hspace{-.1em}s}\).
\end{align}
\end{subequations}
An eigenvector of the chiral operator,
\begin{align}
\Gamma_{\hspace{-.1em}s}^{~}:End(\Vs{\pm}):\xi^\pm\mapsto
\Gamma_{\hspace{-.1em}s}^{~}\hspace{.1em}\xi^\pm=\pm\xi^\pm,\label{GammaxiC}
\end{align}
is called the \emph{chiral spinor}, and its eigenvalue is called the \emph{chirality} in this study.
We assign the spinor with a positive eigenvalue to the \emph{chiral-right} and a negative eigenvalue to the \emph{chiral-left}.
The following sections introduce concrete examples of the Clifford algebra with Euclidean and Lorentzian metrics.

\subsubsection{Clifford algebra in Euclidean metric}
\paragraph{a) representation of dimension-2 :}
The Clifford algebra has the $(\TxT)$-matrix representation, e.g.,
\begin{subequations}
\begin{align}
\bm\gamma^{~}_{\SE}&=
\left\{\gamma^{0}_\SE,\gamma^{1}_\SE,\gamma^{2}_\SE,\gamma^{3}_\SE\right\}^t
=\left\{
\(\begin{array}{cc}
 0 &1\\
 1 & 0
\end{array}\),
\(\begin{array}{cc}
 0 &i\\
 -i & 0
\end{array}\),
\(\begin{array}{cc}
 0 &j\\
 -j & 0
\end{array}\),
\(\begin{array}{cc}
 0 &k\\
 -k & 0
\end{array}\)
\right\}^t\!,\label{2x2Cl}
\end{align}
where $\{i,j,k\}$ are imaginary quaternions fulfiling 
\begin{align}
i^2=j^2=k^2=-1,~~~ij+ji=jk+kj=ki+ik=0,~~\text{and}~~ij=k.\label{quaternion}
\end{align}
\end{subequations}
Each $(\TxT)$-matrix in column vector (\ref{2x2Cl}) fulfils relation (\ref{ClQ}) with the Euclidean metric $\bm\eta=\betaE$.
We note that each Euclidean Clifford algebra is Hermitian as $(\gamma^{a}_\SE)^\dagger=\gamma^{a}_\SE$ for $a\in\{0,1,2,3\}$, where complex conjugate is taken by means of quaternions.
We refer to this representation as the Euclidean \emph{representation of dimension-2} and put the suffix ``\SE'. 

The chiral operator is given as
\begin{align}
\Gamma_{\hspace{-.1em}\SE}&=\sqrt{\frac{1}{\text{det}[\betaE]}}\gamma^0_\SE\gamma^1_\SE
\gamma^2_\SE\gamma^3_\SE=\gamma^0_\SE\gamma^1_\SE\gamma^2_\SE\gamma^3_\SE=
\(\begin{array}{rc}
 -1 & 0\\
  0 & 1
\end{array}\),\label{2x2Gamma}
\end{align} 
yielding the projection operator as
\begin{align}
P^+_{\hspace{-.1em}\SE}:=\frac{1+\Gamma_{\hspace{-.1em}\SE}}{2}=
\(
     \begin{array}{cc}
      0&0\\
      0&1
     \end{array}
\)~~\text{and}~~
P^-_{\hspace{-.1em}\SE}:=\frac{1-\Gamma_{\hspace{-.1em}\SE}}{2}=
\(
     \begin{array}{cc}
      1&0\\
      0&0
     \end{array}
\).\label{Ppm}
\end{align}

\paragraph{b) representation of dimension-4:}
Linear combinations of unit quaternions forms the symplectic group denoted as $Sp(1)$.
The symplectic group has group homomorphism $Sp(1){=}SU(2)$; thus, quaternions have the $(\TxT)$-matrix representation using the Pauli matrices. 
When we replace quaternions with Pauli matrices as
\begin{align}
\Sigma^{~}_\textsc{e}:Sp(1){\rightarrow}SU(2):
\{1,i,j,k\}\mapsto\bsigE&:=\{\bm{1}^{~}_2,-i\hspace{.1em}\bm\sigma\},\label{ijk2E}
\end{align}
they follow the quaternion algebra (\ref{quaternion}), where $\bm\sigma=(\bm\sigma^1,\bm\sigma^3,\bm\sigma^3)$ are the Pauli matrices.
Thus, this replacement on the spinor representation (\ref{2x2Cl})  (in addition, zero and unity are replaced by corresponding $(\TxT)$-matrices) yields a four-component vector consisting of the $(\FxF)$-matrix like
\begin{align}
\Sigma^{~}_\textsc{e}:\bm\gamma^{~}_{\SE}\mapsto\bm\gamma^{~}_{\VE}&=
\(\gamma_\VE^{0},\gamma_\VE^{1},\gamma_\VE^{2},\gamma_\VE^{3}\)^t,\notag\\
&=
\(
\(\begin{array}{rr}
 \bm{0}_2 &\bm{1}_2\\
 \bm{1}_2 & \bm{0}_2
\end{array}\),
\(\begin{array}{rr}
 \bm{0}_2 &-i\bm\sigma^1\\
 i\bm\sigma^1 & \bm{0}_2
\end{array}\),
\(\begin{array}{rr}
 \bm{0}_2 &-i\bm\sigma^2\\
 i\bm\sigma^2 & \bm{0}_2
\end{array}\),
\(\begin{array}{rr}
 \bm{0}_2 &-i\bm\sigma^3\\
 i\bm\sigma^3 & \bm{0}_2
\end{array}\)
\)^t\!,\label{gammaVE}
\end{align}
where transpose is not concerned with matrices but with the four vectors to set them as column ones. 
Direct calculations show $\bm\gamma^{~}_{\VE}$ fulfils relation (\ref{ClQ}) with the Euclidean metric $\betaE$ and
\begin{align*}
\gamma^{a}_{\VE}\gamma^{a\dagger}_{\VE}=\bm{1}_2,~~\text{det}[\gamma^{a}_{\VE}]=1\implies\gamma^{a}_{\VE}{\hspace{.1em}\in\hspace{.1em}}SU(2)
~~\text{for}~~
a\in\{0,1,2,3\}.
\end{align*}
The projection operator is also obtained by $\Sigma^{~}_\textsc{e}$ like
\begin{align}
\Sigma^{~}_\textsc{e}:P^\pm_{\hspace{-.1em}\SE}\mapsto P^\pm_{\hspace{-.1em}\VE}=P^\pm_{\hspace{-.1em}\SE}\big|^{~}_{0\rightarrow\bm0_2,1\rightarrow\bm1_2}.\label{2x2PE}
\end{align}
These operators' operands are four-component $SU(2)$ spinors.
 In this report, we called this representation the \emph{representation of dimension-4} and put the suffix ``\VE''.

\subsubsection{Clifford algebra in Lorentzian metric}
For the Lorentzian Clifford algebra, we start from the ($\FxF$)-representation.

\paragraph{a) representation of dimension-4:}
Instead of map (\ref{ijk2E}), the following replacement of (\ref{2x2Cl}) provides the Clifford algebra with the Lorentzian metric:
\begin{align}
\Sigma^{\pm}_\textsc{l}:Sp(1){\rightarrow}\H(2)\cap\SL2C:\{1,i,j,k\}\mapsto\bm\sigma^{\pm}_\textsc{l}&:=\{\bm{1}^{~}_2,\pm\bm\sigma\},\label{ijk2L}
\end{align}
where $\H(2)$ shows a space of $(\TxT)$-Hermitian matrices.
When we set $\bm\eta=\betaL$ in (\ref{Cl}), (\ref{ijk2L}) yields the Clifford algebra (\ref{ClQ}). 
The Clifford algebra with the Lorentzian metric in the vector representation is provided as
\begin{align}
\bm\gamma^{~}_{\VL}=
\(\gamma_\VL^{0},\gamma_\VL^{1},\gamma_\VL^{2},\gamma_\VL^{3}\)^t=
\(
\(\begin{array}{rr}
 \bm{0}_2 &\bm{1}_2\\
 \bm{1}_2 & \bm{0}_2
\end{array}\),
\(\begin{array}{rr}
 \bm{0}_2 &\bm\sigma^1\\
 -\bm\sigma^1 & \bm{0}_2
\end{array}\),
\(\begin{array}{rr}
 \bm{0}_2 &\bm\sigma^2\\
 -\bm\sigma^2 & \bm{0}_2
\end{array}\),
\(\begin{array}{rr}
 \bm{0}_2 &\bm\sigma^3\\
 -\bm\sigma^3 & \bm{0}_2
\end{array}\)
\)^t,
\end{align}
yielding the chiral representation of the gamma matrices in particle physics.
Direct calculations show $\bm\gamma^{~}_{\VL}$ fulfils relation (\ref{ClQ}) with the Lorentzian metric $\betaL$ and
\begin{align*}
\gamma^{a}_{\VE}=\gamma^{a\dagger}_{\VE},~~\text{det}[\gamma^{a}_{\VL}]=1\implies\gamma^{a}_{\VL}{\hspace{.1em}\in\hspace{.1em}}\H(2)\cap\SL2C
~~\text{for}~~
a\in\{0,1,2,3\}.
\end{align*}
We refer to this representation as the Lorentzian \emph{representation of dimension-4} and put the suffix ``\VL'. 
For this representation, the chiral operator is given as
\begin{align}
\Gamma_{\hspace{-.1em}\VL}&=\sqrt{\frac{1}{\text{det}[\betaL]}}
\gamma_\VL^{0}\gamma_\VL^{1}\gamma_\VL^{2}\gamma_\VL^{3}=
i\hspace{.1em}\gamma_\VL^{0}\gamma_\VL^{1}\gamma_\VL^{2}\gamma_\VL^{3}=
\(\begin{array}{rc}
 -\bm{1}_2 & \bm{0}_2\\
  \bm{0}_2 & \bm{1}_2
\end{array}\).\label{2x2GammaL}
\end{align}
The projection operator is also obtained by $\Sigma^\pm_\textsc{l}$ like
\begin{align}
\Sigma^\pm_\textsc{l}:P^\pm_{\hspace{-.1em}\SE}\mapsto P^\pm_{\hspace{-.1em}\VE}=P^\pm_{\hspace{-.1em}\SE}\big|^{~}_{0\rightarrow\bm0_2,1\rightarrow\bm1_2}.\label{2x2PE}
\end{align}
Chiral operator (\ref{2x2GammaL}) is the same as the vector representation of the Euclidean chiral operator (\ref{2x2Gamma}) thanks to the factor $\sqrt{1/\text{det}[\bm\eta]}$  in (\ref{Gamma}).
Thus, the vector representation of the Lorentzian projection operator is also the same as that in the Euclidean as
\begin{align}
P^\pm_{\hspace{-.1em}\VL}=P^\pm_{\hspace{-.1em}\VE}=\text{(\ref{2x2PE})}.\label{2x2PL}
\end{align}

\paragraph{b) representation of dimension-2 :}
Next, we discuss the ($\TxT$)-representation.
Corresponding to the two irreducible representations, $\sss\lll(2,\C)$ and $\overline{\sss\lll(2,\C)}$, $\bm\gamma^{~}_{\VL}$ is split into two representations, such that
\begin{align}
\bm\gamma^{~}_{\VL}&\rightarrow
\begin{cases}
\bm\gamma^{~}_{\Sl}:=\left\{\gamma^{0}_\SE,\hspace{.8em}\sqrt{-1}\gamma^{1}_\SE,\hspace{.8em}\sqrt{-1}\gamma^{2}_\SE,\hspace{.8em}\sqrt{-1}\gamma^{3}_\SE\right\}^t&\text{for}~~\sss\lll(2,\C),\\
\overline{\bm\gamma}^{~}_{\Sl}:=\left\{\gamma^{0}_\SE,-\sqrt{-1}\gamma^{1}_\SE,-\sqrt{-1}\gamma^{2}_\SE,-\sqrt{-1}\gamma^{3}_\SE\right\}^t=
\bm{O}^\textsc{p}\ncdot\bm\gamma^{~}_{\Sl}&\text{for}~~\overline{\sss\lll(2,\C)}.\label{gammaSE}
\end{cases}
\end{align}
We refer to this representation as the Lorentzian \emph{representation of dimension-2} and put the suffix ``\Sl''. 
These representations have two different imaginary units as $\sqrt{-1}\in\C$ and $i\in\HQ$.
We define the Clifford module $\C\otimes\HQ(2)$ as
\begin{subequations}
\begin{align}
&\alpha_i\in\C, \beta_i\in\HQ\hspace{0.6em}\longrightarrow (\alpha_1\otimes\beta_1)(\alpha_2\otimes\beta_2)=\alpha_1\alpha_2\otimes\beta_1\beta_2,\label{CMCH1}\\
&\alpha\in\C, \beta\in\HQ(2)\longrightarrow (\alpha\otimes\beta)^\dagger=\alpha^*\otimes\beta^\dagger,\label{CMCH2}
\end{align}
\end{subequations}
where $\HQ(2)$  is a set of ($\TxT$) quaternion matrices.
We note that 
\begin{align*}
(\sqrt{-1}\otimes1)\times(1{\otimes}i)=\sqrt{-1}{\otimes}i\neq-1\xrightarrow{\text{s.h.n}.}\sqrt{-1}\hspace{.2em}i\neq-1
\end{align*}
where the last expression is the shorthand notation (s.h.n.).
When we distinguish imaginary unit $\sqrt{-1}\in\C$ from quaternion $i\in\HQ$, we can omit ``$\otimes$'' without any confusion.
Hereafter, we denote the imaginary unit in $\C$ as $\sqrt{-1}\in\C$ to distinguish it from $i\in\HQ$, when necessary.

These two Clifford algebras induce chiral operators as
\begin{align}
\Gamma_{\hspace{-.1em}\Sl}&=\sqrt{-1}
\gamma_\Sl^{0}\gamma_\Sl^{1}\gamma_\Sl^{2}\gamma_\Sl^{3}=\Gamma_{\hspace{-.1em}\SE}=
-\overline\Gamma_{\hspace{-.1em}\Sl},\label{GammaSL}
\end{align}
which yielding projection operators as
\begin{align}
P^+_{\hspace{-.1em}\Sl}=\bar{P}^-_{\hspace{-.1em}\Sl}=P^+_{\hspace{-.1em}\SE}~~\text{and}~~P^-_{\hspace{-.1em}\Sl}=\bar{P}^+_{\hspace{-.1em}\Sl}=P^-_{\hspace{-.1em}\SE}.\label{PpmSL}
\end{align}

The Lorentzian Clifford algebra maintains
\begin{align*}
(\gamma^{\hspace{.4em}a}_{\bullet\textsc{l}})^\dagger=
\gamma^{\hspace{.4em}0}_{\bullet\textsc{l}}\hspace{.1em}
\gamma^{\hspace{.4em}a}_{\bullet\textsc{l}}\hspace{.1em}
\gamma^{\hspace{.4em}0}_{\bullet\textsc{l}},~~~a\in\{0,1,2,3\},
\end{align*}
for both $\bullet=\textsc{v}$ and $\bullet=\textsc{s}$, as well as the Euclidean Clifford algebra $\bm\gamma_{\bullet\textsc{e}}$.

%% file: Sec2-ChiranSpinor.tex
\subsection{Chiral spinors}\label{HWspinor}
We introduced the chiral spinors $\xi^\pm$ as the eigenvector of the chiral operator (\ref{GammaxiC}).
This section discusses them in detail in Euclidean and Lorentzian metric spaces.

\subsubsection{Higgs spinor}
The $\SU(2)$-spinor with spin $s=\hlf$ has a two-dimensional column-vector representation as introduce in (\ref{gEz1z2}) such that
\begin{align}
\lambda^{~}_\hlf(\phiUh,\phiDh)&:=\bphih=\(\!
     \begin{array}{c}
     \phiUh\\\phiDh
     \end{array}
     \!\)
{\in\hspace{.1em}}\VV^{\textsc{e}}_{\hspace{-.1em}\hlf},
     ~~\phiUh,\phiDh{\hspace{.1em}\in\hspace{.1em}}\C.\label{phiE}
\end{align}
We put suffix ``\textsc{h}'' on the Higgs spinor.
Two-component spinor has two eigenvectors concerning to the Euclidean ($SU(2)$) chiral operator as
\begin{subequations}
\begin{align}
\Gamma_{\hspace{-.1em}\SE}\hspace{.2em}\bphiU&=-\bphiU\in\VV^{\textsc{eu}}_\hlf\implies
P^-_{\hspace{-.1em}\SE}\hspace{.1em}\bphih=\bphiU
=\(\!
     \begin{array}{c}
     \phiUh \\ 0
     \end{array}
\!\),\label{phiEm}\\
\Gamma_{\hspace{-.1em}\SE}\hspace{.2em}\bphiD&=+\bphiD\in\VV^{\textsc{ed}}_\hlf\implies
P^+_{\hspace{-.1em}\SE}\hspace{.1em}\bphih=\bphiD
=\(\!
     \begin{array}{c}
     0\\ \phiDh
     \end{array}
\!\),\label{phiEp}
\intertext{then we have}
\bm\phi_\textsc{h}&\in\left\{\bphiU,\bphiD\right\}{\subset}\VV^{\textsc{e}}_{\hspace{-.1em}\hlf}=\VV^{\textsc{eu}}_{\hspace{-.1em}\hlf}\oplus\VV^{\textsc{ed}}_{\hspace{-.1em}\hlf},
\end{align}
which is named a \textit{Higgs-spinor} in this report.
\end{subequations}
We introduce unit Higgs-spinors as
\begin{align}
 \hbphiUh&:={\bphiUh}/{\phiUh}=\(\!
     \begin{array}{c}
     1 \\ 0
     \end{array}
\!\)~~\text{and}~~
 \hbphiDh:={\bphiDh}/{\phiDh}=\(\!
     \begin{array}{c}
     0 \\ 1
     \end{array}
\!\),
\end{align}
as a basis of the Higgs spinor.

We denote a component of the Higgs-spinor $\bphih$ as $[\bphih]^{~}_A=\phi^{~}_A$ with $A\in\{\textsc{u},\textsc{d}\}$.
In contrast to the vector in $\TME$, the column vector representation of the spinor is given with a subscript.
We can denote the $\SU(2)$ transformation of the Higgs-spinor using the component representation as
\begin{align*}
\phi'_{A}&=[\bm{g}_\textsc{e}^{\dag}]_A^{\hspace{.4em}B}\phi^{~}_B.
\end{align*}
The Einstein convention is utilized for repeated indices running ``\textsc{u}'' and ``\textsc{d}''.
The $\SU(2)$-invariant bilinear form is introduced for $\bphih,\bm\varphi^{~}_\textsc{w}{\hspace{.1em}\in\hspace{.1em}}\VV^{\textsc{e}}_{\hspace{-.1em}\hlf}$ as 
\begin{align}
\<\bphih,\bm\varphi^{~}_\textsc{w}\>&:=
\phi^{*}_{A}\hspace{.15em}{\delta}^{AB}\varphi^{~}_B=
\(\phi_{~}^{A}\)^*\varphi^{~}_A=
\bphih^{\!\dagger}\!\cdot\bm\varphi^{~}_\textsc{w},
\quad\text{where}\quad{\delta}^{AB}\:=[\bm1_2]^A_{\hspace{.3em}B}.\label{compPhi2}
\end{align}
In this study, we mainly consider a following subgroup $\pij{\hlf}(\gB{\SOt})\subset\pij{\hlf}(\bm{g}_\textsc{e}^{~})$:
\begin{align}
\pij{\hlf}(\gB{\SOt})=\gBa{\SOt}(\varphi):=
e^{-i({\varphi}/{2})\sigma^a}
=\begin{cases}
  \left(
      \begin{array}{rr}
      \cos{\varphi/2}& -i\sin{\varphi/2}\\
      -i\sin{\varphi/2}&\cos{\varphi/2}
      \end{array}
     \right)&a=1\\
     \hspace{1em}
  \left(
      \begin{array}{rr}
      \cos{\varphi/2}& -\sin{\varphi/2}\\
      \sin{\varphi/2}&\cos{\varphi/2}
      \end{array}
     \right)&a=2\hspace{2.em}\in SU(2)\\
     \hspace{2em}
  \left(
      \begin{array}{cc}
      e^{-i\varphi/2}& 0 \\
     0&      e^{i\varphi/2}
      \end{array}
     \right)&a=3
\end{cases},\label{gE}
\end{align}
where actions with $a=1,2$ are induced by subgroups (\ref{A1K}), and that with $a=3$ is induced by (\ref{A1M}) given in \textbf{Appendix \ref{inducedrep}}.

$Spin(4)$ doubly covers $SO(4)$ as mentioned in the first homomorphism of (\ref{Spin4SU2SU2-2}).
We denote the group homomorphism as
\begin{align}
Spin(4)&=SU^{~}_{\hspace{-.2em}R}(2){\times}SU^{~}_{\hspace{-.2em}L}(2)=SO(4){\otimes}\{R,L\}.
\end{align}
Subscripts ``$R$'' and ``$L$'' stands for the \emph{right} and the \emph{left}, respectively.
An $SU(2)$ has a unique irreducible representation; thus, this categorisation is just a convention in physics.
The Higgs spinors conventionally belong to $SU^{~}_{\hspace{-.2em}R}(2)$ or $SU^{~}_{\hspace{-.2em}L}(2)$ and are transformed to each other under the $SU(2)$ group.

\subsubsection{Weyl spinor}\label{WS}
There are two irreducible representations for an $\SL2C$ group such that $\pi_{\hlf}^{\textsc{l}}(\gB{\textsc{l}})$ and $\pi^{\textsc{l}}_{\overline\hlf}(\gB{\textsc{l}})$, whose Lie algebras are denoted, respectively, as $\sss\lll(1,\C)$ and $\overline{\sss\lll(1,\C)}$. 
The $\SL2C$-spinor with spin $s=\hlf$ has a two-dimensional column-vector representation respectively for $\sss\lll(1,\C)$ and $\overline{\sss\lll(1,\C)}$ as
\begin{subequations}
\begin{align}
\lambda^{~}_\hlf(\xi^{~}_\textsc{u},\xi^{~}_\textsc{d})&:=\bm\xi=\(\!
     \begin{array}{c}
     \xi^{~}_\textsc{u} \\ \xi^{~}_\textsc{d}
     \end{array}\!
     \)
{\in\hspace{.1em}}\VV^{\textsc{l}}_{\hspace{-.1em}\hlf},
     ~~\xi^{~}_\textsc{u},\xi^{~}_\textsc{d}{\hspace{.1em}\in\hspace{.1em}}\C,\label{phiL}
\intertext{and}
\lambda^{~}_{\hspace{-.1em}\overline{\hlf}}(\dot\xi^{~}_{\dot{\textsc{u}}},\dot\xi^{~}_{\dot{\textsc{d}}})&:=\dot{\bm\xi}=\(\!
     \begin{array}{c}
     \dot\xi^{~}_{\dot{\textsc{u}}} \\ \dot\xi^{~}_{\dot{\textsc{d}}}
     \end{array}\!
     \)
{\in\hspace{.1em}}\VV^{\textsc{l}}_{\hspace{-.1em}\overline\hlf},
     ~~\dot\xi^{~}_{\dot{\textsc{u}}},\dot\xi^{~}_{\dot{\textsc{d}}}{\hspace{.1em}\in\hspace{.1em}}\C.\label{DphiL}
\end{align}
\end{subequations}
We call $\SL2C$-spinors as the \textit{Weyl-spinor}.
As shown in (\ref{DphiL}), a spinor belonging to $\VV^{\textsc{l}}_{\hspace{-.1em}\overline\hlf}$ is denoted as $\dot{\bm\xi}$ and called a \textit{dotted spinor}, whose component is also represented by the dotted index as $[\dot{\bpsi}]^{~}_{\dot{A}}$, using the \textit{van der Waerden notation}.
In contrast to the dotted one, spinor $\bm\xi\in\VV_{\hspace{-.1em}\hlf}^{\textsc{l}}$ is called the \textit{undotted spinor}. 
In physics, an undotted (dotted) Weyl spinor is also called a right-handed (left-handed) spinor, respectively.

The undotted and dotted Weyl spinors are, respectively, transformed as the fundamental representation under the $\SL2C$ as
\begin{subequations}
\begin{align}
\bpsi'&=\pij{\hlf}^\textsc{l}(\gB{\textsc{l}})\bpsi
=\g_\textsc{l}^{\dag}(\varphi)\bpsi,\hspace{2.1em}
[\bpsi']_{\!A}=[\g_\textsc{l}^{~}(\varphi)]_{\!A}^{\hspace{.4em}B}[\bpsi]_{\!B},\label{gLpsi}
\intertext{and}
\dot{\bpsi}'&=
\pij{\overline\hlf}^\textsc{l}(\gB{\textsc{l}})\dot{\bpsi}
=\g_\textsc{l}^{*}(\varphi)\dot{\bpsi},~~~~~
\left[\dot\bpsi'\right]_{\!\dot{A}}=
\left[\g_\textsc{l}^{*}(\varphi)\right]_{\!\dot{A}}^{\hspace{.4em}\dot{B}}\left[\dot{\bpsi}\hspace{.1em}\right]_{\!\dot{B}}\!,\label{gLdpsi}
\end{align}
\end{subequations}
where dotted and undotted indices run over $\{\textsc{u},\textsc{d}\}$ and $\{\dot{\textsc{u}},\dot{\textsc{d}}\}$, respectively.
We note that $(\g_\textsc{l}^{\dagger})^*=\g_\textsc{l}^t=\g_\textsc{l}^*$ since $\g_\textsc{l}^{~}$ is the Hermitian matrix.
We raise and lower Weyl-spinor indices using the ($\TxT$)-tensor $\bm{\epsilon}^{~}_2$ as
\begin{align*}
\xi_{~}^A=\epsilon^{\hspace{.3em}AB}_2\xi^{~}_B,~~
\dot\xi_{~}^{\dot{A}}=\epsilon^{\hspace{.3em}\dot{A}\dot{B}}_2\dot{\xi}^{~}_{\dot{B}},
\end{align*}
where 
\begin{align*}
\left[\bm{\epsilon}^{~}_2\right]^{AB}&=
\left[\bm{\epsilon}^{~}_2\right]^{\dot{A}\dot{B}}=\(
\begin{array}{rr}
0&1\\
-1&0
\end{array}
\)=
-\left[\bm{\epsilon}^{~}_2\right]_{AB}=
-\left[\bm{\epsilon}^{~}_2\right]_{\dot{A}\dot{B}}.
\end{align*}
The $\SL2C$-invariant bilinear forms of undotted spinors ($\bpsi,\bm\zeta{\hspace{.2em}\in\hspace{.2em}}\VV_{\hlf}^{\textsc{l}}$) and dotted spinors ($\dot\bpsi,\dot{\bm\zeta}{\hspace{.2em}\in\hspace{.2em}}\VV^{\textsc{l}}_{\overline\hlf})$ are, respectively, given as
\begin{align}
\<\bpsi,\bm\zeta\>&:=
\xi^{~}_A\hspace{.2em}\epsilon^{\hspace{.3em}AB}_2\hspace{.2em}\zeta_B=
\xi^{~}_\textsc{u}\zeta_\textsc{d}-\xi^{~}_\textsc{d}\zeta_\textsc{u},
~~\text{and}~~
\<\dot\bpsi,\dot{\bm\zeta}\>:=
\dot\xi^{~}_{\dot{A}}\hspace{.2em}\epsilon^{\hspace{.3em}\dot{A}\dot{B}}_2\hspace{.2em}\dot\zeta_{\dot{B}}=
\dot\xi^{~}_{\dot{\textsc{u}}}\dot\zeta_{\dot{\textsc{d}}}-\dot\xi^{~}_{\dot{\textsc{d}}}\dot\zeta_{\dot{\textsc{u}}}.\label{LBLF2}
\end{align}
In contrast with the Higgs-spinor case, we do not take complex conjugate for a bra-part of the bilinear form since $\g^{~}_\textsc{l}$ and $\g^{*}_\textsc{l}$ belong to the different representations.
This bilinear form is  $\SL2C$-invariant such as
\begin{align*}
\<\bpsi',\bm\zeta'\>=
\<\gB{\textsc{l}}\bpsi,\gB{\textsc{l}}\bm\zeta\>=&
[\gB{\textsc{l}}\bm\xi]_A\hspace{.2em}\epsilon^{\hspace{.3em}AB}_2\hspace{.2em}[\gB{\textsc{l}}\bm\zeta]_B
=\xi^{~}_A[\bm{g}_{\textsc{l}}^t]^A_{\hspace{.4em}C}\hspace{.2em}
\epsilon^{\hspace{.3em}CD}_2\hspace{.2em}[\gB{\textsc{l}}]_D^{\hspace{.4em}B}\hspace{.1em}\zeta_B=
\xi^A\hspace{.2em}\zeta_A =
\<\bpsi,\bm\zeta\>,
\intertext{where we use}
&[\bm{g}_{\textsc{l}}^t]^A_{\hspace{.4em}C}\hspace{.2em}\epsilon^{\hspace{.3em}CD}_2\hspace{.2em}[\gB{\textsc{l}}]_D^{\hspace{.4em}B}=
\epsilon^{\hspace{.3em}AB}_2.
\end{align*}
The same is true for the dotted spinors.
We denote a skew-transpose for Weyl spinors by $\bullet^T$ using $\bm{\epsilon}^{~}_2$ as 
\begin{align}
\left[\bm\xi^{T}\right]^B:=\xi_A^{~}\hspace{.1em}\epsilon^{\hspace{.3em}AB}_2=\xi^B_{~},~~~
\left[\dot{\bm\xi}^{T}\right]^{\dot{B}}:=\dot{\xi}^{~}_{\dot{A}}\hspace{.1em}\epsilon^{\hspace{.3em}\dot{A}\dot{B}}_2=\dot{\xi}^{\dot{B}}_{~}\implies
\<\bpsi,\bm\zeta\>=\bm\xi^{T}\ncdot\bm\zeta,~~~
\<\dot\bpsi,\dot{\bm\zeta}\>=\dot{\bm\xi}^{T}\ncdot\dot{\bm\zeta}.\label{WeylConjug}
\end{align}

Three group actions (\ref{gE}) are also the isometric transformation for Lorentzian vectors since they belong to the subgroup $SO(3){\subset}SO(1,3)$.
In addition to that, we consider the group actions of  $SO(1,3){\setminus}SO(3)$ such that
\begin{align}
\pij{\hlf}(\gB{b})=\gBa{b}(\chi):=
e^{({\chi}/{2})\sigma^a}
=\begin{cases}
     \hspace{1em}
  \left(
      \begin{array}{rr}
      \cosh{\chi/2}& \sinh{\chi/2}\\
      \sinh{\chi/2}&\cosh{\chi/2}
      \end{array}
     \right)&a=1\\
  \left(
      \begin{array}{rr}
      \cosh{\chi/2}& -i\sinh{\chi/2}\\
      i\sinh{\chi/2}&\cosh{\chi/2}
      \end{array}
     \right)&a=2\hspace{2.em}\in \H(2)\\
     \hspace{2em}
  \left(
      \begin{array}{cc}
      e^{\chi/2}& 0 \\
     0&      e^{-\chi/2}
      \end{array}
     \right)&a=3
\end{cases}.\label{gM}
\end{align}
Actions with $a=1,2$ belong to subgroups (\ref{A1K}), and those with $a=3$ belong to (\ref{A1A}) given in \textbf{Appendix \ref{inducedrep}}.

A space of  Weyl spinors consists of two subspaces owing to the chirality such as
\begin{align}
\VV^\textsc{l}_{\hspace{-.1em}\hlf}=\VV^{\textsc{ul}}_{\hspace{-.1em}\hlf}\oplus\VV^{\textsc{ld}}_{\hspace{-.1em}\hlf}.
\end{align}
For the undotted spinors, we have eigenvectors of the chiral operator as
\begin{subequations}
\begin{align}
\Gamma_{\hspace{-.1em}\Sl}\hspace{.2em}\bm\xi^{~}_{\textsc{u}}&=-\bm\xi^{~}_{\textsc{u}}\in\VV^{\textsc{lr}}_{\hspace{-.1em}\hlf}\implies
P^-_{\hspace{-.1em}\Sl}\hspace{.1em}\bm\xi=\bm\xi^{~}_{\textsc{u}}
=\(\!
     \begin{array}{c}
     \xi^{~}_{\textsc{u}} \\ 0
     \end{array}
\!\)\!,\label{phiLm1}\\
\Gamma_{\hspace{-.1em}\Sl}\hspace{.2em}\bm\xi^{~}_{\textsc{d}}&=+\bm\xi^{~}_{\textsc{d}}\in\VV^{\textsc{ld}}_{\hspace{-.1em}\hlf}\implies
P^+_{\hspace{-.1em}\Sl}\hspace{.1em}\bm\xi=\bm\xi^{~}_{\textsc{d}}
=\(\!
     \begin{array}{c}
     0\\ \xi^{~}_{\textsc{d}}
     \end{array}
\!\)\!,\label{phiLp1}
\intertext{then we have}
\bm\xi&\in\left\{\bm\xi^{~}_{\textsc{u}},\bm\xi^{~}_{\textsc{d}}\right\}{\subset}
\VV^\textsc{l}_{\hspace{-.1em}\hlf}=\VV^{\textsc{lu}}_{\hspace{-.1em}\hlf}\oplus\VV^{\textsc{ld}}_{\hspace{-.1em}\hlf}.
\end{align}
\end{subequations}
We call the Weyl spinor $\bm\xi^{~}_{\textsc{u}}$ ($\bm\xi^{~}_{\textsc{d}}$) spin-up (spin-down) state of fermion in physics.
We also obtain the dotted chiral Weyl-spinors; although the result is trivial, we show them for completeness.
\begin{subequations}
\begin{align}
\overline\Gamma_{\hspace{-.1em}\Sl}\hspace{.2em}\dot{\bm\xi}^{~}_{\textsc{u}}&
=+\dot{\bm\xi}^{~}_{\textsc{u}}\in\VV^{\textsc{lu}}_{\hspace{-.1em}\overline{\hlf}}\implies
\bar{P}^+_{\hspace{-.1em}\Sl}\hspace{.1em}\dot{\bm\xi}=\dot{\bm\xi}^{~}_\textsc{u}
=\(\!
     \begin{array}{c}
     \dot{\xi}^{~}_{\dot{\textsc{u}}} \\ 0
     \end{array}
\!\),\label{phiLm2}\\
\overline\Gamma_{\hspace{-.1em}\Sl}\hspace{.2em}\dot{\bm\xi}^{~}_{\textsc{d}}&
=-\dot{\bm\xi}^{~}_{\textsc{d}}\in\VV^{\textsc{ld}}_{\hspace{-.1em}\overline{\hlf}}\implies
\bar{P}^-_{\hspace{-.1em}\Sl}\hspace{.1em}\dot{\bm\xi}=\dot{\bm\xi}^{~}_\textsc{d}
=\(\!
     \begin{array}{c}
     0\\ \dot{\xi}^{~}_{\dot{\textsc{d}}}
     \end{array}
\!\),\label{phiLp2}
\intertext{then we have}
\dot{\bm\xi}&\in\left\{\dot{\bm\xi}^{~}_{\textsc{u}},\dot{\bm\xi}^{~}_{\textsc{d}}\right\}{\subset}
\VV^\textsc{l}_{\hspace{-.1em}\overline\hlf}=\VV^{\textsc{lu}}_{\hspace{-.1em}\overline\hlf}\oplus\VV^{\textsc{ld}}_{\hspace{-.1em}\overline\hlf}.
\end{align}
\end{subequations}
We note that the chirality of dotted and undotted spinors with the same spin state has an opposite to each other; $\bm\xi^{~}_\textsc{d}$ and $\dot{\bm\xi}^{~}_\textsc{u}$ have a positive chirality and $\bm\xi^{~}_\textsc{u}$ and $\dot{\bm\xi}^{~}_\textsc{d}$ have a negative chirality.

The Lie algebra homomorphism (\ref{ismor2}) allows us to equate each representation of (\ref{piLslsr}) to the angular momentum of a physical field, which justifies terms ``\textit{chiral-right}'' and ``\textit{chiral-right}'' as a helicity state of a particle.
Group $Spin(1,3)$ doubly covers $SO(1,3)$ as shown in  (\ref{Spin4SU2SU2-2}).
We denote the Lie-algebra homomorphism as
\begin{align}
\ggg^{~}_{Spin(1,3)}&=\sss\lll(2,\C)\oplus\overline{\sss\lll(2,\C)}=\sss\ooo(1,3){\otimes}\{R,L\},
\end{align}
similarly to the Higgs spinor.
However, chiral-right and chiral-left Weyl spinors are not transformed to each other under the $\SL2C$ group in contrast with the Higgs spinor. 

\subsubsection{Dirac spinor}\label{DS}
We introduce the four-component spinor, namely the Dirac spinor, owing to the Clifford algebra representation of dimension-4 introduced in \textbf{section \ref{Cliffordalgebra}}, as
\begin{align}
\psi&:=\(\!
     \begin{array}{c}
     \dot\bpsi \\ \bpsi
     \end{array}
\!\)\in
\VV^{{\textsc{l}}}_{\hspace{-.1em}\hlf\otimes\overline\hlf}:=
\VV^{\textsc{l}}_{\hspace{-.1em}\hlf}\otimes\VV^{\textsc{l}}_{\hspace{-.1em}\overline\hlf},
~~\text{with}~~~
\bpsi\in\VV^{\textsc{l}}_{\hspace{-.1em}\hlf},~~\dot{\bpsi}\in\VV^{\textsc{l}}_{\hspace{-.1em}\overline\hlf}.
\label{psiM}
\end{align}
In this representation, chiral and projection operators are provided as the same as the Euclidean case as
\begin{subequations}
\begin{align}
\psi_{~}^{L}&:=P^-_{\hspace{-.1em}\VL}\hspace{.3em}\psi=\(\!
     \begin{array}{l}
     \dot\bpsi \\\bm{0}_\Sl
     \end{array}
\!\),~~\Gamma^{~}_{\hspace{-.1em}\VL}\hspace{.2em}\psi_{~}^L=-\psi_{~}^L,\label{phil}\\
\psi_{~}^R&:=P^+_{\hspace{-.1em}\VL}\hspace{.1em}\psi=\hspace{.1em}\(\!
     \begin{array}{l}
     \dot{\bm0}_\Sl\\ \bpsi
     \end{array}
\!\),~~\Gamma^{~}_{\hspace{-.1em}\VL}\hspace{.2em}\psi_{~}^R=+\psi_{~}^R.\label{phir}
\end{align}
\end{subequations}
Superscripts ``$R$'' and ``$L$'' on $\psi$ stand for the ``\textit{right-handed}'' and ``\textit{left-handed}'' according to those of the Weyl spinors.

%% file: Sec2-3.tex
%
\subsection{Vectors}\label{SRV}
An ordered list of four real numbers provides the standard representation of a four-vector.
In four-dimensional Euclidean (Lorentzian) metric space, we can represent it by $(\TxT)$ complex matrices using the irreducible representation of $SU(2)$ ($\SL2C$).
This section introduces the representation of dimension-2 for Euclidean and Lorentzian vectors.

\subsubsection{Euclidean vectors}\label{SRVE}
Suppose $\hat{\v}_\VE=(\hat{v}^0,\hat{v}^1,\hat{v}^2,\hat{v}^3)^t{\hspace{.1em}\in\hspace{.1em}}V(\TME)$ is a real unit-vector $\|\hat{\v}_\VE\|^{~}_\textsc{e}=1$ in the four-dimensional Euclid manifold $(\M,\betaE)$.
A space of unit vectors is equivalent to $S^3$ as shown in (\ref{S3}).
The transformation,
\begin{align}
\Sigma^{~}_\textsc{e}:S^3{\rightarrow}\SU(2):\hat{\v}_\VE\mapsto
  \hat{\v}^{~}_\SE:=\bm{\sigma}_\textsc{e}^t\cdot\hat{\v}_\VE&=
  \(\!
      \begin{array}{rr}
        \hat{v}^0-i\hspace{.1em}\hat{v}^3 &-i\hspace{.1em}\hat{v}^1 -\hspace{.45em}\hat{v}^2\\
         -i\hspace{.1em}\hat{v}^1+\hspace{.45em}\hat{v}^2 &\hat{v}^0 +i\hspace{.1em}\hat{v}^3\\
      \end{array}
     \!\),\label{SigmaEv}
\end{align}
maps a unit vector in $V(\TME)$ to an $\SU(2)$ matrix.
There is a one-to-one correspondence between Euclid vectors and $\SU(2)$-spinors as
\begin{align*}
\VV^\textsc{e}_{\hspace{-.2em}\hlf,\hlf}\ni
\(\!
     \begin{array}{c}
     \phiU\\\phiD
     \end{array}
\!\)\otimes
\(\!
     \begin{array}{c}
     \phiU\\\phiD
     \end{array}
\!\)^{\!*}
 \mapsto
\(
     \begin{array}{rr}
     \phiU&-\phiD^* \\
     \phiD&\phiU^*
     \end{array}
 \)\!.
\end{align*}
Thus, Euclidean vectors belong to the spinor space (\ref{lambda}).
The $\SU(2)$ invariant bilinear form and the norm are given as
\begin{subequations}
\begin{align}
\<\hat\u_\SE|\hat\v_\SE\>:=&
\frac{1}{2}\(\Tr[\hat\u^{~}_\SE]\hspace{.2em}\Tr[\hat\v^{~}_\SE]
-\Tr[\hat\u^{~}_\SE\hspace{.2em}\hat\v^{~}_\SE]\)
=\hat{u}^0_{~}\hspace{.1em}\hat{v}^0_{~}+\hat{u}^1_{~}\hspace{.1em}\hat{v}^1_{~}
+\hat{u}^2_{~}\hspace{.1em}\hat{v}^2_{~}+\hat{u}^3_{~}\hspace{.1em}\hat{v}^3_{~}\label{<vu>VE}
,\\
\|\hat{\u}^{~}_\SE\|^2:=&\<\hat{\u}^{~}_\SE|\hat{\u}^{~}_\SE\>=\text{det}[{\hat{\u}}^{~}_\SE]=1,
\end{align}
\end{subequations}
in the representation of dimension-2.
Any non-zero vector $\u^{~}_\SE$ is represented as
\begin{align*}
\u^{~}_\SE&=\|\u^{~}_\SE\|\hspace{.1em}\hat{\bm v}^{~}_\SE,
~~\text{with}~~
\|\u^{~}_\SE\|^2>0.
\end{align*}
Thus, the transformation (\ref{SigmaEv}) maps a pint in $\SO(4)$ to the $\SU(2)$ matrix as 
\begin{align*}
\Sigma^{~}_\textsc{e}:V(\SO(4)){\rightarrow}SU(2)\otimes\R:\bm\v^{~}_\VE\mapsto\bm\v^{~}_\SE.
\end{align*}

Group action (\ref{gE}) acts on a vector as the adjoint representation such that:
\begin{align}
\hat{\bm v}'_\SE&=\tau\(\gBa{\SOt}(\varphi)\)\(\hat{\bm v}^{~}_\SE\)
=\bm{\sigma}_\textsc{e}^t\cdot
\begin{cases}
\hspace{.4em}
\(\!
     \begin{array}{c}
       \hat{v}^0 \\ 
       \hat{v}^1 \\
       \hat{v}^2\cos{\varphi}-\hat{v}^3\sin{\varphi}\\
       \hat{v}^2\sin{\varphi}+\hat{v}^3\cos{\varphi}\\ 
      \end{array}
\!\)&a=1\\
\(\!
     \begin{array}{c}
       \hat{v}^0 \\ 
       \hspace{.7em}\hat{v}^1\cos{\varphi}+\hat{v}^3\sin{\varphi}\\
       \hat{v}^2 \\
      -\hat{v}^1\sin{\varphi}+\hat{v}^3\cos{\varphi}\\ 
      \end{array}
\!\)&a=2\\
\hspace{.4em}
\(\!
     \begin{array}{c}
       \hat{v}^0 \\ 
       \hat{v}^1\cos{\varphi}-\hat{v}^2\sin{\varphi}\\
       \hat{v}^1\sin{\varphi}+\hat{v}^2\cos{\varphi}\\ 
       \hat{v}^3
      \end{array}
\!\)&a=3
\end{cases}\!,\label{Erot}
\intertext{where}
\tau(\g^{~}_r)(\v_\SE)&:=\g^{~}_r\hspace{.1em}\v_\SE\hspace{.1em}\g_r^\dagger.
\end{align}
Group actions (\ref{gE}) are isometric transformations keeping the bilinear form (\ref{<vu>VE}) invariant under the $SU(2)$ transformation.
In reality, (\ref{Erot}) induces an anti-clockwise $\varphi$-rotation of $\hat{\v}$ around the $\hat{v}^a$-axis.
Actions $\gBa{\SOt}(\varphi)$ and $-\gBa{\SOt}(\varphi)$ map vector $\hat{\v}$ to the same vector $\hat{\bm v}'$; thus, this is a \emph{two-to-one} transformation.

The subgroups $G$, which have continuous homomorphism fulfiling
\begin{align*}
g:\R{\rightarrow}G:\varphi{\mapsto}g(\varphi),~~g(\vartheta+\varphi)=g(\vartheta)\hspace{.1em}g(\varphi)
~~\text{and}~~g(0)=\bm{1}_\text{id},
\end{align*}
is called a \emph{one-parameter subgroup}.
Group elements in (\ref{gE}) are the one-parameter subgroup of $SU(2)=S^3=Spin(3)=SO(3)\otimes\{\pm1\}$.
The general element of the $SU(2)$ group is given as (\ref{generalrepSU2}) in \textbf{Appendix \ref{inducedrep}}.
This one-parameter subgroup belongs to the subgroup $SO_{~}^{\perp}\hspace{-.1em}(3){\subset}SO(4)$, where  $SO_{~}^{\perp}\hspace{-.1em}(3)$ is the rotation group of the three-dimensional Euclidean space perpendicular to the $\hat{v}^0$-axis.

\subsubsection{Lorentzian vectors}\label{SRL}
For Lorentzian vector $\hat{\u}_\VL=(\hat{u}^0,\hat{u}^1,\hat{u}^2,\hat{u}^3)^t{\hspace{.1em}\in\hspace{.1em}}V(\TML)$ with $\|\hat{\u}_\VL\|^{~}_\textsc{l}=\pm1$, we have the representation of dimension-2 for $\hat{\u}_\VL$ as
\begin{align}
\Sigma^{\pm}_\textsc{l}:\text{Hyp}^3{\rightarrow}
\H(2)\cap(\SL2C\otimes\{-1,1\}):\hat{\u}_\VL\mapsto
  \hat{\u}^{\pm}_\Sl:=\(\bm\sigma^{\pm}_\textsc{l}\)^t\cdot\hat{\u}_\VL&=
  \(\!
      \begin{array}{rr}
        \hat{u}^0\pm\hspace{.45em}\hat{u}^3 & \pm\hat{u}^1 \mp i\hspace{.1em}\hat{u}^2\\
        \pm\hat{u}^1\pm i\hspace{.1em}\hat{u}^2 &\hat{u}^0 \mp\hspace{.45em}\hat{u}^3\\
      \end{array}
   \!\),\label{SL2C1}
\end{align}
where $\text{Hyp}^3$ is a three-dimensional hyperbolic plane given by $\|\hat{\u}_\VL\|^{~}_\textsc{l}=\pm1$.
$\SL2C\otimes\{-1,1\}$ shows a space of complex $(\TxT)$-matrices with their determinant being $\pm1$.
Transformation (\ref{SL2C1}) maps a Lorentzian vector to an $\H(2)\cap(\SL2C\otimes\{-1,1\})$ matrix.
The Lorentzian vectors belong to the bi-spinor space (\ref{LbispinorS}) as $V(\TML)=\VV^\textsc{l}_{\hspace{-.2em}\hlf,\overline\hlf}$.
We can construct any vector in the Lorentzian metric space using a bi-spinor as
\begin{align*}
\VV^\textsc{l}_{\hspace{-.2em}\hlf,\overline\hlf}\ni
\(\!
     \begin{array}{c}
     \xi_\textsc{u}\\ \xi_\textsc{d}
     \end{array}
\!\)\otimes
\(\!
     \begin{array}{c}
     \xi_\textsc{u}\\ \xi_\textsc{d}
     \end{array}
\!\)^{\!*}
 \mapsto
\(\!
     \begin{array}{ccr}
     (\xi_\textsc{u}+\xi^*_\textsc{u})/2&\xi^*_\textsc{d} \\
     \xi_\textsc{d}&i(\xi^*_\textsc{u}-\xi_\textsc{u})/2
     \end{array}
\!\).
\end{align*}
The $\SL2C$ invariant bilinear form is provided for $\hat{\u}_\VL,\hat{\v}_\VL\in\H(2)\cap(\SL2C\otimes\{-1,1\})$ as
\begin{subequations}
\begin{align}
\<\hat{\u}^{~}_\Sl,\hat{\v}^{~}_\Sl\>&:=
\frac{1}{2}\(\Tr[\hat{\u}^{~}_\Sl]\hspace{.2em}\Tr[\hat{\v}^{~}_\Sl]
-\Tr[\hat{\u}^{~}_\Sl\hspace{.2em}\hat{\v}^{~}_\Sl]\)=
\hat{v}^0\hspace{.1em}\hat{v}^0-\hat{u}^1\hspace{.1em}\hat{v}^1
-\hat{u}^2\hspace{.1em}\hat{v}^2-\hat{u}^3\hspace{.1em}\hat{v}^3,\label{<uv>L}\\
\|\hat{\u}^{~}_\Sl\|^2&:=\<\hat{\u}^{~}_\Sl|\hat{\u}^{~}_\Sl\>=
(\hat{u}^0)^2-(\hat{u}^1)^2-(\hat{u}^2)^2-(\hat{u}^3)^2=
\text{det}[\hat{\u}^{~}_\Sl];
\end{align}
thus, the Lorentzian metric is correctly embedded in the map $\Sigma^{\pm}_\textsc{l}$.
\end{subequations}
Transformation (\ref{SL2C1}) maps a vector in $\SO(1,3)$ to $(\TxT)$-Hermitian matrices:
\begin{align*}
\Sigma^{\pm}_\textsc{l}:V(\SO(1,3)){\rightarrow}\H(2)\otimes\R:\bm\u^{~}_\VL\mapsto\bm\u^{~}_\Sl.
\end{align*}

Action (\ref{gM}) acts on the vector ${\bm u}^{+}_\Sl$, and ${\pi}_{\overline\hlf}^{~}(\gB{b})=\gBs{b}^{*}=(\gBs{b}^\dag)^{-1}$ acts on ${\bm u}^{-}_\Sl$ as
\begin{align}
{\bm u}^{\pm'}_\Sl&=\left\{\hspace{-.4em}
\begin{array}{l}
\tau(\hspace{.1em}\gBa{b}(\chi))\({\bm u}^{+}_\Sl\)\\
\tau({\gBa{b}}^*(\chi))\({\bm u}^{-}_\Sl\)
\end{array}\right.
={\bm\sigma^{\pm}_\textsc{l}}^t\cdot
\begin{cases}
\left(
     \begin{array}{c}
       {u}^0\cosh{\chi}+{u}^1\sinh{\chi}\\
       {u}^0\sinh{\chi}+{u}^1\cosh{\chi}\\ 
       {u}^2 \\ 
       {u}^3 
      \end{array}
\right)&a=1,\\
\left(
     \begin{array}{c}
       {u}^0\cosh{\chi}+{u}^2\sinh{\chi}\\
       {u}^1 \\ 
       {u}^0\sinh{\chi}+{u}^2\cosh{\chi}\\ 
       {u}^3 
      \end{array}
\right)&a=2,\\
\left(
     \begin{array}{c}
       {u}^0\cosh{\chi}+{u}^3\sinh{\chi}\\
       {u}^1 \\ 
       {u}^2 \\
       {u}^0\sinh{\chi}+{u}^3\cosh{\chi}
      \end{array}
\right)&a=3,
\end{cases}\label{Lboost}
\intertext{where}
\tau(\g_b^{~})(\v_\Sl)&:=\g_b^{~}\hspace{.1em}\v_\Sl\hspace{.1em}\g_b^\dagger,
\end{align}
which induce a $\chi$-boost along the ${u}^a$-axis.
This group action is also a \emph{two-to-one} transformation as the Euclidean case.
These representation belongs to the subgroup of $SO(1,3)$ concerning the boost operation.
We denote the group action of $\SL2C$ as 
\begin{align}
\gBs{L}^+(\varphi,\chi):=\gB{\SOt}(\varphi)\hspace{-.1em}\otimes\hspace{.1em}\gB{b}(\chi)~~&\text{and}~~
\gBs{L}^-(\varphi,\chi):=\gB{\SOt}(\varphi)\hspace{-.1em}\otimes\hspace{.1em}\gBs{b}^*(\chi).\label{gLorentz}
\end{align}
We nothe that they are commutable as
\begin{align*}
\gB{\SOt}(\varphi)\bcdot\hspace{.1em}\gBs{b}^{(*)}(\chi)=
\gBs{b}^{(*)}(\chi)\bcdot\hspace{.1em}\gB{\SOt}(\varphi).
\end{align*}
Both $\SL2C$ and $SO^{\uparrow}\hspace{-.1em}(1,3)$ are simply connected and $\SL2C$ entirely covers $SO^{\uparrow}\hspace{-.1em}(1,3)$.
The representation of dimension-2 for $Spin(1,3)\hspace{-.2em}=\hspace{-.2em}\SL2C$ has two fundamental representations as (\ref{pipibar}) and is labeled by two half-integers $\{s^{~}_1,s^{~}_2\}$ as (\ref{piLslsr}).

%% file: Sec2-GRfields.tex
%
%
\subsection{Gravitational fields}
This section discusses irreducible representations of the gravitational fields, i.e., space-time curvature, spin connection, and vierbein fields, in the Lorentzian inertial manifold.
This section omits the suffix ``$\textsc{l}$'' of tensors for simplicity.
 
\paragraph{a) Riemann curvature:} 
 The Riemann curvature tensor, $R^{\bullets}_{\hspace{.5em}\bullets}$, is a rank-two in the tangent space and a rank-two in the cotangent space of the inertial manifold as
\begin{align}
\Omega^2(\TsM){\otimes}V^2(\TM)\ni\RRR=\frac{1}{2}\RRR^{ab}(\partial_a\times\partial_b)=\frac{1}{4}R^{ab}_{\hspace{.9em}cd}\hspace{.2em}(\eee^c\wedge\eee^d)(\partial_a\times\partial_b).
\label{Rtensor}
\end{align}
The Riemann curvature is a reducible representation:  an irreducible decomposition of the Riemann curvature has the following visualization\cite{fre2012gravity} using Young tableaux as
\begin{align*}
\ytableausetup{aligntableaux=top}
\ytableausetup{boxsize=1.25em}
R^{ab}_{\hspace{.9em}cd}\Rightarrow
   \ytableaushort{a,b}\otimes\ytableaushort{c,d}&= \ydiagram{2,2}\oplus\ydiagram{2,1,1}\oplus\ydiagram{1,1,1,1}
   =\widehat{\ydiagram{2,2}}\oplus\widehat{\ydiagram{2}}\oplus\bullet\oplus\widehat{\ydiagram{2,1,1}}\oplus\ydiagram{1,1}\oplus\ydiagram{1,1,1,1},
\end{align*}
where horizontal (vertical) boxes show  (anti-)symmetric indices, respectively.
A hat over two horizontal tableaux indicates the traceless nature of the symmetric tensor.
In general, the symmetric tensor gives decomposition like
\begin{align*}
\ydiagram{2}&=\widehat{\ydiagram{2}}\oplus\bullet,
\end{align*}
where $\bullet$ shows a scalar of the tensor's trace.
The Bianchi identity eliminates the last three of six irreducible representations of the Riemann curvature tensor.
As a result, the Riemann curvature is decomposed into three irreducible curvature tensors as\begin{align*}
R^{ab}_{\hspace{.9em}cd}&=\mathscr{W}^{ab}_{\hspace{.9em}bd}+\frac{2}{3}\delta^{[a}_{[c}\eta^{b]\bcdot}\hat{R}_{d]\bcdot}+\frac{1}{12}\(\delta^{a}_{c}\delta^{b}_{d}+\delta^{a}_{d}\delta^{b}_{c}\)R,
\intertext{where $\mathscr{W}^{ab}_{\hspace{.9em}cd}$ is the Weyl curvature tensor and $\hat{R}_{ab}$ is the symmetric traceless Ricci curvature tensor related with the Ricci curvature tensor as}
\hat{R}_{ab}&={R}_{ab}-\frac{1}{4}\eta^{~}_{ab}R.
\end{align*}
In reality, the Einstein equation consists of irreducible representations: the Ricci curvature tensor, the scalar curvature, and the cosmological constant.

\paragraph{b) vierbein and perturbed metric filed:}
The vierbein form has a coefficient representation as
\begin{align}
\Omega^1(\TsMM){\otimes}V^1(\TML)\ni\eee=
\eee^{a}\partial_a=\Varepsilon^{\hspace{.4em}a}_\mu(x\in\MM)\hspace{.1em}dx^\mu\hspace{.2em}\partial_a,\label{Eamutensor}
\end{align}
where $\eee$ is the cotangent vector (differential form) in the global manifold and, at the same time,  the tangent vector in the local inertial bundle.
The coefficient function $\Varepsilon^a_\mu$ is defined in the global manifold; thus, we treat it as the global vector.
Irreducible representations of the corresponding metric tensor are given as
\begin{align}
g_{\mu\nu}=\eta^{~}_\bcdots\hspace{.1em}\Varepsilon^\bcdot_\mu\hspace{.1em}\Varepsilon^\bcdot_\nu
\Rightarrow
   \ytableaushort{\mu}\otimes\ytableaushort{\nu}&=\ydiagram{2}\oplus\ydiagram{1,1}
   =\widehat{\ydiagram{2}}\oplus\bullet\oplus\ydiagram{1,1}.
\end{align}
Among three irreducible representations, the antisymmetric representation is zero for the symmetric tensor, and representation $\ytableausetup{boxsize=0.8em}\widehat{\ydiagram{2}}$ gives a graviton.

When we set metric tensor perturbation as 
\begin{align}
g_{\mu\nu}(x\in\MM)&=\eta_{\mu\nu}+h_{\mu\nu}(x\in\MM)~~\text{with}~~h^{~}_{\mu\nu}(^{\hspace{.1em}\forall}{\hspace{-.2em}x}\in\MM)\ll1,
\intertext{and introduce the symmetric traceless tensor as}
\gamma_{\mu\nu}(x)&:=h_{\mu\nu}(x)-\frac{1}{\bm\eta\cdot\bm\eta}\eta_{\mu\nu}\(h_{\rho\sigma}(x)\eta^{\rho\sigma}\)~~\text{with}~~\bm\eta\cdot\bm\eta:=\eta_{\mu\nu}\hspace{.1em}\eta^{\mu\nu}
\implies\eta^{\mu\nu}\gamma_{\mu\nu}(x)=0,
\end{align}
the perturbed field $\gamma_{\mu\nu}(x)$ belongs to  $\widehat{\ydiagram{2}}$ and gives the dynamic field.
It has two degrees of freedom corresponding to a massless spin-two boson, i.e., the graviton.

Next, we look at metric tensor representations from the angular momentum coupling point of view.
The vierbein is a spin-1 vector in the global manifold; thus, it has three spin stats, i.e., $s^{~}_3=0,\pm1$.
Angular momentum coupling of two vierbein fields provide three multiplets of $\SU(2)$ as
\begin{align}
\bm3\otimes\bm3&=\bm5\oplus\bar{\bm3}\oplus\bm1,
\end{align}
where $\bm5$ is a symmetric spin-2 quintet, $\bar{\bm3}$ is an antisymmetric spin-1 triplet and $\bm1$ is a spin-0 symmetric singlet.
The triplet stats do not appear since the metric tensor is the symmetric production of vierbeins.
The spin-2 quintet corresponds to the graviton.
In reality, among five states of $s^{~}_3=\pm2,\pm1,0$, spin-two components $s^{~}_3=\pm2$ are survived for the massless graviton.
 
\paragraph{c) spin connection:}
The spin connection has a coefficient representation as
\begin{align}
\Omega^1(\TsM){\otimes}V^2(\TM)\ni\www=\frac{1}{2}\www^{ab}(\partial_a\times\partial_b)=\frac{1}{2}\omega^{\hspace{.4em}ab}_{c}\hspace{.1em}\eee^c(\partial_a\times\partial_b),
\label{wabctensor}
\end{align}
which has two antisymmetric upper-indices in $\TM$ and one vector index in $\TsM$.
The irreducible decomposition of the spin connection representation is given as
\begin{align*}
\ytableausetup{boxsize=1.25em}
\omega^{\hspace{.4em}ab}_{c}\Rightarrow
   \ytableaushort{a,b}\otimes\ytableaushort{c}&= \ydiagram{2,1}\oplus\ydiagram{1,1,1}
   =\widehat{\ydiagram{2,1}}\oplus\ydiagram{1}\oplus\ydiagram{1,1,1}.
\end{align*}
Basis $\eee^a{\in}\Omega^1(\TsM)$  and $\partial_b{\in}V^1(\TM)$ are dual to each other as 
\begin{align*}
\eee^a\partial_b=\Varepsilon^a_\mu\hspace{.1em}\Varepsilon_b^\nu dx^\mu\(\frac{\partial}{\partial x^\nu}\)
=\delta^a_b;
\end{align*}
thus, the last representation is null.
We denote corresponding tensors as
\begin{subequations}
\begin{align}
\ytableaushort{a}&\Rightarrow\omega^a:=\omega^{\hspace{.4em}a\bcdot}_\bcdot\label{omegaa},\\
\widehat{\ytableaushort{{\scriptstyle\hat{b}}{\scriptstyle\hat{b}},a}}&\Rightarrow\omega^{a\hat{b}}:=\omega^{\hspace{.4em}a\hat{b}}_{\hat{b}}
&\text{(do not take the summation for the hatted index)},\label{omegab}\\
\ytableaushort{a,b,c}&\Rightarrow\overline{\omega}^{\hspace{.4em}ab}_c=0&\text{(completely anti-symmetric concerning three indices)},\label{totalasymm}
\end{align}
where $\omega^{a\hat{b}}$ is ${\omega}^{\hspace{.4em}ab}_c$ with $b=c$ eliminating null components of $\overline{\omega}^{\hspace{.4em}ab}_c$ and diagonal components.
\end{subequations}

Number of spin connection's total degrees of freedom (d.o.f$.$) is 12 after requiring constraint (\ref{totalasymm}).
We require four components of the spin connection (\ref{omegaa}) are independent among them.
A covariant gauge fixing condition
\begin{align}
\partial_\bcdot\omega^\bcdot=0\label{gaugeomega}
\end{align}
and  the massless condition eliminates two d.o.f. 
Thus, the spin-connection has two physical d.o.f$.$ corresponding to the right-handed and left-handed circular polarisation.

When the vierbein is given, the spin-connection can be determined by algebraically solving the torsion-less equation.
Reversely, when the spin-connection is given, the vierbein can be determined by solving the torsion-less equation as a first-order partial differential equation.

%% file: Sec2-6.tex
%
%
\subsection{Dirac operator}\label{Diracoperator}
The standard theory of particle physics utilises the differential operator contracting with the Clifford algebra, namely the Dirac operator, to describe the electron's equation of motion.
Dirac invented it as the ``\emph{square-root of the Laplacian}''.
The Dirac operator flips the chirality of the spinor; it induces the Atiyah--Singer index theorem\cite{Atiyah:1968mp}.
The Dirac operator is generally introduced in the $\Z_2$-grading vector space owing to the Clifford algebra. 
This study defines the Dirac operator following Ref.\cite{berline2003heat} and provides representations of dimension-2 and dimension-4.
We provide an application of the index theorem in \textbf{Appendix \ref{IAMS}}.
\begin{definition}[Dirac operator]\label{DiracOp}
Suppose $\VV^{\pm}$ is the $\Z_2$-grading vector space and $\xi^\pm{\in}\VV^{\pm}$.
The Dirac operator $\slash{\partial}$ is a first-order differential operator flipping a chirality such that:
\begin{align*}
\slash{\partial}:{\VV^{\pm}}\rightarrow {\VV^{\mp}}:
\xi^\pm\mapsto\slash{\partial}\xi^\pm\in{\VV^{\mp}}.
\end{align*}
A Dirac operator squared, $\slash{\partial}^2$, is called a generalized Laplacian. 
\end{definition}
\noindent
We construct the Dirac operator using the contraction concerning the Clifford algebra as
\begin{align}
\ds&:=\iota^{~}_{\gamma}d:
\Gamma\left(\M,\Wedge^0\right)
\xrightarrow{d}
\Gamma\left(\M,\Wedge^{1}\right)
\xrightarrow{\iota^{~}_{\gamma}}
\Gamma\left(\M,\Wedge^0\right),\label{DSG}
\end{align}
where $\iota^{~}_{\gamma}$ is the contraction with the corresponding Clifford algebra.
Owing to the Clifford algebra, the Dirac operator flips an undotted spinor to the dotted one in the Lorentzian metric space, and vice versa, like
\begin{align}
\bm\xi\in\VV^\textsc{l}_{\hlf}\rightarrow\ds\bm\xi\in\VV^\textsc{l}_{\overline\hlf}~~\text{and}~~
\dot{\bm\xi}\in\VV^\textsc{l}_{\overline\hlf}\rightarrow\ds\dot{\bm\xi}\in\VV^\textsc{l}_{\hlf},\label{dotundotflip}
\end{align}
which follows from (\ref{GggG}) and (\ref{PpmSL}).
Thus, it flips the chirality and maintains the definition.

\subsubsection{representation of dimension-2}\label{2x2Dirac}
\paragraph{a) Euclidean metric:}
The representation of dimension-2 of the Dirac operator in the Euclidean metric space is provided as 
\begin{subequations}
\begin{align}
\iota^{~}_{\gamma}\ds_\SE&=
\slash{\partial}_\SE:=
\gamma_\SE^\bcdot\hspace{.1em}\partial_\bcdot=\(
\begin{array}{cc}
0&h^\bcdot_\textsc{e}\hspace{.1em}\partial_\bcdot \\
\bar{h}^\bcdot_\textsc{e}\hspace{.1em}\partial_\bcdot &0
\end{array}
\),\label{gammadelE}
\intertext{where}
h^a_\textsc{e}&:=\(1,i,j,k\)~~\text{and}~~\bar{h}^a_\textsc{e}:=\(1,-i,-j,-k\),
\end{align}
\end{subequations}
owing to the Clifford algebra (\ref{2x2Cl}).
Direct calculations yield that
\begin{align*}
P^\pm_s{\slash{\partial}}_\SE P^\pm_s=0
~~\text{and}~~
P^\mp_s{\slash{\partial}}_\SE P^\pm_s={\slash{\partial}}_\SE P^\pm_s.
\end{align*}
Thus, the differential operator (\ref{gammadelE}) is the Dirac operator. 
We can also confirm $\({\slash{\partial}}_\SE\)^2=\Delta_\textsc{e}$ by direct calculations.

\paragraph{b) Lorentzian metric:}
Depending on the operand is undotted or dotted spinors, there are two types of the ($\TxT$)-Dirac operator in the Lorentzian metric space such as:
\begin{align}
\ds_\textsc{sl}\bm\xi&=
\iota^{~}_{\gamma}d_\textsc{sl}\bm\xi:=
\gamma_\textsc{sl}^\bcdot(\partial_\bcdot\bm\xi)=:{\slash{\partial}}_\Sl,~~~~
\overline{\ds}_\textsc{sl}\dot{\bm\xi}=
\iota^{~}_{\overline\gamma}\overline{d}_\textsc{sl}\dot{\bm\xi}:=
\overline\gamma_\textsc{sl}^\bcdot(\partial_\bcdot\dot{\bm\xi})=:\overline{\slash{\partial}}_\Sl\bm\xi,\label{gammadelL}
\end{align}
where $\bm{\gamma}^{~}_\textsc{sl}$ and $\overline{\bm{\gamma}}^{~}_\textsc{sl}$ are given in (\ref{gammaSE}).
The Dirac operator (\ref{gammadelL}) yields $\({\slash{\partial}}_\textsc{sl}\)^2=\(\overline{\slash{\partial}}_\textsc{sl}\)^2=\Delta_\textsc{l}\bm{1}_2$.
We can confirm the chirality flip owing to the Dirac operator in this representation by direct calculations from (\ref{gammadelL}).

\paragraph{c) ($\TxT$) differential operator:}
We introduce the ($\TxT$)-differential operators as
\begin{subequations}
\begin{align}
{\bm\partial}_\Sl&:={\bsigE}^t\cdot\bm\partial\hspace{.4em}
=\bm{1}^{~}_2\hspace{.1em}\partial_0-i\bm\sigma^1\partial_1-i\bm\sigma^2\partial_2-i\sigma^3\bm\partial_3=
  \left(\hspace{-.5em}
      \begin{array}{rr}
        \partial_0-i\hspace{.1em}\partial_3 &-i\hspace{.1em}\partial_1 -\hspace{.45em}\partial_2\\
         -i\hspace{.1em}\partial_1+\hspace{.45em}\partial_2 &\partial_0 +i\hspace{.1em}\partial_3\\
      \end{array}
     \right),\label{SigmadelE}\\
{\bm\partial}_\textsc{sl}&:={\bm\sigma^{+}_\textsc{l}}^t\cdot\bm\partial=
\bm{1}^{~}_2\hspace{.1em}\partial_0+\hspace{.3em}\bm\sigma^1\partial_1+\hspace{.3em}\bm\sigma^2\partial_2
+\hspace{.3em}\bm\sigma^3\partial_3=\hspace{.2em}
  \left(
      \begin{array}{rr}
        \partial_0+\hspace{.45em}\partial_3 &\partial_1 - i\hspace{.1em}\partial_2\\
        \partial_1+ i\hspace{.1em}\partial_2 &\partial_0 -\hspace{.45em}\partial_3\\
      \end{array}
     \right),\label{SigmadelM}
\end{align}
those are \textbf{not} the Dirac operator.
Operators (\ref{SigmadelE}) and  (\ref{SigmadelE}) act on the ($\TxT$)-vector as
\end{subequations}
\begin{align*}
\<\bm\partial_{\textsc{s}\bullet},\v_{\textsc{s}\bullet}\>&=
\partial_0v^0\pm\partial_1v^1\pm\partial_2v^2\pm\partial_3v^3,
\end{align*}
where an upper (lower) sign for the Euclidean (Lorentzian) metric, respectively.
The inner product is defined by (\ref{<vu>VE}) for the Euclidean metric and (\ref{<uv>L}) for the Lorentzian metric, respectively.

\subsubsection{representation of dimension-4}\label{4x4Dirac}
\paragraph{a) Lorentzian metric:}
The ($\FxF$) Dirac operator acts on the Dirac spinor from the left as
\begin{align}
\ds_{\VL}\psi&=\iota^{~}_{\gamma}d\psi=\iota^{~}_{\gamma}\((\partial_j\psi)\eee^j\)
=\gamma^\bcdot_{\VL}\hspace{.1em}\partial_\bcdot\psi=:\slash{\partial}_{\VL}\psi,\label{diracd}
\end{align}
In the $\Z_2$-grading vector space introduced in (\ref{GammaxiA}) and (\ref{GammaxiC}), we obtain
\begin{align}
\psi^\pm{\hspace{.1em}\in\hspace{.1em}}\VV^{{\textsc{l}}^\pm}_{\hspace{-.2em}\hlf\otimes\overline\hlf}\hspace{.2em}
\Longrightarrow\hspace{.2em}
\slash{\partial}_{\VL}\psi^\pm&=\slash{\partial}_{\VL}\(\frac{\bm{1}_\Sp\pm\Gamma_{\hspace{-.1em}\VL}}{2}\psi\)=
\frac{\bm{1}_\Sp\mp\Gamma_{\hspace{-.1em}\VL}}{2}\(\slash{\partial}_{\VL}\psi\)\hspace{.3em}{\hspace{.1em}
\in\hspace{.1em}}\hspace{.3em}\VV^{{\textsc{l}}^\mp}_{\hspace{-.2em}\hlf\otimes\overline\hlf},\label{Vbullet}
\end{align}
where $\{\psi^+_{~},\psi^-_{~}\}=\{\psi^R,\psi^L\}$ given by (\ref {phil}) and (\ref {phir}).
Thus, differential operator $\slash{\partial}_\VL$ in (\ref{diracd}) fulfil the definition of the Dirac operator.

A square of the Dirac operator provides
\begin{align*}
\slash{\partial}^2_{\VL}&=
\(
\(\partial_0\)^2-\(\partial_1\)^2-\(\partial_2\)^2-\(\partial_3\)^2
\)\bm{1}_\Sp=:\Delta^{~}_{{\VL}},
\end{align*}
thus, $\Delta^{~}_\VL$ is the Laplacian in the four-dimensional Lorentzian metric space.

\paragraph{b) Euclidean metric:}
The representation of dimension-4 of the Euclidean Dirac operator can be obtained by replacing the Lorentzian Clifford algebra and the metric tensor to the Euclidean ones denoted by $\bullet_\VE$  as
\begin{align}
\ds_{\VE}&=\iota^{~}_{\gamma}d=\iota^{~}_{\gamma}\((\partial_j)\eee^j\)
=\gamma^\bcdot_{\VE}\hspace{.1em}\partial_\bcdot=:\slash{\partial}_{\VE},\label{diracdE}
\intertext{yielding}
\slash{\partial}^2_{\VE}&=
\(
\(\partial_0\)^2+\(\partial_1\)^2+\(\partial_2\)^2+\(\partial_3\)^2
\)\bm{1}_\Sp=:\Delta^{~}_{{\VE}}.\notag
\end{align}

\paragraph{c) Lorentzian Dirac operator on Higgs spinor:}
We prepare the Lorentzian Dirac operator for the Higgs spinor and its bilinear form here.
In the standard theory of particle physics, the Higgs field is the $\SU(2)$ spinor defined in the Lorentzian manifold.
Thus, we treat the Lorentzian Dirac operator applying on the $\SU(2)$ spinor in the Yang--Mills theory.
The Lorentzian Dirac operator (\ref{gammadelL}) acts on the Higgs spinor as
\begin{align}
\ds^{~}_\Sl\bphih&=\slash{\partial}^{~}_\Sl\bphih,
~~
\text{yielding}
~~
\slash{\partial}^{~}_\Sl\(P^\pm_{\hspace{-.1em}\SE}\bphih\)=P^\mp_{\hspace{-.1em}\SE}\(\slash{\partial}^{~}_\Sl\bphih\),
\end{align}
owing to the formal relation  $P^\pm_{\hspace{-.1em}\SE}=P^\pm_{\hspace{-.1em}\Sl}$.
Therefore, the Lorentzian Dirac operator fulfils \textbf{Definition \ref{DiracOp}} for the Higgs spinor.

We introduce a co-Dirac operator as 
\begin{align}
i\cds^{~}_\Sl&:=i\ds^\dagger_\Sl,\quad\text{yielding a weak relation}\quad
 \<i\cds^{~}_\Sl\bphih,\bphih\>=\<\bphih,i\ds^{~}_\Sl\bphih\>
\intertext{such that}
\int_\ML\<i\cds^{~}_\Sl\bphih,\bphih\>&=
-i\int_\ML\({\gamma^\bcdot_\Sl}^\dagger\hspace{.1em}\partial_\bcdot\bphih\)^\dagger\bphih=
-i\int_\ML\(\partial_\bcdot\bphih^\dagger\){\gamma^\bcdot_\Sl}{\bphih},\notag\\&=
-i\int_{\ML}\partial_\bcdot\(\bphih^\dagger{\gamma^\bcdot_\Sl}{\bphih}\)
+i\int_\ML\bphih^\dagger{\gamma^\bcdot_\Sl}(\partial_\bcdot{\bphih}),\notag\\&=
\int_\ML\<\bphih,i\ds^{~}_\Sl\bphih\>,\notag
\end{align}
where we utilise the integration by parts, and assume the vanishing Higgs spinor at the boundary. 
The Dirac operator acts on the Euclidean bilinear form as
\begin{align*}
\ds^{~}_\Sl\<\bphih,\bphih\>&:=\<\cds^{~}_\Sl\bphih,\bphih\>+\<\bphih,\ds^{~}_\Sl\bphih\>=
2\<\bphih,\ds^{~}_\Sl\bphih\>.
\end{align*}

%% file: Sec-3-YMU.tex
\section{Yang--Mills--Utiyama (\YMU\!) theory}\label{YMUtheory}
We refer to the Yang--Mills theory treating the local space-time $\SO(1,3)$ on equal footing with the internal gauge symmetry as the \emph{Yang--Mills-Utiyama} (\YMU\!) theory in this report.
This section constructs the \YMU theory in the inertial \textbf{Lorentzian} manifold based on mathematical tools prepared up to the previous section.
The subscript ``\textsc{l}'' is omitted for simplicity through this section.
The \YMU theory treated in this study is the gauge theory with internal $\SU(2)$ symmetry and local Lorentz symmetry.
We formulate the theory in the inertial manifold; after fixing the frame and treating the theory to the local inertial frame, the global $GL(4)$ symmetry is broken while keeping the local $\SO(1,3)$ symmetry.
The local information of the global manifold is encapsulated in the spin-connection form $\www$ in this formalism. 

%
%
\input{Sec2-spinorbundle}

%
%
\input{Sec2-gaugebundle}

%
%
\input{Sec2-spinorgaugebundle}

%
%
\input{Sec2-Lagrangian}

%
%
\input{Sec2-TSE}

This section constructs the \YMU theory in the inertial \textbf{Lorentzian} manifold based on mathematical tools prepared up to the previous section.
Thus, we omit the subscript ``\textsc{l}'' for simplicity.
The \YMU theory treated in this study is the gauge theory with internal $\SU(2)$ symmetry and local Lorentz symmetry.
We formulate the theory in the inertial manifold; after fixing the frame and treating the theory to the local inertial frame, the global $GL(4)$ symmetry is broken while keeping the local $\SO(1,3)$ symmetry.
The local information of the global manifold is encapsulated in the spin-connection form $\www$ in this formalism.

%% file: Sec2-spinorbundle.tex
\subsection{Principal bundles}\label{Principalbundle}
We introduce principal bundles with the structure groups $Spin(1,3)$ and $\SU(2)$ and twist them together as the Whitney sum. 
The \YMU theory is formulated in the twisted spinor-gauge bundle, where internal and space-time symmetries are treated equally.

\subsubsection{Spinor bundle}\label{spinerbundle}
Suppose Lorentzian inertial manifold $\M$ is globally equipping a spin structure.
We define the principal spinor bundle as 
\begin{align*}
\(\M{\otimes}\(\VV^{\textsc{l}}_{\hspace{-.2em}\hlf}\oplus\VV^{\textsc{l}}_{\hspace{-.2em}\overline\hlf}\),\pi_\Sp,\M, G^{~}_{\hspace{-.1em}\cP}{\otimes}Spin(1,3)\).
\end{align*}
A total space is the trivially combined Weyl-spinor spaces (or the Dirac-spinor space) with Lorentzian manifold.
Projection map $\pi_\Sp$ is provided owing to the covering map,
\begin{align*}
&\tau_\cov:Spin(1,3)\rightarrow SO(1,3)\otimes\{R,L\},
\intertext{yeilding}
&\pi_\Sp:=\tau_\cov/\{R,L\}:
\M{\otimes}\(\VV^{~}_{\hspace{-.2em}\hlf}\oplus\VV^{~}_{\hspace{-.2em}\overline\hlf}\)
\rightarrow\M/\{R,L\}:
\left.{\bm\xi}\right|_p\hspace{.1em}\mapsto p\in\M,
\end{align*}
where $\bm\xi$ is the Weyl spinor representing both dotted and undotted spinors defined in \textbf{section \ref{WS}} (or the Dirac spinor defined in \textbf{section \ref{DS}}), and it is the section $\bm\xi\in\Gamma_\Sp\hspace{-.3em}\left(\M,\VV^{~}_{\hspace{-.2em}\hlf}\oplus\VV^{~}_{\hspace{-.2em}\overline\hlf},Spin(1,3)\right)$.
The structure group of the inertial bundle, $SO(1,3)$, is lifted to $Spin(1,3)$ owing to projection map $\pi_\Sp$.

The connection of the spinor bundle for dotted and undotted Weyl spinors are provided, respectively,  as
\begin{align}
\overline{\AAA}_\Sp:=\Tr^{~}_\Cl[\www\hspace{.1em}\overline{S}]
&=\frac{i}{2}
\www_{~}^{\bcdots}\hspace{.1em}{\overline{S}}_{\bcdots},~~~~
\AAA_\Sp:=\Tr^{~}_\Cl[\www{S}]
=\frac{i}{2}
\www^{\bcdots}{S}_{\bcdots},\label{connectSp}
\end{align}
where $\overline{S}$ and $S$ are generators of $Spin(1,3)$ given as
\begin{align}
\overline{S}^{ab}_{~}&:=
i\hspace{.1em}\left[\frac{\overline{\gamma}^a_{~}}{2},\frac{\overline{\gamma}^b_{~}}{2}\right],~~~~
S^{ab}_{~}:=
i\hspace{.1em}\left[\frac{\gamma^a_{~}}{2},\frac{\gamma^b_{~}}{2}\right].\label{Asp}
\end{align}
The connection $\overline{\AAA}_\Sp$ and $\AAA_\Sp$  provide curvatures through the structure equation as
\begin{align}
\overline{\FFF}_\Sp&:=d\overline{\AAA}_\Sp+i\hspace{.1em}\cG\overline{\AAA}_\Sp\wedge\overline{\AAA}_\Sp=
\frac{1}{2}\overline{S}^{~}_{\bcdots}\hspace{.1em}\RRR_{~}^{\bcdots},~~~~
\FFF_\Sp:=d\AAA_\Sp+i\hspace{.1em}\cG\AAA_\Sp\wedge\AAA_\Sp=
\frac{1}{2}S^{~}_{\bcdots}\hspace{.1em}\hspace{.1em}\RRR_{~}^{\bcdots}.
\end{align}
Covariant Dirac operators concerning $Spin(1,3)$ on the dotted and undotted Weyl spinor are provided, respectively, as
\begin{align}
\overline{\ds}_{\Sp}^{~}\dot{\bm\xi}&:=\iota^{~}_{\gamma}\overline{d}_\Sp^{~}\dot{\bm\xi}=\(\overline{\slash{\partial}}
-i\hspace{.1em}\cG\hspace{.1em}\overline{\slash{\AAA}}_\Sp\)\dot{\bm\xi},~~
\dSp\bm\xi:=\iota^{~}_{\gamma}d_\Sp\bm\xi=\(\slash{\partial}
-i\hspace{.1em}\cG\hspace{-.2em}\slash{\AAA}_\Sp\)\bm\xi.\label{DiracSp}
\end{align}
Although the Lorentzian Clifford algebra consists of quaternions, we denote the imaginary unit as $i\in\C$ instead of $\sqrt{-1}$ here since Clifford-algebraic calculations without the concrete quaternion representation are enough to discuss details of the \YMU theory.
When we perform direct calculations using the component representations, we use $\sqrt{-1}\in\C$ treating $\bm\gamma^{~}_\Sl$ as the Clifford module $\C\otimes\HQ(2)$ given in (\ref{CMCH1}) and (\ref{CMCH2}).

%% file: Sec2-gaugebundle.tex
\subsubsection{Gauge bundle}\label{YMgauegbundle}
In the inertial manifold, we treat a principal bundle with a Lie group, namely the principal gauge bundle.
We exploit the $\SUW$ group as the structure group of the principal bundle.
We denote this gauge group as $\SUW$ to distinguish that owing to Euclidean space-time. 
The Higgs spinor appears as a section of the gauge bundle.

A principal gauge bundle is defined as a tuple such that:
\begin{align*}
\(\M\otimes\VV^{\textsc{e}}_{\hlf},\pi^{~}_{\SU},\M,\SUW\).
\end{align*}
We introduce the gauge group in the inertial bundle and treat two fundamental fields, the gauge connection $\AAA^{~}_\SU$ and the spinor section field $\bphih$.
A space of the $\SUW$-spinor is introduced as a section $\bphih{\hspace{.1em}\in\hspace{.1em}}\Gamma\left(\M,\VV^{\textsc{e}}_{\hlf},{\SUW}\right)$ belonging to the fundamental representation of the $\SUW$ symmetry.
The Higgs spinor is the $\SUW$-spinor and is two independent scalar functions concerning the local Lorentz group.

A gauge connection $\AAA^{~}_{\SU}:=\AAA_\SU^I\hspace{.2em}\tau^{~}_I\in\Omega^1(\TsM){\otimes}Ad(\sss\uuu(2))$ is a Lie-algebra valued one-form object, where $\tau^{~}_I$ is a generator of the $\SUW$ group and has the ($\TxT$) Hermitian matrix representation using the Pauli matrix, such as $\tau_I^{~}=\bm\sigma^I/2$.
Connection $\AAA^{~}_\SU$ belongs to an adjoint representation of the gauge group transformed like
\begin{align}
\Gsu\hspace{-.2em}\left(\AAA^{~}_\SU\right)&=
\bm{g}^{-1}_\SU\hspace{.2em}\AAA^{~}_\SU\hspace{.2em}\bm{g}_\SU^{~}
+i\hspace{.1em}{c^{-1}_\SU}\gdg=\bm{g}^{-1}_\SU\hspace{.2em}\AAA^{~}_\SU\hspace{.2em}\bm{g}_\SU^{~},\label{pisu}
\end{align}
where $\gdg=0$ is used.
At the same time, $\AAA^{~}_\SU$ is a vector in $\TsML$ and is Lorentz transformed as
\begin{align}
\Gso:End\(\Omega^1(\TsM)\):
\AAA^{~}_\SU\mapsto\Gso(\AAA^{~}_\SU)&=
\LLambda\hspace{.1em}\AAA^{~}_\SU.\label{SUcd}
\end{align}
Corresponding gauge curvature two-form $\FFF_\SU^{~}$ is defined through a structure equation such that:
\begin{align}
\FFF_\SU^{~}=\FFF_\SU^I\hspace{.2em}\tau^{~}_I&:=d\AAA^{~}_\SU
-i\hspace{.1em}{{c^{~}_\SU}}\hspace{.1em}\AAA^{~}_\SU\wedge\AAA^{~}_\SU
=\left(d\AAA_\SU^I
+\frac{{c^{~}_\SU}}{2}f^I_{~JK}\hspace{.1em}\AAA_\SU^J\wedge\AAA_\SU^K
\right)\tau^{~}_I{\hspace{.2em}\in\hspace{.2em}}\Omega^2(\TsM){\otimes}Ad(\sss\uuu(2)).
\label{SUCRV}
\end{align}
The covariant differential on $p$-form object $\aaa\in\Omega^p(\TsML)$  concerning the $\SUW$ symmetry is provided as
\begin{align}
d^{~}_{\hspace{-.1em}g}\hspace{.1em}\aaa\hspace{.1em}:=
\bm1_\SU\hspace{.1em}d\aaa-\frac{i}{2}\hspace{.1em}{{c^{~}_\SU}}\hspace{.1em}[\AAA^{~}_\SU,\aaa]_\wedge=
\bm1_\SU\hspace{.1em}d\aaa-i\hspace{.1em}{{c^{~}_\SU}}\hspace{.2em}\AAA^{~}_\SU\wedge\aaa.\label{cdSU}
\end{align}
Real constant ${c^{~}_\SU}\in\R$ is a coupling constant of the gauge interaction in physics.
The first and second Bianchi identities are
\begin{align}
\(d^{~}_{\hspace{-.1em}g}{\wedge}d^{~}_{\hspace{-.1em}g}\)\aaa^a=
-i{c^{~}_\SU}\hspace{.1em}\FFF_{\SU}\wedge\aaa^a
\quad\textrm{and}\quad
d^{~}_{\hspace{-.1em}g}\hspace{.1em}\FFF_{\SU}=0.\label{BianchiSU}
\end{align}
The gauge connection and curvature are represented using the standard bases in $\TsM$, respectively, as
\begin{align}
\AAA^{~}_\SU&=\AAA_\SU^I\hspace{.2em}\tau^{~}_I=:\Aa^I_{a}\eee^a\hspace{.2em}\tau^{~}_I,\quad\textrm{and}\quad
\FFF_\SU^{~}=\FFF_\SU^I\hspace{.2em}\tau^{~}_I=:\frac{1}{2}\f^I_{ab}\hspace{.1em}\eee^a\hspace{-.1em}\wedge\eee^b\hspace{.2em}\tau^{~}_I.\label{Fan}
\end{align}
In this component representation, tensor coefficients of the gauge curvature is provided using the gauge connection as 
\begin{align}
(\ref{SUCRV})\implies\f^I_{ab}&=\partial_a\Aa^I_b-\partial_b\Aa^I_a+
{{c^{~}_\SU}}\hspace{.1em}f^I_{~JK}\Aa^J_a\Aa^K_b
,\label{FabISU}
\end{align}
The $\SUW$ covariant Dirac operator acts on the $\SUW$-spinor as
\begin{align}
\ds^{~}_{\hspace{-.1em}g}\hspace{.1em}\bphih&:=\(\bm1_\SU\!
\slash{\partial}-i\hspace{.1em}c^{~}_\SU\slash{\AAA}^{~}_\SU\)\bphih,\quad\text{where}\quad
\slash{\AAA}^{~}_\SU=\gamma_{~}^{\bcdot}\hspace{.2em}\Aa^I_\bcdot\tau^{~}_I.\label{cdSU}
\end{align}
The $\SUW$ generator acts on the Higgs spinor as
\begin{align}
[\tau^{~}_I\bm\phi_\textsc{h}]^{~}_A:=[\tau_I]^{\hspace{.4em}B}_{A}\hspace{.2em}\phi^{~}_B,\quad A,B\in\{\textsc{u},\textsc{d}\}.\label{TauH}
\end{align}

%% file: Sec2-spinorgaugebundle.tex
\subsubsection{Spinor-gauge bundle}\label{sgb}
A spinor-gauge bundle is a Whitney sum of spinor and gauge bundles given as
\begin{align*}
\left(\M{\otimes}\(
\VV^{\textsc{l}}_{\hlf}\oplus\VV^{\textsc{l}}_{\overline\hlf}
\){\otimes}\VV^{\textsc{e}}_{\hlf},\pi^{~}_{\Sp}\oplus\pi^{~}_{\SU},\M, G^{~}_\cP{\otimes}Spin(1,3)\otimes{\SUW}\right).
\end{align*}
The total space of the gauge bundle is lifted to the spin manifold owing to the bundle map $\pi^{~}_{\Sp}$.
A connection and a curvature are provided, respectively, as
\begin{align}
\AAA_\SG&=\AAA_\SU\otimes\bm{1}_\Sp+\bm{1}_\SU^{~}\otimes\AAA_\Sp,\quad\textrm{and}\quad
\FFF_\SG=\FFF_\SU^{~}\otimes\bm{1}_\Sp+\bm{1}_\SU^{~}\otimes\FFF_\Sp,\label{SGcc}
\end{align}
yielding the spinor-gauge covariant differential as
\begin{align}
d_{\SG}^{~}:=\(\bm{1}_\SU^{~}\otimes\bm{1}_\Sp\)d
-i\hspace{.1em}{c^{~}_\SU}({\AAA}_{\SU}\otimes\bm{1}_\Sp)
-i\hspace{.1em}\cG(\bm{1}_\SU^{~}\otimes{{\AAA}}_\Sp),\label{codiffSp1}
\end{align}

The standard theory of particle physics has the $\SUW$ doublet of Dirac spinors, e.g., the left-handed electron and neutrino.
In this report, we introduce a tensor product of Weyl- and Higgs-spinorrs, namely the \textit{dual-spinor} and define the binary operation between them as
\begin{subequations}
\begin{align}
&\bm\Xi^1=\bpsi^1\otimes\bphih^{\!1},\quad\bm\Xi^2=\bpsi^2\otimes\bphih^{\!2}\implies
\left\{
\begin{array}{ccl}
\<\bm\Xi^1,\bm\Xi^2\>&=&\<\bpsi^1,\bpsi^2\>\hspace{.2em}\<\bphih^{\!1},\bphih^{\!2}\>\\
\<\bm\xi^1,\bm\Xi^2\>&=&\<\bpsi^1,\bpsi^2\>\hspace{.2em}\bphih^{\!2}\\
\<\bm\Xi^1,\bm\xi^2\>&=&\<\bpsi^1,\bpsi^2\>\hspace{.2em}\bphih^{\!1}\\
\<\bm\Xi^1,\bphih^{\!2}\>&=&\<\bphih^{\!1},\bphih^{\!2}\>\hspace{.2em}\bpsi^1\\
\<\bphih^{\!1},\bm\Xi^2\>&=&\<\bphih^{\!1},\bphih^{\!2}\>\hspace{.2em}\bpsi^2
\end{array}\!,
\right.\label{XXxXpX}\\
&\bpsi\otimes\bphih^{\!1}+\bpsi\otimes\bphih^{\!2}=\bpsi\otimes(\bphih^{\!1}+\bphih^{\!2}),\quad
\bpsi^1\otimes\bphih+\bpsi^2\otimes\bphih=(\bpsi^1+\bpsi^2)\otimes\bphih.
\end{align}
\end{subequations}
\begin{subequations}
We introduce a weak pair of undotted dual-spinors
\begin{align}
{\bm\Xi}^{~}_{\textsc{w}}&:={\bm\xi}^1_{~}\otimes\hbphiUh+{\bm\xi}^2_{~}\otimes\hbphiDh
\in\Gamma(\TML,\VV^\textsc{l}_{\!\hlf}\otimes\VV^\textsc{e}_{\!\hlf},Spin(1,3)\otimes{\SUW}),\label{VVXIenu1}\\ 
\intertext{and that of dotted dual-spinors}
\dot{\bm\Xi}^{~}_{\textsc{w}}&:=\dot{\bm\xi}^1_{~}\otimes\hbphiUh+\dot{\bm\xi}^2_{~}\otimes\hbphiDh
\in\Gamma(\TML,\VV^\textsc{l}_{\!\overline\hlf}\otimes\VV^\textsc{e}_{\!\hlf},Spin(1,3)\otimes{\SUW}),\label{VVXIenu2}
\end{align}
\end{subequations}
where superscripts ``$1$'' and ``$2$'' are labels to identify the vector spaces they belong to.
Their invariant bilinear form is
\begin{align}
\<\dot{\bm\Xi}^{~}_{\textsc{w}},\dot{\bm\Xi}^{~}_{\textsc{w}}\>&=
\<\dot{\bpsi}^{1}\!,\bpsi^1\>\hspace{.1em}\<\hbphiUh,\hbphiUh\>
+\<\dot{\bpsi}^{2}\!,\bpsi^1\>\hspace{.1em}\<\hbphiDh,\hbphiUh\>
+\<\dot{\bpsi}^{1}\!,\bpsi^2\>\hspace{.1em}\<\hbphiUh,\hbphiDh\>
+\<\dot{\bpsi}^{2}\!,\bpsi^2\>\hspace{.1em}\<\hbphiDh,\hbphiDh\>,\notag\\&=
\<\dot{\bpsi}^{1}\!,\bpsi^1\>+\<\dot{\bpsi}^{2}\!,\bpsi^2\>.
\end{align}
The $\SUW$ group action $\gB{r}$ acts on the Higgs-spinor in the dual-spinor as
\begin{align}
\pij{\hlf}(\gB{\textsc{e}})\hspace{.1em}\dot{\bm\Xi}^{~}_{\textsc{w}}&=
\dot{\bm\xi}^1_{~}\otimes({\gB{\textsc{e}}}^{\hspace{-.3em}\dag}\hbphiUh)+
\dot{\bm\xi}^2_{~}\otimes({\gB{\textsc{e}}}^{\hspace{-.3em}\dag}\hbphiDh),\notag\\&=
[{\gB{\textsc{e}}}^{\hspace{-.3em}\dag}]_1^{\hspace{.3em}1}\dot{\bm\xi}^1_{~}\otimes
\(\hspace{-.4em}
\begin{array}{c}
1\\0
\end{array}
\hspace{-.4em}\)+
[{\gB{\textsc{e}}}^{\hspace{-.3em}\dag}]_2^{\hspace{.3em}1}\dot{\bm\xi}^1_{~}\otimes
\(\hspace{-.4em}
\begin{array}{c}
0\\1
\end{array}
\hspace{-.4em}\)+
[{\gB{\textsc{e}}}^{\hspace{-.3em}\dag}]_1^{\hspace{.3em}2}\dot{\bm\xi}^2_{~}\otimes
\(\hspace{-.4em}
\begin{array}{c}
1\\0
\end{array}
\hspace{-.4em}\)+
[{\gB{\textsc{e}}}^{\hspace{-.3em}\dag}]_2^{\hspace{.3em}2}\dot{\bm\xi}^2_{~}\otimes
\(\hspace{-.4em}
\begin{array}{c}
0\\1
\end{array}
\hspace{-.4em}\)\!,\notag\\&=
\([{\gB{\textsc{e}}}^{\hspace{-.3em}\dag}]_1^{\hspace{.3em}1}\dot{\bm\xi}^1_{~}+
[{\gB{\textsc{e}}}^{\hspace{-.3em}\dag}]_1^{\hspace{.3em}2}\dot{\bm\xi}^2_{~}\)\otimes
\(\hspace{-.4em}
\begin{array}{c}
1\\0
\end{array}
\hspace{-.4em}\)+
\([{\gB{\textsc{e}}}^{\hspace{-.3em}\dag}]_2^{\hspace{.3em}1}\dot{\bm\xi}^1_{~}+
[{\gB{\textsc{e}}}^{\hspace{-.3em}\dag}]_2^{\hspace{.3em}2}\dot{\bm\xi}^2_{~}\)\otimes
\(\hspace{-.4em}
\begin{array}{c}
0\\1
\end{array}
\hspace{-.4em}\)\!,\notag\\&=
[{\gB{\textsc{e}}}^{\hspace{-.3em}*}]^{1}_{\hspace{.4em}A}[\dot{\bpsi}^A]_1\otimes\hbphiUh+
[{\gB{\textsc{e}}}^{\hspace{-.3em}*}]^{2}_{\hspace{.4em}A}[\dot{\bpsi}^A]_2\otimes\hbphiDh.\label{pigXi}
\end{align}
When $\bm{g}$ is a unitary matrix, $\bm{g}^t$ is also a unitary matrix; thus, we obtain
\begin{align}
\<\dot{\bm\Xi}'_{\textsc{w}},\dot{\bm\Xi}'_{\textsc{w}}\>&=
\<\dot{\bm\Xi}^{~}_{\textsc{w}},\dot{\bm\Xi}^{~}_{\textsc{w}}\>,
\quad\text{where}\quad
\dot{\bm\Xi}'_{\textsc{w}}=\pij{\hlf}(\gB{\textsc{e}})\dot{\bm\Xi}^{~}_{\textsc{w}}.
\end{align}
The $\SUW$ generator $\bm\tau$ and group action $\gB{r}$ act on the Higgs-spinor in the dual-spinor as
\begin{align}
\bm{\tau}^{~}_I\cdot\dot{\bm\Xi}^{~}_\textsc{w}&:=\dot{\bm\xi}^1_{~}\otimes\bm{\tau}^{~}_I\cdot\hbphiUh+
\dot{\bm\xi}^2_{~}\otimes\bm{\tau}^{~}_I\cdot\hbphiDh,\label{tauXi}
\end{align}
Consequently, the dual-spinor emulates the $\SUW$ symmetry for a pair of Weyl-spinors.
The above operations with a dotted dual-spinor can easily be converted to an undotted one.
The Clifford algebra with the Lorentzian metric acts on the Weyl-spinor in the dual-spinor as  
\begin{align}
\gamma^{a}\bm\Xi=(\gamma^{a}_\Sl\bpsi)\otimes\bphih&\implies
\<\dot{\bm\Xi}^*_\textsc{w},\gamma^{a}_\Sl\bm\Xi_\textsc{w}\>=
\<\dot{\bpsi}^{1*}\!,\gamma^{a}_\Sl\bpsi^1\>+\<\dot{\bpsi}^{2*}\!,\gamma^{a}_\Sl\bpsi^2\>
=\dot{\bpsi}^{1\dagger_{\hspace{-.1em}\epsilon}}\gamma^{a}_\Sl\bpsi^1+
\dot{\bpsi}^{2\dagger_{\hspace{-.1em}\epsilon}}\gamma^{a}_\Sl\bpsi^2\!,\label{GgG}
\end{align}
where $\bpsi^{\dagger_{\hspace{-.1em}\epsilon}}_{~}:=(\bpsi^{*}_{~})^T$\!.

The spinor-gauge covariant Dirac operator acts on the dotted dual-spinor like
\begin{align}
\overline{\ds}_{\SG}^{~}:&End(\VV^\textsc{l}_{\!\overline\hlf}\otimes\VV^\textsc{e}_{\!\hlf}):\dot{\bm\Xi}_\textsc{w}\mapsto
\overline{\ds}_{\SG}^{~}\dot{\bm\Xi}_\textsc{w}:=(\iota_{\bar\gamma}{d}_{\SG}^{~})\dot{\bm\Xi}_\textsc{w},\notag\\&=
{\slash\partial}\dot{\bm\Xi}_\textsc{w}
-i\hspace{.1em}\cG\hspace{.1em}\(({\overline{\slash{\AAA}}}_\Sp\hspace{.3em}\dot{\bpsi}^1_{~})\otimes\hbphiUh+
({\overline{\slash{\AAA}}}_\Sp\hspace{.3em}\dot{\bpsi}^2_{~})\otimes\hbphiDh\)
-i\hspace{.1em}{c^{~}_\SU}\!\(
\dot{\bpsi}^1_{~}\otimes(\slash{\AAA}_\SU\hspace{.1em}\hbphiUh)+
\dot{\bpsi}^2_{~}\otimes(\slash{\AAA}_\SU\hspace{.1em}\hbphiDh)\)\!,\notag\\&=
%
({\slash\partial}\dot{\bm\xi}^1_{~})\otimes\hbphiUh+({\slash\partial}\dot{\bm\xi}^2_{~})\otimes\hbphiDh\notag\\
&\hspace{1em}
-i\hspace{.1em}\cG\hspace{.1em}\(({\overline{\slash{\AAA}}}_\Sp\hspace{.3em}\dot{\bpsi}^1_{~})\otimes\hbphiUh+
({\overline{\slash{\AAA}}}_\Sp\hspace{.3em}\dot{\bpsi}^2_{~})\otimes\hbphiDh\)
-i\hspace{.1em}{c^{~}_\SU}\hspace{.1em}\Aa^I_\bcdot\overline{\gamma}^\bcdot\!\(
\dot{\bpsi}^1_{~}\otimes(\bm{\tau}_I\cdot\hbphiUh)+
\dot{\bpsi}^2_{~}\otimes(\bm{\tau}_I\cdot\hbphiDh)
\)\!,\label{codiracSp1}
\end{align}
where $\bm{\gamma_\Sl}^t=\overline{\bm{\gamma}}_\Sl$ is used.

The $\SUW$ group is the internal symmetry, and the Higgs spinor does not interact directly with the spin connection. 
Consequently, the gauge-covariant Dirac operator in the spinor-gauge bundle for the Higgs spinor is the same as in the gauge bundle given by (\ref{cdSU}).

%% file: Sec2-Lagrangian.tex
\subsection{\YMU Lagrangian and equation of motion}\label{YMUL}
The \YMU Lagrangian form consists of five sub Lagrangians as
\begin{align}
\LLL^{~}_{\textsc{y\hspace{-.1em}mu}}=\LLL^{~}_{\textsc{y\hspace{-.1em}m}}+\LLL^{~}_{\textsc{g}\hspace{-.05em}\textsc{r}}
+\LLL^{~}_{\textsc{f}_{\hspace{-.1em}\textsc{m}}}+\LLL^{~}_\textsc{h}+\LLL^{~}_{\textsc{f}_{\hspace{-.1em}\textsc{m}}\text{-}\textsc{h}},
\end{align}
respectively, the Yang--Mills gauge part, the gravitation part, the fermionic matter part, the Higgs part and the fermion-Higgs interaction part.
The first two parts represent gauge bosons of the Yang-Mills gauge group and the local Lorentz group. 
The action integral is provided in common as
\begin{align}
\I^{~}_{\hspace{-.2em}\bullet}&:=\int_\TsML\LLL^{~}_{\bullet},\quad\bullet\in
\{\textsc{y\hspace{-.1em}m},\textsc{g}\hspace{-.05em}\textsc{r},\textsc{f}_{\hspace{-.1em}\textsc{m}},\textsc{h},\textsc{m}\text{-}\textsc{h}\},
\end{align}
for each Lagrangian.
We set the Lagrangian-form and action integral to null physical dimension as $[\LLL]=[\I]=1$ in this study.

We treat the inertial {Lorentzian} manifold and continue to omit the subscript ``\textsc{l}''  in this section.
%
%
\subsubsection{Gauge bosons}\label{YMUGB}
\paragraph{a) general formula:}
The gauge-boson part of the \YMU Lagrangian has a form in common as
\begin{align}
\LLL^{~}_{G}(\AAA^{~}_G)&:=C^{~}_{\hspace{-.1em}\textsc{dim}}\hspace{.1em}\Tr^{~}_{G}\hspace{-.2em}\left[\FFF^{~}_G\wedge\hat{\FFF}^{~}_G\right]
=i\hspace{.1em}C^{~}_{\hspace{-.1em}\textsc{dim}}\hspace{.1em}\Tr^{~}_{G}\hspace{-.2em}\left[\left\|\FFF^{~}_G\right\|^2\right]\vvv,\label{LagrangianG}
\end{align}
where $C^{~}_{\hspace{-.1em}\textsc{dim}}$ is a dimensional constant to adjust the Lagrangian to have null physical dimension. 
The curvature form is defined from the connection form owing to the structure equation (\ref{adjointG}). 
The Lagrangian form is a functional of the connection, and the equation of motion is given as the stationary point of the action integral concerning the functional variation such as
\begin{align}
&\frac{\delta\I_{\hspace{-.2em}G}(\AAA^{~}_G)}{\delta\AAA^{~}_G}=0\implies 
\hat{d}_G\FFF^{~}_G:=\hat{d}\FFF^{~}_G-i\hspace{.1em}c_G\left[\hat\AAA_G,\FFF^{~}_G\right]^{~}_{\hspace{-.1em}\wedge}=0,\label{ELEoM}
\intertext{where}
&\hat{d}\aaa=\text{(\ref{codiff})},~\hat\AAA^{~}_G\wedge\aaa:=
\text{det}[\bm\eta]^{-1}(-1)^{p(n-p+1)}\hspace{.2em}\HD\(\AAA^{~}_G\wedge\HD(\aaa)\)\label{defahat}
\implies\int\<{d}_G\aaa,\bbb\>\vvv=\int\<\aaa,\hat{d}_G\bbb\>\vvv.
\end{align}
We define $\hat\AAA$ through (\ref{defahat}).
Except co-differential $\hat{d}$ and dual-connection $\hat\AAA$, $\hat\bullet:=\HD(\bullet)\in\Omega^{n-p}$ shows Hodge-dual of $\bullet\in\Omega^p$.

Simple calculations provide \ref{ELEoM}) as follows:
\begin{align}
\frac{1}{iC^{~}_{\hspace{-.1em}\textsc{dim}}}\frac{\delta\I_{\hspace{-.2em}G}^{~}(\AAA^{~}_G)}{\delta\AAA^{~}_G}
&:=\frac{1}{iC^{~}_{\hspace{-.1em}\textsc{dim}}}\lim_{t\rightarrow0}\frac{d}{dt}\I_{\hspace{-.2em}G}(\AAA^{~}_G+t\aaa)
=\lim_{t\rightarrow0}\frac{d}{dt}\int\|\FFF^{~}_G(\AAA^{~}_G+t\aaa)\|^2\vvv,\notag\\&
=2\lim_{t\rightarrow0}\int
\la\frac{d}{dt}\FFF^{~}_G(\AAA^{~}_G+t\aaa),\FFF^{~}_G(\AAA^{~}_G+t\aaa)\ra\vvv
=2\lim_{t\rightarrow0}\int\la d_G\aaa,\FFF^{~}_G(\AAA^{~}_G)\ra\vvv,\notag\\&
=2\lim_{t\rightarrow0}\int\la\aaa,\hat{d}_{G}\FFF^{~}_G(\AAA^{~}_G)\ra\vvv=0\implies(\ref{ELEoM}),\label{dFAdt}
\end{align}
where we require the boundary condition as $\delta\AAA^{~}_G=0$ at the integration boundary.
If and only if the curvature $\FFF^{~}_G$ fulfils the equation of motion (\ref{ELEoM}), (\ref{dFAdt}) is maintained for any $\aaa\in\Omega^1{\otimes}Ad(\ggg^{~}_G)$.
In contrast with (\ref{ELEoM}), the similar relation $d_{G}\FFF^{~}_G=0$ is not the equation of motion but the Bianchi identity.

\paragraph{b) Yang--Mills gauge boson:}
There are two gauge-bosons in the \YMU theory; the Yang--Mills gauge boson and the graviton.
We start from the Yang--Mills gauge boson part;
\begin{align}
\LLL^{~}_{\textsc{y\hspace{-.1em}m}}(\AAA^{~}_\SU)&:=
i\hspace{.2em}\Tr^{~}_{\SU}\hspace{-.2em}
\left[\|\FFF^{~}_\SG(\AAA^{~}_\SU)\|^2\right]\vvv=
\Tr^{~}_{\SU}\hspace{-.2em}
\left[\FFF^{~}_\SG(\AAA^{~}_\SU)\wedge\hat{\FFF}^{~}_\SU(\hat\AAA^{~}_\SU)\right],\label{YMLL}
\end{align}
Owing to Lagrangian (\ref{YMLL}), we obtain the equation of motion as
\begin{align}
\frac{\delta\I^{~}_{\hspace{-.2em}\textsc{y\hspace{-.1em}m}}(\AAA^{~}_\SU)}{\delta\AAA^{~}_\SU}=\dot{0}\implies
\hat{d}_\SU\FFF^{~}_\SU=\dot{0},\label{YMGBEoM}
\end{align}
where $\dot{0}$ shows it is zero considering one of five sub-Lagrangians.
Since $\I_{\hspace{-.2em}\textsc{f}_{\hspace{-.1em}\textsc{m}}}$ also has the $SU(2)$ gauge field;thus, equation (\ref{YMGBEoM}) is not completed yet.
After preparing all equations of motion using the whole Lagrangian, we sum them up according to the field differentiating the action. 
The Yang--Mills gauge boson Lagrangian has a component representation as
\begin{align}
\text{(\ref{YMLL})}&=\frac{1}{4}\f^I_\bcdots\f_I^\bcdots\hspace{.1em}\vvv,~~\text{where}~~\f^I_{ab}=\text{(\ref{FabISU})}.\label{YMGBEoM2}
\end{align}
We omit a subscript $\SU$ on the component representation for simplicity.
Consequently, the equation of motion for the Yang--Mills gauge field in a vacuum is provided with the component representations as
\begin{align}
\etaL^\bcdots\(\hspace{.1em}\partial_\bcdot\f^I_{\hspace{.3em}\bcdot a}+
c^{~}_\SU\hspace{.1em}f^{I}_{\hspace{.3em}JK}\hspace{.1em}\Aa^J_{\hspace{.2em}\bcdot}\f^K_{\hspace{.3em}\bcdot a}\)\tau_I&=\dot{0}.\label{YMGBEoM3}
\end{align}

Physical dimension of the Yang--Mills gauge boson Lagrangian is computed as
\begin{align*}
\left[\partial\right]=\left[\Aa\right]=\left[\TT\right]=L^{-1},~~\left[\vvv\right]=L^4
\implies\left[\LLL^{~}_{\textsc{y\hspace{-.1em}m}}(\AAA^{~}_\SU)\right]=1;
\end{align*}
thus, we set $C^{~}_{\hspace{-.1em}\textsc{dim}}=1$ in (\ref{YMLL}).

\paragraph{c) gravitational gauge boson:}
Next, we treat the Lagrangian of the gravitational gauge boson, namely the graviton;
\begin{align}
\LLL_{\textsc{g}\hspace{-.05em}\textsc{r}}(\www,\eee)&:=
-\frac{i}{\hbar\kE}\Tr^{~}_{\cP}\hspace{-.2em}
\left[\|\FFF^{~}_\cP\|^2\right]\vvv=
-\frac{1}{\hbar\kE}\Tr^{~}_\cP\hspace{-.2em}\left[\FFF_\cP\wedge\hat\FFF_\cP\right],\label{EHLL}
\end{align}
where $\kE$ is the Einstein constant of  gravitation.
The cosmological constant term is omitted for simplicity in this study.
The gravitational action yields the standard Einstein--Hilbert Lagrangian as shown in \textbf{Theorem 4.2} in Ref.\cite{doi:10.1063/1.4990708} (see also, \textbf{Remark 2.1} in Ref.\cite{Kurihara_2020}).
Here, we shortly show the result from (\ref{EHLformula}) using (\ref{cPLieA1}) and (\ref{cPLieA2}) as
\begin{align*}
{\hbar\kE}\hspace{.2em}\LLL_{\textsc{g}\hspace{-.05em}\textsc{r}}&=
-\Tr_{\cP}\left[\FFF^{~}_{\cP}\wedge\hat\FFF^{~}_{\cP}\right]=
-\FF^K_{\hspace{.3em}I\hspace{-.1em}J}\Tr^{~}_\SO\left[\FFF_\cP^I\wedge\hat\FFF_\cP^J\right]\Theta_K,\\
&=-\Tr_\SO[\RRR(\www)\wedge\SSS(\eee)]
=\frac{1}{2}\RRR^{\bcdots}(\www)\wedge\SSS_{\bcdots}(\eee),
\end{align*}
which is nothing other than the standard Einstein--Hilbert Lagrangian.
Thus, we obtain 
\begin{align}
\text{(\ref{EHLL})}&=\frac{1}{\hbar\kE}\Tr_\SO[\RRR(\www)\wedge\SSS(\eee)]=\frac{1}{\hbar\kE}R\hspace{.1em}\vvv,
\end{align}
where the rightmost expression is a component representation using the scalar curvature.
Physical dimension of the gravitational gauge boson Lagrangian is
\begin{align*}
\left[R\right]=\left[\hbar\kE\right]^{-1}=L^{-2},~~\left[\vvv\right]=L^4\implies\left[\LLL_{\textsc{g}\hspace{-.05em}\textsc{r}}\right]=1.
\end{align*}
Thus, we set $C^{~}_{\hspace{-.1em}\textsc{dim}}=(\hbar\kE)^{-1}$ in (\ref{EHLL}).
The gravitational Lagrangian has the Planck constant to set its physical dimension null. 
Although the Planck constant appears in the Lagrangian,  the theory is still classical.

The gauge boson in the Yang--Mills theory corresponds to the connection in geometry; thus, we identify the spin connection $\www^{ab}$ as the gravitational gauge-boson, namely the graviton.
In reality, $\www^{ab}$ is a massless rank-two tensor field corresponding to the spin-two boson.
In addition to the connection, Lagrangian (\ref{EHLL}) is the functional of $\eee^a$, which is a section in the spinor-gauge bundle. 
Equations of motion concerning $\www$ and $\eee$ are, respectively, provided as
\begin{subequations}
\begin{align}
\frac{\delta\I_{\hspace{-.2em}\textsc{g}\hspace{-.05em}\textsc{r}}(\www,\eee)}{\delta\eee}=\dot{0}&\implies 
\frac{1}{\kE}\GGG^{~}_{a}=\dot{0},
~~\GGG^{~}_{a}:=\frac{1}{2}\epsilon_{a\bcdot\bcdots}\hspace{.2em}\RRR^{\bcdots}
\hspace{-.2em}\wedge\eee^{\hspace{.1em}\bcdot},\label{EinsteinEq}\\
\frac{\delta\I_{\hspace{-.2em}\textsc{g}\hspace{-.05em}\textsc{r}}(\www,\eee)}{\delta\www}=\dot{0}&\implies 
\frac{1}{\hbar\kE}d_\www\eee^a=\frac{1}{\hbar\kE}\TTT^{a}_{~}=\dot{0},\label{TorsionlessEq}
\end{align}
where the first is the Einstein equation, and the second is the torsion equation.
\end{subequations}
The three-form object, $\GGG^{~}_{a}$, is called the Einstein form.
We also provide component representations of the above equations as
\begin{align*} 
\frac{1}{\hbar\kE}\GGG^{~}_{a}=\dot{0}&\implies 
\frac{1}{\hbar\kE}G^{ab}_{\hspace{-.1em}\textsc{e}}=\dot{0},{\quad\text{with}\quad}G^{ab}_{\hspace{-.1em}\textsc{e}}:=R^{ab}-\frac{1}{2}\eta^{ab}\hspace{.1em}R,\\
\frac{1}{\hbar\kE}\TTT^{a}_{~}=\dot{0}&\implies
	\frac{1}{2\hbar\kE}\(\partial_\mu\Varepsilon^a_\nu+\cG\hspace{.1em}\omega^{\hspace{.3em}a}_{\mu\hspace{.2em}\bcdot}\hspace{.1em}\Varepsilon^\bcdot_\nu
	- (\mu\leftrightarrow\nu)\)=\dot{0},
\end{align*}
where suffix ``\textsc{e}'' stands for Einstein and $G^{ab}_{\hspace{-.1em}\textsc{e}}$ called the Einstein tensor.
We note that these equations are under the vacuum condition, ignoring Lagrangians other than $\LLL^{~}_{\textsc{g}\hspace{-.05em}\textsc{r}}$.
In reality, the Yang-Mills fields induce the torsion and curvature of space-time.
Later in this report, we will provide concrete expressions of the torsion and the energy-stress tensor induced by the Yang-Mills fields.
Although it is unnecessary for the vacuum equation, we keep factor $1/\hbar\kE$ in front of equations (\ref{EinsteinEq}) and (\ref{TorsionlessEq}) for future convenience.

\subsubsection{Fermions}
In the standard theory of particle physics, an electron is represented using the Dirac spinor belonging to the representation $\psi\in\VV^{~}_{\hlf}\otimes\VV^{~}_{\overline\hlf}$.
On the other hand, there is only a dotted-spinor neutrino in nature, and it is natural to represent it using the Weyl spinor.
This report primarily treats an electron and a neutrino as massless Weyl spinors, ignoring a quark sector in the \YMU theory.

The fermion Lagrangian consists of chiral-right and chiral-left parts individually, as
\begin{align}
\LLL^{~}_{\textsc{f}_{\hspace{-.1em}\textsc{m}}}:=
\LLL^{R}_{\textsc{f}_{\hspace{-.1em}\textsc{m}}}+
\LLL^{L}_{\textsc{f}_{\hspace{-.1em}\textsc{m}}}=
\(\LL^{R}_{\textsc{f}_{\hspace{-.1em}\textsc{m}}}+\LL^{L}_{\textsc{f}_{\hspace{-.1em}\textsc{m}}}\)\vvv,\label{WspinLagLR}
\end{align} 
since the chiral-right Lagrangian consists of the Weyl spinor and the chiral-left one has a weak pair of dual spinors.
The Lagrangian density must be the Hermitian and $\SL2C$ invariant scalar.

We start from the chiral-right part, which only has an electron as a fermion.
The chiral-right Lagrangian density consists of undotted spinors as
\begin{subequations}
\begin{align}
&\LL^{R}_{\textsc{f}_{\hspace{-.1em}\textsc{m}}}\!\(\bpsi^e_{~},\bpsi^{e\dagger_{\hspace{-.1em}\epsilon}}_{~}\):=
\<\bpsi^{e*}_{~},i{\ds^{~}_{\Sp}}\hspace{.1em}\bpsi^e_{~}\>
=i\hspace{.1em}\bpsi^{e\dagger_{\hspace{-.1em}\epsilon}}_{~}\(\hspace{-.2em}{\ds^{~}_{\Sp}}\hspace{.1em}\bpsi^e_{~}\)\!,
\label{WspinLagR}
\intertext{yielding the equation of motion as}
&\frac{\delta\I^{R}_{\textsc{f}_{\hspace{-.1em}\textsc{m}}}(\bpsi^e_{~},\bpsi^{e\dagger_{\hspace{-.1em}\epsilon}}_{~})}{\delta\bpsi^{e\dagger_{\hspace{-.1em}\epsilon}}_{~}}=\dot{0}
\implies
i{\ds^{~}_{\Sp}}\hspace{.2em}\bpsi^e_{~}=\dot{0}\implies
{\gamma}^\bcdot_{~}\(
i\hspace{.1em}\partial_\bcdot-\frac{1}{2}\cG\hspace{.1em}\omega^{\hspace{.3em}\stars}_\bcdot\hspace{.1em}{S}^{~}_{\stars}
\){\bpsi}^e_{~}=\dot{0},\label{WspinEoMR}
\end{align}
\end{subequations}
where we treat the spinor $\bpsi^{e\dagger_{\hspace{-.1em}\epsilon}}_{~}$ as an independent field of $\bpsi^e_{~}$.
The equation of motion is called the Weyl equation in physics.
Lagrangian (\ref{WspinLagR}) consists of a bilinear form between the electron Weyl spinors; a bra-part is the complex conjugated spinor, and a ket-part is the spinor with the Dirac operator, which is Hermitian in total.
Since $\bm\xi^*\in(\VV^-=\VV_{\overline{\hlf}})\ni\ds\bm\xi$, the bilinear form is $\SL2C$ invariant.

For the chiral-left fermion Lagrangian, we utilise the weak pair of dual-spinors (\ref{VVXIenu2}) as
\begin{subequations}
\begin{align}
\LL^{L}_{\textsc{f}_{\hspace{-.1em}\textsc{m}}}\!\(\dot{\bm\Xi}^{~}_{\textsc{w}},\dot{\bm\Xi}^{\dagger_{\hspace{-.1em}\epsilon}}_{\textsc{w}}\)&:=
\hspace{.1em}\la\dot{\bm\Xi}^{*}_{\textsc{w}},i\bar{\ds^{~}_{\SG}}\dot{\bm\Xi}^{~}_{\textsc{w}}\ra=
i\hspace{.1em}\dot{\bm\Xi}^{\dagger_{\hspace{-.1em}\epsilon}}_{\textsc{w}}\(\hspace{-.2em}\bar{\ds^{~}_{\SG}}\hspace{.1em}\dot{\bm\Xi}^{~}_{\textsc{w}}\)\!,
\label{WspinLagL}
\intertext{yielding the equation of motion as}
\frac{\delta\I^{L}_{\textsc{f}_{\hspace{-.1em}\textsc{m}}}(\dot{\bm\Xi}^{~}_{\textsc{w}},\dot{\bm\Xi}^{\dagger_{\hspace{-.1em}\epsilon}}_{\textsc{w}})}
{\delta\dot{\bm\Xi}^{\dagger_{\hspace{-.1em}\epsilon}}_{\textsc{w}}}=\dot{0}&\implies
i{\bar{\ds^{~}_\SG}}\hspace{.2em}\dot{\bm\Xi}^{~}_{\textsc{w}}=
\overline{\gamma}^\bcdot_{~}\(\bm1_\SU
\(i\hspace{.1em}\partial_\bcdot
-\frac{1}{2}\cG\hspace{.2em}\omega^{\hspace{.4em}\stars}_{\bcdot}\hspace{.1em}\overline{S}^{~}_{\stars}\)
+{c^{~}_\SU}\hspace{.2em}\Aa^I_\bcdot\bm\tau^{~}_I
\)\dot{\bm\Xi}^{~}_{\textsc{w}}=\dot{0}.\label{WspinEoML}
\end{align}
\end{subequations}
The spinor-gauge covariant differential acts on the weak pair of dual-spinors as (\ref{codiracSp1}).
The chiral-left fermion Lagrangian provides an electron-neutrino weak-interaction after quantisation.
The chiral-left spinor Lagrangian is functional concerning the Yang--Mills gauge field, too; thus, we have the additional equation of motions such that
\begin{align}
\frac{\delta\I^{L}_{\textsc{f}_{\hspace{-.1em}\textsc{m}}}(\AAA^{~}_\SU)}{\delta\AAA^{~}_\SU}=\dot{0}&\implies
-c^{~}_\SU\hspace{.1em}\dot{\bm\Xi}^{\dagger_{\hspace{-.1em}\epsilon}}_{\textsc{w}}\gamma^\bcdot_{~}\hspace{.1em}\dot{\bm\Xi}^{~}_{\textsc{w}}\VVV_\bcdot\stackrel{\text{(\ref{GgG})}}{=}
-c^{~}_\SU\(\dot{\bm\xi}^{\nu\dagger_{\hspace{-.1em}\epsilon}}_{~}\gamma^\bcdot_{~}\hspace{.1em}\dot{\bm\xi}^\nu_{~}+
\dot{\bm\xi}^{e\dagger_{\hspace{-.1em}\epsilon}}_{~}\gamma^\bcdot_{~}\hspace{.1em}\dot{\bm\xi}^e_{~}
\)\VVV_\bcdot=\dot{0}.\label{xiiEoM}
\end{align}
The right-hand side of (\ref{xiiEoM}) appears as a source term of the Yang--Mills gauge boson such that
\begin{align}
\text{(\ref{YMGBEoM})}|&\implies
\hat{d}_\SG\FFF^{~}_\SG=   
c^{~}_\SU\hspace{.2em}\dot{\bm\Xi}^{\dagger_{\hspace{-.1em}\epsilon}}_{\textsc{w}}
\gamma^\bcdot_{~}\hspace{.1em}\dot{\bm\Xi}^{~}_{\textsc{w}}\VVV_\bcdot,
\implies
\eta^{a\bcdot}\left[\text{l.h.s. of (\ref{YMGBEoM3})}\right]^{~}_\bcdot=
c^{~}_\SU\hspace{.2em}\dot{\bm\Xi}^{\dagger_{\hspace{-.1em}\epsilon}}_{\textsc{w}}\gamma^a_{~}\hspace{.1em}\dot{\bm\Xi}^{~}_{\textsc{w}}.
\label{YMGBEoMDWS}
\end{align}


\subsubsection{Higgs spinor}\label{Higgsspinor}
%
The standard form of the scalar Lagrangian consists of a differentiated scalar field squared.
We introduce the Higgs Lagrangian as
\begin{subequations}
\begin{align}
\LLL^{~}_{\textsc{h}}({\bphi}^{~}_\textsc{h},{\bphi}^{\dagger}_\textsc{h}):=&
\LLL^{kin}_{\textsc{h}}({\bphi}^{~}_\textsc{h},{\bphi}^{\dagger}_\textsc{h})-V\({\bphi}^{~}_\textsc{h},{\bphi}^{\dagger}_\textsc{h}\)\vvv\label{LHiggs1},
\intertext{where $V({\bphi}^{~}_\textsc{h},{\bphi}^{\dagger}_\textsc{h}){\in}\Omega^0(\TsM){\otimes}\VV^\textsc{e}_{\hspace{-.2em}\hlf}{\otimes}Ad(\sss\uuu(2))$ and}
\LLL^{kin}_{\textsc{h}}({\bphi}^{~}_\textsc{h},{\bphi}^{\dagger}_\textsc{h}):=&
\frac{1}{4}(i{\ds}_\SU){\bcdot}(i{\ds}_\SU)
\<{\bphi}^{*}_\textsc{h},{\bphi}^{~}_\textsc{h}\>\vvv
=\frac{1}{2}(i{\ds}_\SU)
\<{\bphi}^{*}_\textsc{h},i{\ds}_\SU{\bphi}^{~}_\textsc{h}\>\vvv,\notag\\=&
\<{\bphi}^{*}_\textsc{h},\(i{\ds}_\SU\)^2{\bphi}^{~}_\textsc{h}\>\vvv=
{\bphi}^{\dagger}_\textsc{h}\cdot\(\(i{\ds}_\SU\)^2{\bphi}^{~}_\textsc{h}\)\vvv,
\intertext{yielding the equation of motion as}
\frac{\delta\I_{\textsc{h}}({\bphi}_\textsc{h},{\bphi}_\textsc{h}^\dagger)}{\delta{\bphi}_\textsc{h}^\dagger}&=\dot{0}\implies
\(\(i{\ds}_\SU\)^2-
\frac{\delta V({\bphi}_\textsc{h},{\bphi}_\textsc{h}^\dagger)}{\delta{\bphi}^\dagger_\textsc{h}}\){\bphi}^{~}_\textsc{h}=\dot{0}.\label{HiggsEoM}
\end{align}
\end{subequations}
The Dirac operator squared is called the Lichnerowicz formula\cite{Lichnerowicz1963,berline2003heat}.
The $\SUW$ symmetry is internal one in the Lorentzian \YMU theory; thus, the Higgs spinor does not directly couple to the spin connection.
In this case, the Dirac operator squared can be written as
\begin{align}
-\(i{\ds}_\SU\)^2&=
\bm1_\SU\hspace{.1em}\Delta+\widehat{\sslash{\FFF}}^{~}_\SU,\label{Lf0}
\intertext{where $\widehat{\sslash{\FFF}}^{~}_\SU$ is a differential operator defined as}
\widehat{\sslash{\FFF}}^{~}_\SU\hspace{.1em}&:=
\iota_{\gamma}\iota_{\gamma}(d^{~}_\SU{\wedge}d^{~}_\SU).\label{Lf1}
\end{align}
Formula (\ref{Lf0}) is called the Lichnerowicz formula in a flat space-time, which is provided by direct calculations as 
\begin{align}
(\ref{Lf1})&=
\iota_{\gamma}\iota_{\gamma}\(\(\bm1_\SU\hspace{.1em}d-i\hspace{.1em}{c^{~}_\SU}\AAA^{~}_\SU\)
\wedge\(\bm1_\SU\hspace{.1em}d-i\hspace{.1em}{c^{~}_\SU}\AAA^{~}_\SU\)\),\notag\\
&=
-i{c^{~}_\SU}\(
\slash{\AAA}^{~}_\SU\hspace{-.2em}\slash{\partial}
+\(\hspace{-.2em}\slash{\partial}\hspace{-.3em}\slash{\AAA}^{~}_\SU\)
-i\hspace{.1em}{c^{~}_\SU}\hspace{-.1em}\slash{\AAA}^{~}_\SU\hspace{-.1em}\slash{\AAA}^{~}_\SU
\),\notag\\&=-
(i\bm1_\SU{\slash\partial^{~}}+\hspace{.1em}{c^{~}_\SU}\hspace{-.3em}\slash{\AAA}_{\SU})
\(i\bm1_\SU\hspace{-.2em}\slash\partial^{~}+{c^{~}_\SU}\hspace{-.2em}\slash{\AAA}_{\SU}\)
-\bm1_\SU\hspace{.1em}\Delta=
-\(i{\ds}_\SU\)^2-\bm1_\SU\hspace{.1em}\Delta,\label{HssF}\\
&\implies(\ref{Lf0}).\notag
\end{align}
The Lichnerowicz formula gives a representation of the Laplacian operator using the covariant differential and the double contraction as 
\begin{align}
\bm1_\SU\hspace{.1em}\Delta&=
(\iota_{\gamma}d^{~}_\SU)(\iota_{\gamma}d^{~}_\SU)-
\iota_{\gamma}\hspace{.1em}\iota_{\gamma}(d^{~}_\SU{\wedge}\hspace{.1em}d^{~}_\SU).\label{Lf2}
\end{align}
Owing to the relation 
\begin{align*}
\(\AAA^{~}_\SU{\wedge}d+d\AAA^{~}_\SU\){\bphi}^{~}_\textsc{h}&=
\AAA^{~}_\SU{\wedge}d{\bphi}^{~}_\textsc{h}+d(\AAA^{~}_\SU{\bphi}^{~}_\textsc{h})
=\AAA^{~}_\SU{\wedge}d{\bphi}^{~}_\textsc{h}+d(\AAA^{~}_\SU){\bphi}^{~}_\textsc{h}-\AAA^{~}_\SU{\wedge}d{\bphi}^{~}_\textsc{h}=
(d\AAA^{~}_\SU){\bphi}^{~}_\textsc{h},
\end{align*}
the curvature operator can be written from the second line of (\ref{HssF}) as 
\begin{align*}
\widehat{\sslash{\FFF}}^{~}_\SU&=-i\hspace{.1em}{c^{~}_\SU}\hspace{.1em}\iota_{\gamma}\iota_{\gamma}
\((d\AAA^{~}_\SU)-i\hspace{.1em}{c^{~}_\SU}\hspace{.1em}\AAA^{~}_\SU{\wedge}\AAA^{~}_\SU\)
=-i\hspace{.1em}{c^{~}_\SU}\sslash{\FFF}^{~}_\SU,~~\text{where}~~
\sslash{\FFF}^{~}_\SU:=\frac{1}{2}\f_\bcdots^{I}\gamma^\bcdot_{~}\gamma^\bcdot_{~}\tau_I.
\end{align*}
Consequently, we obtain the Lagrangian and the equation of motion as
\begin{subequations}
\begin{align}
\text{(\ref{LHiggs1})}&\implies
\LLL^{~}_{\textsc{h}}({\bphi}^{~}_\textsc{h},{\bphi}^{\dagger}_\textsc{h})=-{\bphi}^{\dagger}_\textsc{h}
\(\bm1_\SU\hspace{.1em}\Delta+i{c^{~}_\SU}\sslash{\FFF}^{~}_\SU\){\bphi}^{~}_\textsc{h}-V({\bphi}_\textsc{h})\label{LHiggs2},\\
\text{(\ref{HiggsEoM})}&\implies
\(\bm1_\SU\hspace{.1em}\Delta+i{c^{~}_\SU}\sslash{\FFF}^{~}_\SU+
\frac{\delta V({\bphi}_\textsc{h},{\bphi}_\textsc{h}^\dagger)}
{\delta{\bphi}^\dagger_\textsc{h}}\){\bphi}^{~}_\textsc{h}=\dot{0}.\label{HiggsEoM2}
\end{align}
\end{subequations}
The curvature contracted with the Clifford algebra $\sslash{\FFF}^{~}_\SU$ is called the \emph{twisting curvature of the Clifford module}\cite{berline2003heat}

\subsubsection{Fermion mass term}\label{intrHW}
Fermionic matter Lagrangian (\ref{WspinLagLR}) provides an equation of motion for massless electrons and neutrinos and their interaction.
In reality, an electron has a finite mass; thus, we introduce the mass term into the \YMU Lagrangian.
We discuss two methods: a direct method and a spontaneous symmetry breaking (SSB) method.
\paragraph{a) direct method:}
A Hermitian term
\begin{align}
\LL^{m_e}_{\textsc{f}_{\hspace{-.1em}\textsc{m}}}&=m_e\<\dot\bpsi^{e*}_{~},\bpsi^e_{~}\>=
m_e\hspace{.2em}({\dot\bpsi}^e_{~})^{\dagger_{\hspace{-.1em}\epsilon}}\cdot\bpsi^e_{~}=
m_e\hspace{.2em}[\dot{\bm{\xi}}^{e*}]^{~}_{A}\hspace{.1em}\epsilon_2^{\hspace{.3em}AB}\left[\bm\xi^e_{~}\right]^{~}_B\!,\label{mxixi}
\end{align}
provides an electron mass term of the fermion Lagrangian.
We note that $\dot{\bm{\xi}}^{e*}\in\VV_\hlf$; thus it has undotted index.
Since both a bra-part and a ket-part belong to the $\VV_{\hspace{-.2em}\hlf}$ space,  bi-linear form (\ref{mxixi}) is ${\SL2C}$ invariant.
When we utilise the chiral basis for an electron spinor as
\begin{subequations}
\begin{align}
\bpsi^e_{~}=\bpsi^e_\textsc{u}+\bpsi^e_\textsc{d},&~~\dot\bpsi^e_{~}=\dot\bpsi^e_\textsc{u}+\dot\bpsi^e_\textsc{d},
\intertext{we obtain the electron mass term as}
\<\dot\bpsi^{e*}_\textsc{u},\bpsi^e_\textsc{u}\>=\<\dot\bpsi^{e*}_\textsc{d},\bpsi^e_\textsc{d}\>=0&\implies
\LL^{m_e}_{\textsc{f}_{\hspace{-.1em}\textsc{m}}}=-m_e
\(\<\dot\bpsi^{e*}_\textsc{u},\bpsi^e_\textsc{d}\>-
   \<\dot\bpsi^{e*}_\textsc{d},\bpsi^e_\textsc{u}\>\)\!.\label{me1}
\end{align}
\end{subequations}
The mass term consists of the bilinear form between chiral-left and chiral-right spinors, at the same time, chiral-left and chiral-right spinors, contrasting to the fermion Lagrangian (\ref{WspinLagLR}).

\paragraph{b) SSB method:}
The standard \YMU Lagrangian has the interaction term between Higgs and Weyl spinors, which is cast into the fermion-mass term after the spontaneous symmetry breaking.
This interaction term keeps local $\SL2C$ and internal $\SUW$ symmetries.
This report describes the fermion mass term owing to the spontaneous symmetry breaking using the weak-pair of dual-spinors.

We introduce a Higgs-fermion bilinear form using dual-spinors as
\begin{align}
\<\dot{\bm\Xi}^{~}_{\textsc{w}},\bphiw\>\overset{(\ref{XXxXpX})}{=}
\<\bphiUw,\hbphiUw\>\hspace{.1em}\dot{\bm\xi}^\nu_{~}+\<\bphiDw,\hbphiDw\>\hspace{.1em}\dot{\bm\xi}^e_{~}
=\phi^{~}_\textsc{u}\hspace{.1em}\dot{\bm\xi}^\nu_{~}+\phi^{~}_\textsc{d}\hspace{.1em}\dot{\bm\xi}^e_{~}.\label{XwPhiw}
\end{align}
Then, we provide the fermion-mass Lagrangian as 
\begin{align}
\LL^{m_e}_{\textsc{f}_{\hspace{-.1em}\textsc{m}}}&=
\lambda\<\<\dot{\bm\Xi}^{*}_{\textsc{w}},\bphiw\>,{\bm\Xi}^{~}_{\textsc{w}}\>=
\lambda\hspace{.1em}\phi^{~}_\textsc{u}\<\dot{\bm\xi}^{\nu*}_{~},{\bm\xi}^\nu_{~}\>+
\lambda\hspace{.1em}\phi^{~}_\textsc{d}\<\dot{\bm\xi}^{e*}_{~},{\bm\xi}^e_{~}\>,\notag\\
&=-\lambda\hspace{.1em}\phi^{~}_\textsc{d}
\(\<\dot\bpsi^{e*}_\textsc{u},\bpsi^e_\textsc{d}\>-
   \<\dot\bpsi^{e*}_\textsc{d},\bpsi^e_\textsc{u}\>\)-\lambda\hspace{.1em}\phi^{~}_\textsc{u}
\(\<\dot\bpsi^{\nu*}_\textsc{u},\bpsi^\nu_\textsc{d}\>-
   \<\dot\bpsi^{\nu*}_\textsc{d},\bpsi^\nu_\textsc{u}\>\)\!.\label{me2}
\end{align}
When we set the Higgs spinor as
\begin{align}
\(
\phi^{~}_\textsc{u}=0,~\phi^{~}_\textsc{d}=v_0/\sqrt{2}\in\R,~m_e=\lambda\hspace{.1em} v_0/\sqrt{2}
\)\implies
(\ref{me2})=(\ref{me1}),
\end{align}
where $v_0$ is a vacuum expectation value, we have a finite electron mass proportional to the vacuum expectation value.

When the neutrino is massless, the above Lagrangian is enough to describe a leptonic part of the standard theory of particle physics since the chiral-right singlet neutrino does not interact with any other fields than the graviton.
On the other hand, in the quark sector, we know that the up-type quarks also have a mass. 
This report does not treat quarks in the \YMU theory.

%% file: Sec2-TSE.tex
%
%
\subsection{Torsion and stress-energy forms}\label{TandSE}
In \textbf{section \ref{YMUGB}}, we presented two equations of motion for the gravitational field: the torsion equation and the Einstein equation.
A static point of the action integral concerning the vierbein provides the torsion equation, and that concerning the spin-connection provides the Einstein equation.
Other equations of motion are obtained as the static point concerning matter, Higgs and Weyl-spinor fields. 
In reality, the fermion Lagrangian has the spin-connection and all Lagrangians have the vierbein; thus, the static point concerning the vierbein and spin-connection contributes to the torsion and Einstein equations.
We introduce the torsion and stress-energy forms in the \YMU theory.

Torsion two-form $\TTT^{~}_{\bullet}$ and stress-energy three-form $\EEE^{~}_{\bullet}$ of the \YMU Lagrangian for $\bullet\in\{\textsc{y\hspace{-.1em}m\hspace{-.1em}u},\textsc{y\hspace{-.1em}m},\textsc{g}\hspace{-.05em}\textsc{r},\textsc{f}_{\hspace{-.1em}\textsc{m}},\textsc{h},\textsc{m}\text{-}\textsc{h}\}$ are respectively defined as
\begin{align}
&\frac{1}{2}\epsilon_{ab\bcdots}\TTT_{\hspace{-.1em}\bullet}^{\hspace{.2em}\bcdot}\wedge\eee^\bcdot_{~}:=\frac{\delta\LLL^{~}_{\bullet}}{\delta\www^{ab}}
-d\left(\frac{\delta\LLL^{~}_{\bullet}}{\delta(d\www^{ab})}\right),~~
\EEE_{\hspace{.2em}a}^{\bullet}:=\hbar\(\frac{\delta\LLL^{~}_{\bullet}}{\delta\eee^a}
-d\left(\frac{\delta\LLL^{~}_{\bullet}}{\delta(d\eee^a)}\right)\)\!.
\end{align}
We note that the above definition has opposite sign to that of the Einstein form and torsion form.
These form objects have physical dimensions as
\begin{align*}
&\left[\TTT^{~}_{\hspace{-.1em}\bullet}\right]=L^{-1}~~\mathrm{and}~~
\left[\EEE_{~}^{\bullet}\right]=E,
\intertext{and their tensor coefficients are denoted as}
&\TTT_{\hspace{-.1em}\bullet}^{\hspace{.2em}a}=:
\frac{1}{2}T_{\hspace{-.1em}\bullet\hspace{.3em}\bcdots}^{\hspace{.2em}a}\hspace{.1em}\eee^\bcdot\wedge\eee^\bcdot
~~\text{and}~~
\EEE_{\hspace{.2em}a}^{\bullet}=:E^{\bullet}\hspace{.1em}\VVV_a;
\intertext{thus, these tensors have physical dimensions as}
&\left[T_{\hspace{-.1em}\bullet}\right]=L^{-1}/L^{2}~~\text{and}~~
\left[E^\bullet\right]=E/L^3.
\end{align*}
Tensor coefficient $T_{\hspace{-.1em}\bullet}$ is a torsion density per unit two-dimensional surface, and $E^\bullet$ is an energy-stress density per unit three-volume.  

\paragraph{a) torsion two-form:}
First, we calculate a torsion induced by the fermion Lagrangian.
We denote a variation of the Lagrangian concerning the spin-connection as
\begin{align}
\frac{1}{2}\epsilon_{ab\bcdots}\hspace{.1em}\TTT_{\hspace{-.1em}\textsc{f}_{\hspace{-.1em}\textsc{m}}}^{\hspace{.3em}\bcdot}
\wedge\eee^\bcdot_{~}&=
\frac{\delta\LLL_{\textsc{f}_{\hspace{-.1em}\textsc{m}}}}{\delta\www^{ab}}=
\cG\(\bpsi^{\dagger_{\hspace{-.1em}\epsilon}}_{~}\gamma_{~}^\bcdot{\Ss}^{~}_{ab}\hspace{.2em}\bpsi\)\VVV_\bcdot.\label{kform}
\end{align}
The torsion form can be obtained by solving the algebraic equation (\ref{kform})  as
\begin{align}
\text{(\ref{kform})}\implies
T_{\hspace{-.1em}{\textsc{f}_{\hspace{-.1em}\textsc{m}}}\hspace{.3em}bc}^{\hspace{.2em}a}=
\bpsi^{\dagger_{\hspace{-.1em}\epsilon}}_{~}\hspace{-.1em}\left(\gamma_{~}^a{\Ss}^{~}_{bc}+
2\delta^a_{\hspace{.2em}[b}{\eta}^{~}_{c]\bcdot}\gamma_{~}^\bcdot
\right)\hspace{-.2em}\bpsi\implies
\TTT^a_{\textsc{f}_{\hspace{-.1em}\textsc{m}}}=
\frac{1}{2}T_{\hspace{-.1em}{\textsc{f}_{\hspace{-.1em}\textsc{m}}}\hspace{.3em}\bcdots}^{\hspace{.2em}a}
\hspace{.1em}\eee^\bcdot\wedge\eee^\bcdot. \label{TYMFm}
\end{align}
Since only the fermion Lagrangian includes the spin connection, this is all of the non-zero torsion forms in the \YMU theory.

\paragraph{b) stress-energy three-form:}
In contrast with the spin-connection, all sub-Lagrangians in the \YMU Lagrangian include the vierbein form as a part of the volume form.
We can construct the stress-energy three-form commonly from the Lagrangian of a type 
\begin{align*}
\LLL_\bullet(\vvv)=\Phi_\bullet(f)\vvv
\quad\text{as}\quad
\EEE_a^{\bullet}&=\hbar\frac{\delta\LLL_\bullet(\vvv)}{\delta\eee^a_{~}}=
\hbar\Phi_\bullet(f)\frac{\delta~}{\delta\eee^a_{~}}
\(\frac{1}{4!}\epsilon_{\bcdots\bcdots}\hspace{.1em}\eee^\bcdot\wedge\eee^\bcdot\wedge\eee^\bcdot\wedge\eee^\bcdot\),\\
&=\hbar\hspace{.1em}\Phi_\bullet(f)\frac{\hbar}{3!}\epsilon_{a\bcdot\bcdots}\hspace{.1em}\eee^\bcdot\wedge\eee^\bcdot\wedge\eee^\bcdot
=\hbar\hspace{.1em}\Phi_\bullet(f)\hspace{.1em}\VVV^{~}_{a},\\
\implies E^\bullet&=\hbar\hspace{.1em}\Phi_\bullet(f),
\end{align*}
where $\Phi(f)$ is a scalar functional of physical fields $f$.
Consequently, the stress-energy form owing to the Yang--Mills field and the fermion Lagrangians is provided as
\begin{subequations}
\begin{align}
(\ref{YMLL})&\implies
E^{\textsc{y\hspace{-.1em}m}}=-\frac{\hbar}{4}\f^I_\bcdots\f_I^\bcdots,\label{esYM}\\
(\ref{WspinEoML})&\implies
E^{\textsc{f}_{\hspace{-.1em}\textsc{m}}}=\hbar\hspace{.1em}
\bpsi^{\dagger_{\hspace{-.1em}\epsilon}}_{~}\hspace{.1em}\gamma_{~}^\bcdot\(i\hspace{.1em}\partial_\bcdot-\frac{1}{2}\cG\hspace{.1em}
\hspace{.1em}\omega_\bcdot^{~\stars}{\Ss}^{~}_\stars+
{c^{~}_\SU}\hspace{.2em}\Aa^I_\bcdot\tau^{~}_I
\)\!\bpsi\label{esfm}\\
(\ref{LHiggs2})&\implies
E^{\textsc{h}}_{a}=\hbar
\(\eta^\bcdots(\partial^{~}_\bcdot\bm\phi^\dagger_\textsc{h})(\partial^{~}_\bcdot\bm\phi^{~}_\textsc{h})
-i\hspace{.1em}{c^{~}_\SU}\hspace{.1em}\frac{1}{2}\f_\bcdots^{I}\gamma^\bcdot_{~}\gamma^\bcdot_{~}\tau_I
(\bm\phi^\dagger_\textsc{h}\bm\phi^{~}_\textsc{h})
-V(\bm\phi^{~}_\textsc{h},\bm\phi^\dagger_\textsc{h})\)\label{esHiggs}\\
(\ref{me1})&\implies
E^{m_e}=-\hbar\hspace{.1em}m_e
\(\<\dot\bpsi^{e*}_\textsc{u},\bpsi^e_\textsc{d}\>-
   \<\dot\bpsi^{e*}_\textsc{d},\bpsi^e_\textsc{u}\>\).\label{eseMass}
\end{align}
\end{subequations}
In (\ref{esHiggs}), we use integration by parts.

In summary, the Einstein equation and the torsion equation in the \YMU theory are respectively provided as
\begin{subequations}
\begin{align}
\GGG_a^{\hspace{-.1em}\textsc{g}\hspace{-.05em}\textsc{r}}&=
\EEE^{\textsc{y\hspace{-.1em}m}}_a+
\EEE^{\textsc{f}_{\hspace{-.1em}\textsc{m}}}_a+
\EEE^{\textsc{h}}_{a}+
\EEE^{m_e}_{a}
~~\text{and}~~
\TTT^a_{\hspace{-.1em}\textsc{g}\hspace{-.05em}\textsc{r}}=\TTT^a_{\textsc{f}_{\hspace{-.1em}\textsc{m}}},\label{EoM3}
\intertext{where}
\GGG_a^{\hspace{-.1em}\textsc{g}\hspace{-.05em}\textsc{r}}&:=\frac{1}{\kE}\GGG_a=\frac{1}{2\kE}\epsilon_{a\bcdot\bcdots}\hspace{.2em}\RRR^{\bcdots}
\hspace{-.2em}\wedge\eee^{\hspace{.1em}\bcdot}
~~\text{and}~~
\TTT^a_{\hspace{-.1em}\textsc{g}\hspace{-.05em}\textsc{r}}:=\frac{1}{\hbar\kE}\TTT^a=\frac{1}{\hbar\kE}d_\www\eee^a.
\end{align}
\end{subequations}

%% file: Sec3.tex
\section{Amphometric space}\label{AIndex}
We construct the \YMU theory in the Lorentzian metric space in the previous section.
This section develops a method to simultaneously treat Euclidean and Lorentzian metrics in the \YMU theory.
In the quantum field theory, we utilise an analytic continuation concerning the time coordinate, known as the Wick rotation.
A primary purpose of the Wick rotation is to improve integration convergence.
Concomitantly, it determines analyticity of a particle propagator.
The analytic continuation is the complexification of the time coordinate with a pure imaginary number, and a point in Minkowski space is mapped to that in Euclidean space.
We propose to refine the method: two metric spaces are continuously connected using a one-parameter family of metrics, namely the \emph{amphometric}, which induces a one-parameter subgroup of $\SL2C$.
Though the Wick rotation is a technical issue when performing integration in the Minkowski space, the amphometric method provides a deep understanding of the physical relation between Euclidean and Lorentzian metrics.

%
%
\input{Sec3-1}

%
%
\input{Sec3-2}

%
%
\input{Sec3-4}

%
%
\input{Sec3-3}

%% file: Sec3-1.tex
%
%
\subsection{Amphometric}\label{q-metric}
\begin{figure}[t] 
\centering
\includegraphics[width={7cm}]{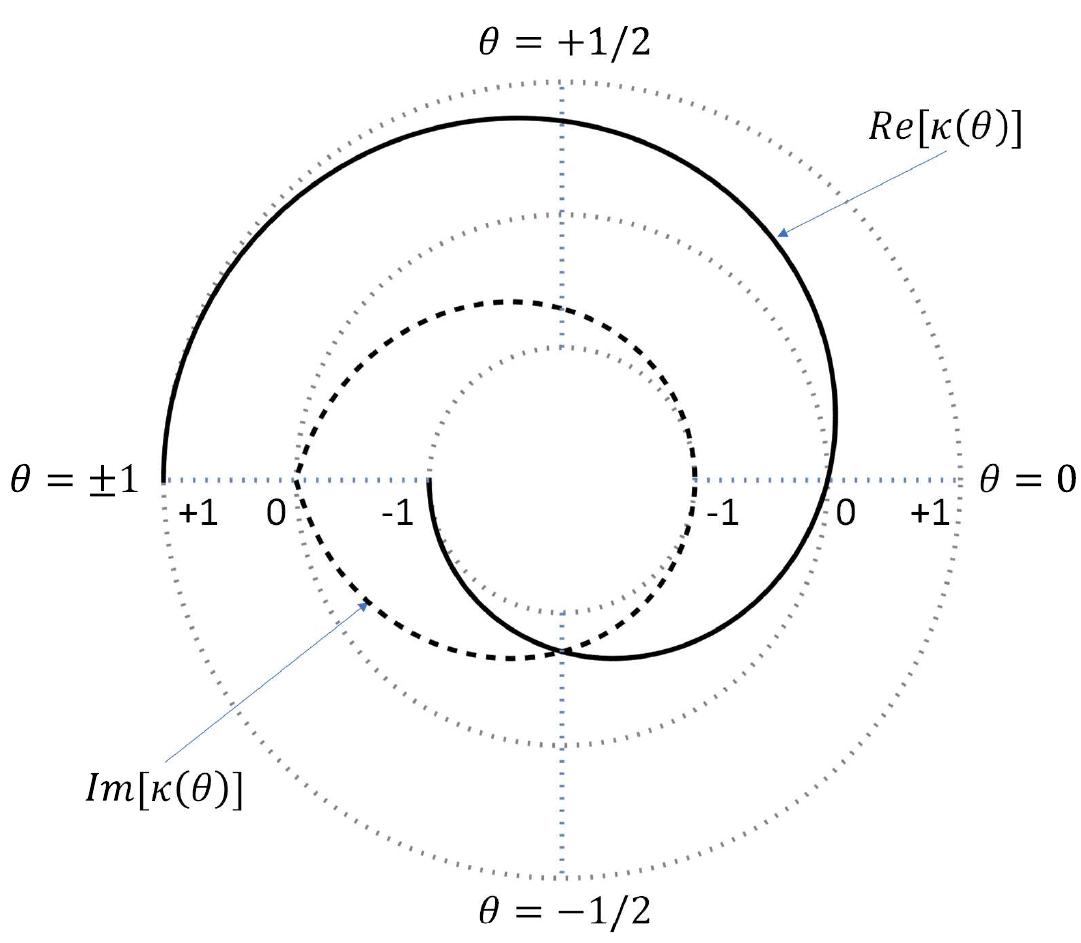}%
 \caption{
$\kappa(\theta)$ is drawn in $-1\leq\theta\leq1$.
Solid- and dashed-lines show real- and imaginary-parts of the function, respectively.
} 
\label{figKappa} 
\end{figure}

This section proposes a refinement of the Wick-rotation to simultaneously treat Euclidean and Lorentzian metric spaces as two boundaries of the one-parameter subgroup in $\SL2C$.
In \textbf{Section \ref{Cliffordalgebra}}, we introduced a Clifford algebra in both Euclidean and Lorentzian metric spaces and defined the chiral operator to be shared in both spaces.
Moreover, we introduced the maps (\ref{ijk2E}) and (\ref{ijk2L}), transforming the quaternion representation to the representation of dimension-4 in Euclidean and Lorentzian metric spaces. 
We unify these maps to a single map using the smooth function $\kappa(\theta)\in\C$ such that
\begin{align}\left.
\begin{array}{c}
(\ref{ijk2E}) \\
(\ref{ijk2L})
\end{array}
\right\}&
\Longrightarrow\Sigma^{~}_\theta:Sp(1){\rightarrow\hspace{.1em}}\GLTC:
\{1,i,j,k\}^t\mapsto\bm\sigma^{~}_\theta:=\{\bm{1}^{~}_2,\kappa(\theta)\bm\sigma\}^t\!.
\label{ijk2Q}
\end{align}
When we require it to behave $\kappa(\theta=0)=-i$ and $\kappa(\theta=\pm1)=\pm1$, it provides dimension-2 representation of the Euclidean Clifford algebra at $\theta=0$ and the Lorentzian Clifford algebra at $\theta=\pm1$.
Moreover, we require $\kappa(\theta)$ to be a non-zero function in $\theta\in[-1,1]$.
We propose the following function for $\kappa(\theta)$:
\begin{align}
\kappa(\theta)&:=e^{i\pi(\theta-1)/2}=
\begin{cases}
+1&\theta=+1\\
-i&\theta=\hspace{.8em}0\\
-1&\theta=-1
\end{cases}\!.\label{kappatheta}
\end{align}
Figure \ref{figKappa} depicts a behaviour of $\kappa(\theta)$ in $\theta\in[-1,1]$.

Suppose $\M^{~}_\theta$ is a four-dimensional smooth manifold equipping the Clifford algebra concerning the symmetric $(\FxF)$ complex matrix $\betathe$.
The map (\ref{SL2C1}) with $\bm\sigma^{~}_\theta$ induces the representation of dimension-2 for real vector $\v_{\textsc{v}\theta}=(v^0,v^1,v^2,v^3)^t{\hspace{.1em}\in\hspace{.1em}}V(\TM^{~}_\theta:\R)$ as
\begin{align}
\Sigma^{~}_\theta:V(\R^4){\rightarrow\hspace{.1em}}V(\SL2C):\v_{\textsc{v}\theta}\mapsto
  \v_{\textsc{s}\theta}:=\bm{\sigma}^t_\theta\cdot\v_{\textsc{v}\theta}&=
\(
      \begin{array}{cc}
        v^0+\kappa(\theta)v^3 &\kappa(\theta)\(v^1 -i v^2\)\\
         \kappa(\theta)\(v^1 +i v^2\)&v^0-\kappa(\theta)v^3 \\
      \end{array}
\)\!.\label{SigmathetaV}
\end{align}
We refer to a complex ($\TxT$)-matrix of the shape $(\ref{SigmathetaV})$ as the \emph{apmphovector}, which is a unitary matrix at $\theta=0$ and the Hermitian matrix at $\theta=\pm1$.
We determine the complex-valued symmetric tensor $\betathe$ owing to the relation 
\begin{subequations}
\begin{align}
\<\u_{\textsc{s}\theta},\v_{\textsc{s}\theta}\>&:=
\frac{1}{2}\(\Tr[\u_{\textsc{s}\theta}]\hspace{.1em}\Tr[\v_{\textsc{s}\theta}]
-\Tr[\u_{\textsc{s}\theta}\hspace{.1em}\v_{\textsc{s}\theta}]\)=
\u_{\textsc{s}\theta}^t\hspace{.1em}\betathe\hspace{.1em}\v_{\textsc{s}\theta}=
u^0v^0+e^{i\pi\theta}\(u^1v^1+u^2v^2+u^3v^3\)\!,\label{<vu>Q}
\intertext{which provides}
(\ref{<vu>Q})&\implies\betathe=\text{diag}[1,e^{i\pi\theta},e^{i\pi\theta},e^{i\pi\theta}],
~~\text{yielding}~~
\betathe=
\begin{cases}
\betaE,&\theta=\hspace{.8em}0\\
\betaL,&\theta=\pm1
\end{cases}\!.\label{etatheta}
\end{align}
Complex matrix $\betathe$ is the Lorentzian metric at both ends of the domain $\theta\in[-1,1]$; thus, we call it as  the \emph{amphometric tensor}, and the pair $(\M^{~}_\theta,\betathe)$ is referred to as the \emph{amphometric space} in this report.
We define the \emph{amphonorm} of the vector $\v_{\textsc{s}\theta}$ as
\begin{align}
\|\v_{\textsc{s}\theta}\|^{2}&:=\<\v_{\textsc{s}\theta},v_{\textsc{s}\theta}\>=
\text{det}\left[\v_{\textsc{s}\theta}\right]=({v}^0)^2+e^{i\pi\theta}\(({v}^1)^2+({v}^2)^2+({v}^3)^2\)\!.\label{etatheta0}
\end{align}
\end{subequations}

Consider four-dimensional \textit{rotation} operator $\gB{\theta}$ whose $(\FxF)$-matric representation $\bm\Lambda_\theta$ fulfiling
\begin{align*}
\text{det}[\bm\Lambda_\theta]=1,~~\bm\Lambda_{\theta=1}\in\SO(1,3)~\text{and}~\bm\Lambda_{\theta=0}\in\SO(4). 
\end{align*}
A set of $\gB{\theta}$ is denoted $Ampho(2)$.
The the representation of dimension-2 relates with the representation of dimension-4 as 
\begin{align}
\v'_{\textsc{v}\theta}=\bm{\Lambda}_\theta\cdot\v_{\textsc{s}\theta}^{~}\leftrightarrow
\bm{\sigma}^t_\theta\cdot\v'_{\textsc{v}\theta}=\v'_{\textsc{v}\theta}=\gB{\theta}\cdot\v^{~}_{\textsc{v}\theta}\cdot\g_{\theta}^{\dagger}
\implies
\text{det}[\v'_{\bullet\theta}]=\text{det}[\v^{~}_{\bullet\theta}],~
\bullet\in\{\textsc{v},\textsc{s}\}\!.\label{LamdaG2}
\end{align}
Thus, $Ampho(2)$ is a one-parameter subgroup of $\SL2C$.

The amphometric connects the space-time symmetries of $SO(4)$ and $SO(1,3)$ groups, forming the one-parameter subgroup in $\GLTC$.
The amphometric space $(\M^{~}_\theta,\betathe)$ provides the Euclidean space $(\ME,\betaE)$ at $\theta=0$ and the Minkowski space $(\ML,\betaL)$ at $\theta=\pm1$. 
On the other hand, it is not the Riemannian manifold at points other than $\theta=0,\pm1$.

%% file: Sec3-2.tex
%
%
\subsection{Causal subgroup and invariant bilinear form}\label{cSubGroup}
When we allow all $\GLTC$ group actions on $Ampho(2)$ at $\theta\neq\pm1$, they violate the causality at $\theta=\pm1$.
This section introduces a subgroup of $Ampho(2)$, namely the causal subgroup,  preserving the causality of the physical phenomenon in the Lorentzian metric space.
The restriction of group actions in the causal subgroup corresponds to choosing one axis as the ``time'' axis and prohibiting mixing the ``time'' direction with the ``spatial'' directions in Euclidean space.

Although the Euclidean manifold is the simply connected space owing to the $SO(4)$ transformation (four-dimensional rotation), the Minkowski manifold has four disconnected subspaces under the Lorentz transformation: the null-distance region, the space-like region, the time-like past region, and the time-like future region.
When we mix $SO(1,3)$ and $SO(4)$ symmetries, time-like separated two points in the Minkowski space can be mapped to the space-like separation or reversed time-ordered position by the $SO(4)$ rotation and it violates the causality between those two points.
We restrict transformations in the amphometric space to preserve the causality for physical theories.

\begin{definition}[Causal subgroup]\label{Defamphoortho}
Suppose matrix $\g^{~}_c(\theta){\hspace{.1em}\in\hspace{.1em}}\Ca(2){\hspace{.1em}\subset\hspace{.1em}}Ampho(2)$ fulfils 
\begin{align}
&\g^{~}_c(\theta=\pm1)\in\SO^{\uparrow}\hspace{-.1em}(1,3) {\subset}SO(1,3)
~~\text{and}~~
\g^{~}_c(\theta=0){\hspace{.1em}\in\hspace{.1em}}SO_{~}^{\perp}\hspace{-.1em}(3){\subset}SO(4),\label{gcSO+}
\intertext{which acts on the amphovector as}
&\gB{c}:End\(\Ca(2)\):\v_{\textsc{s}\theta}\mapsto\v_{\textsc{s}\theta}':=\tau\(\gB{c}\)\(\v_{\textsc{s}\theta}\).
\end{align}
The $\mathrm{(\FxF)}$-matrix groups in \textup{(\ref{gcSO+})} mean the corresponding $\mathrm{(\TxT)}$-matrix ones shown \textup{(\ref{LamdaG2})}.
We call $\g^{~}_\theta$ as the causal operator and $\Ca(2)$ as the causal subgroup of $Ampho(2)$.
\QED
\end{definition}
\noindent
We consider the following $\Ca(2)$ group actions in the amphometric space:
%
%
\begin{subequations}
\begin{align}
\g^a_c(\theta;\varphi,\chi)&:=\left.\gBa{\SOt}(\varphi)\bcdot\hspace{.1em}\gBa{b}(\chi)\right|^{~}_{\bm\sigma\rightarrow(\chi\hspace{.2em}\theta-i\varphi)\bm\sigma}=
\exp{\hspace{-.2em}\(-i\sigma^a\frac{\varphi}{2}\)}
\exp{\hspace{-.2em}\(\sigma^a\frac{\chi\hspace{.2em}\theta}{2}\)},\label{gaq}
\intertext{which acts on an amphovector  as}
\v'_{\textsc{s}\theta}&=\tau\(\g^a_c(\theta;\varphi,\chi)\)\(\v_{\textsc{s}\theta}\)
= \Sigma^{~}_\theta
\begin{cases}
\left(
     \begin{array}{c}
       v^0\cosh(\chi\hspace{.2em}\theta)
       +v^1_\theta\sinh(\chi\hspace{.2em}\theta)\\
       v^0\sinh(\chi\hspace{.2em}\theta)
       +v^1_\theta\cosh(\chi\hspace{.2em}\theta)\\ 
       v^2_\theta\cos{(\varphi)}-v^3_\theta\sin{(\varphi)}\\
       v^2_\theta\sin{(\varphi)}+v^3_\theta\cos{(\varphi)}\\ 
      \end{array}
\right)&a=1,\\
\left(
     \begin{array}{c}
       v^0\cosh(\chi\hspace{.2em}\theta)
       +v^2_\theta\sinh(\chi\hspace{.2em}\theta)\\
       v^1_\theta\cos{(\varphi)}+v^3_\theta\sin{(\varphi)}\\
       v^0\sinh(\chi\hspace{.2em}\theta)
       +v^2_\theta\cosh(\chi\hspace{.2em}\theta)\\ 
      -v^1_\theta\sin{(\varphi)}+v^3_\theta\cos{(\varphi)}\\ 
      \end{array}
\right)&a=2,\\
\left(
     \begin{array}{c}
       v^0\cosh(\chi\hspace{.2em}\theta)
       +v^3_\theta\sinh(\chi\hspace{.2em}\theta)\\ 
       v^1_\theta\cos{(\varphi)}-v^2_\theta\sin{(\varphi)}\\
       v^1_\theta\sin{(\varphi)}+v^2_\theta\cos{(\varphi)}\\ 
       v^0\sinh(\chi\hspace{.2em}\theta)
       +v^3_\theta\cosh(\chi\hspace{.2em}\theta)
      \end{array}
\right)&a=3,
\end{cases}\label{Qrot}
\intertext{where}
\(v^0,v^1_\theta,v^2_\theta,v^3_\theta\)&:=\(v^0,\kappa(\theta)v^1,\kappa(\theta)v^2,\kappa(\theta)v^3\).
\end{align}
\end{subequations}
This action simultaneously induces the rotation around the $v^a$-axis in the three-dimensional sub-manifold perpendicular to the $v^0$-axis and the \textit{amphoboost} along the $v^a$-axis, where the amphoboost is the true boost only at $\theta\hspace{-.2em}=\hspace{-.2em}\pm1$.
We can confirm that the determinant of the action (\ref{gaq}) is unity as
\begin{align*}
\text{det}\left[\g^a_c(\theta;\varphi,\chi)\right]&=
\text{det}\left[\exp{\hspace{-.2em}\(-i\sigma^a\frac{\varphi}{2}\)}\right]
\text{det}\left[\exp{\hspace{-.2em}\(\sigma^a\frac{\chi\hspace{.2em}\theta}{2}\)}\right]=
\exp{\hspace{-.2em}\(\Tr\left[-i\sigma^a\frac{\varphi}{2}\right]\)}
\exp{\hspace{-.2em}\(\hspace{-.1em}\Tr\left[\sigma^a\frac{\chi\hspace{.2em}\theta}{2}\right]\)}=1,
\end{align*}
since the Pauli matrices are traceless.
Consequently, (\ref{gaq}) preserves bilinear form (\ref{<vu>Q}).
The causal operator keeps causality by preserving the sign of the $0$'th component of amphovectors.
The three-dimensional rotation $\gBa{\SOt}(\varphi)$ and boost $\gBa{b}(\chi\hspace{.2em}\theta)$ are endomorphisms on $SO^{\uparrow}\hspace{-.1em}(1,3)$ and $SO^\perp(3)$; thus,  (\ref{gaq}) is the causal operator.
Moreover, the complex conjugate of the causal operator is given as
\begin{align}
\g^a_c(\theta;\varphi,\chi)^*
&=\gBs{\SOt}^{a=2}(\pi)
\hspace{.2em}\g^a_c(-\theta;\varphi,\chi)\hspace{.2em}
\gBs{\SOt}^{a=2}(\pi)^\dagger
~~\text{for}~~a\in\{1,2,3\}.
\end{align}
Thus, two irreducible representations $\gB{\theta}$ and $\gBs{\theta}^*$ in $\SL2C$ are homotopy equivalence in the amphometric space embedding $\SL2C$ at both ends.
Therefore, $\Ca(2)$ has a unique irreducible representation. 
The general causal operator can be written as
\begin{align}
\gB{c}&=
\g^{a_1}_c(\theta;\varphi^{~}_1,\chi^{~}_1)
\cdots
\g^{a_n}_c(\theta;\varphi^{~}_n,\chi^{~}_{n}).\label{defgc}
\end{align}
The amphovector has the invariant bilinear form under the causal subgroup at any values of $\theta$.
We schematically show a relation between the Euclidean and the Lorentzian vectors in the amphometric space as
\begin{align}
\begin{array}{ccccc}
\theta=-1&{\longleftrightarrow}&\theta=0&{\longleftrightarrow}&\theta=1\\
\downarrow&~&\downarrow&~&\downarrow\\
\v_\Sl^-&{\longleftrightarrow}&\v_\SE&{\longleftrightarrow}&\v_\Sl^+\\
\downarrow&~&\downarrow&~&\downarrow\\
\tau\(\g_{\hspace{-.1em}\textsc{l}}^*\)\(\v_\Sl^-\)&{\longleftrightarrow}&
\tau\(\gB{\textsc{e}}\)\(\v_\SE\)&{\longleftrightarrow}&
\tau\(\gB{\textsc{l}}\)\(\v_\SE^+\)
\end{array}\!.
\end{align}

%
%
The irreducible representation of the amphospinor is given similarly to (\ref{piElr}):
In the amphometric space, $z_1$ and $z_2$ in (\ref{S3}) are not on $S_3$ in general, but are written using $(\tilde{z}_1,\tilde{z}_2){\in}(S^3=SU(2))\subset\C^2$ as 
\begin{align}
z_1&=\Re[\tilde{z}_1]+\kappa(\theta)\Im[\tilde{z}_1],~~z_2=-i\kappa(\theta)(\Re[\tilde{z}_2]+i\hspace{.1em}\Im[\tilde{z}_2]),
~~\text{where}~~
\left|\tilde{z}_1\right|^2+\left|\tilde{z}_2\right|^2=1.\label{amphoq}
\end{align}
The polynomial defined in (\ref{lambda}) with the amphometric and its vector space are denoted as $\lambda_{s}^{\theta}$ and $\VV^{\theta}_{\hspace{-.1em}s}$, respectively.
If we consider the definition (\ref{amphoq}) to be a transformation from $\{\tilde{z}_1,\tilde{z}_2\}$ to $\{z_1,z_2\}$, it is a linear isomorphism
\begin{align}
\text{Amp}:SU(2)\otimes[-1,1]{\rightarrow}\Ca(2):(\tilde{z}_1,\tilde{z}_2)\otimes\theta
{\mapsto}\text{Amp}(\theta)(\tilde{z}_1,\tilde{z}_2)=(\ref{amphoq})~{\in\hspace{.4em}}\Ca(2){\subset}Ampho(2).
\end{align}
This transformation provides the representation for $\Ca(2)$ from the $SU(2)$ representation given by (\ref{gEz1z2}) as
\begin{align}
\text{Amp}(\theta)\(\left.(\ref{gEz1z2})\right|^{~}_{z_i\rightarrow\tilde{z}_i}\)&\implies
\(\pij{s}^\theta(\gB{\theta})\bcdot\lambda_{s}^{\theta}\)(z_1,z_2)=\lambda_{s}^{\theta}(z'_1,z'_2),~~~
\text{with}~~~
\(
     \begin{array}{c}
      z'_1\\
      z'_2
     \end{array}
\)=\g^{-1}_\theta
\(
     \begin{array}{c}
      z^{~}_1\\
      z^{~}_2
     \end{array}
\).
\end{align}
This is one-to-one transformation and $\pij{s}$ is irreducible; thus. the representation $\pij{s}^\theta$ is also irreducible.
We obtain group and Lie-algebra representation as
\begin{subequations}
\begin{align}
\pi^\theta_s(\gB{c})&=\gB{c},~~\text{and}~~
\dot\pi^\theta_s(\ggg_c)=\ggg_{c}\in\LCa_\theta(2),\label{pigc}
\intertext{yielding}
\pi^\theta_s(\gB{c})\big|^{~}_{\theta=0}&=\gB{\textsc{e}},~~
\pi^\theta_s(\gB{c})\big|^{~}_{\theta=1}=\gB{\textsc{l}},~~\text{and}~~
\pi^\theta_s(\gB{c})\big|^{~}_{\theta=-1}=\pi^{~}_{\bar{s}}(\gB{\textsc{l}})=\bm{g}_{\textsc{l}}^*,
\end{align}
where $\LCa|\theta(2)$ is a space of the Lie-algebra for the causal subgroup.
\end{subequations}

Next, we consider two Lie-group homomorphisms (\ref{ismorSO4}) and (\ref {ismorSO13}) under the causal subgroup. 
The causal subgroup $\Ca(2)$ at $\theta=0$ is a subgroup of $SU(2)$, and that at $\theta=1$ is equivalent to $\SL2C$, respectively.
At $(-1,1)\supset\theta\neq0$, we obtain the Lie-algebra $\LCa_\theta(2)\subset\ggg\lll(2,\C)$ as (\ref{pigc}).
We schematically summarize the relations among them as  
\begin{align}
\begin{array}{ccccccc}
(\ref{ismorSO4})&\longrightarrow&\sss\ooo(4)&=&\sss\uuu(2)&\oplus&\sss\uuu(2)\\
&&&&\rotatebox[origin=c]{90}{$\subset$}&&\rotatebox[origin=c]{90}{$\subset$}\\
&&&&\left.\LCa_\theta(2)\right|^{~}_{\theta=0}&\oplus&\left.\LCa_\theta(2)\right|^{~}_{\theta=0}\\
&&&&\updownarrow&&\updownarrow\\
&&\LCa_\theta(2)\otimes[-1,1]&=&\LCa_\theta(2)\otimes[0,1]&\oplus&\LCa_\theta(2)\otimes[-1,0]\\
&&&&\updownarrow&&\updownarrow\\
&&&&\left.\LCa_\theta(2)\right|^{~}_{\theta=1}&\oplus&~\left.\LCa_\theta(2)\right|^{~}_{\theta=-1}\\
&&&&\rotatebox[origin=c]{90}{=}&&\rotatebox[origin=c]{90}{=}\\
(\ref{ismorSO13})&\longrightarrow&\sss\ooo(1,3)\otimes\C&=&\sss\lll(2,\C)&\oplus&\overline{\sss\lll(2,\C)}
\end{array}.\label{SU2SU2SL2CSL2C}
\end{align}
We introduced the tensor product of two vector spaces in the amphometric space with the internal (non space-time) group $G$  as
\begin{align}
\VV_{\hspace{-.1em}\textsc{s}}(\theta^{~}_1,\theta^{~}_2;G)&:=
\VV_{\!\hlf}^{\theta_1}\left[\gB{c}{\in}\Ca(2)\right]
\otimes_G
\VV_{\!\hlf}^{\theta_2}\left[\gB{c}{\in}\Ca(2)\right],\label{Sq22}
\end{align}
where $\VV_{\!\hlf}^{\theta}\left[\bm{g}{\in}G\right]$ means the vector space $\VV_{\!\hlf}^{\lambda}\in\C^2$ having the invariant bilinear form under the group action $\bm{g}$ of the group $G$ in the amphometric space.
This representation gives the Euclidean one (\ref{piElr}) at $\theta=0$ and the Lorentzian one (\ref{piLslsr}) at $\theta=1$ as
\begin{align*}
\VV_{\!\hlf}^{\theta}\left[\gB{\theta}{\in}\Ca(2)\right]\big|_{\theta=0}&=\VV_{\!\hlf}^\textsc{e},~~~
\VV_{\!\hlf}^{\theta}\left[\gB{\theta}{\in}\Ca(2)\right]\big|_{\theta=1}=\VV_{\!\hlf}^\textsc{l},~~\text{and}~~
\VV_{\!\hlf}^{\theta}\left[\gB{\theta}{\in}\Ca(2)\right]\big|_{\theta=-1}=\VV_{\!\overline{\hlf}}^\textsc{l}.
\end{align*}

A bilinear form for spinors in the amphometric space (amphospinor)  $\bm\xi(\theta),\bm\zeta(\theta){\hspace{.1em}\in\hspace{.1em}}\VV^{\theta}_{\hlf}$ is defined by combining (\ref{compPhi2}) and (\ref{LBLF2}) as
\begin{subequations}
\begin{align}
\<\bm\xi(\theta),\bm\zeta(\theta)\>&=
\left[\xi(-\theta)\right]^*_A\left[\bm{\epsilon}(\theta)\right]^{AB}\left[\bm\zeta(\theta)\right]_B~~\text{where}~
\bm{\epsilon}(\theta):=\(\!
\begin{array}{rr}
\cos(\pi\theta/2)&\sin(\pi\theta/2)\\
-\sin(\pi\theta/2)&\cos(\pi\theta/2)
\end{array}
\!\)\!,\label{<xixi>S}
\intertext{yielding}
\bm{\epsilon}(0)=\bm1^{~}_2\hspace{.7em}&\implies
\<\bm\xi(0),\bm\zeta(0)\>=\<\bm\phi^{~}_\textsc{h},\bm\varphi^{~}_\textsc{h}\>,
\end{align}
\end{subequations}
where $\bm\phi^{~}_\textsc{h}$ and $\bm\varphi^{~}_\textsc{h}$ are Higgs spinors corresponding to $\bm\xi(\theta=0)$ and $\bm\zeta(\theta=0)$, respectively.
We obtain that 
\begin{align*}
&\bm\xi(\theta)\big|_{\theta=+1}\in\VV^\textsc{l}_{\hlf}\iff\bm\xi^*(-\theta)\big|_{\theta=+1}\in\VV^\textsc{l}_{\hlf},\\
&\bm\xi(\theta)\big|_{\theta=-1}\in\VV^\textsc{l}_{\overline\hlf}\iff\bm\xi^*(-\theta)\big|_{\theta=-1}\in\VV^\textsc{l}_{\overline\hlf};
\intertext{thus, we can set in (\ref{<xixi>S}) as}
&\bm\xi^*(-\theta)=\bm\xi(\theta)\implies\bm\xi^*(-1)=\bm\xi(1)=\bpsi_\textsc{l},~~
\bm\xi^*(1)=\bm\xi(-1)=\dot{\bpsi}_\textsc{l},
\intertext{yielding}
\bm{\epsilon}(\pm1)=\pm\bm\epsilon^{~}_2&\implies
\<\bm\xi(1),\bm\zeta(1)\>=\<\bm\xi_\textsc{l},\bm\zeta_\textsc{l}\>,~~
\<\bm\xi(-1),\bm\zeta(-1)\>=-\<\dot{\bm\xi}_\textsc{l},\dot{\bm\zeta}_\textsc{l}\>.
\end{align*}
The epsilon-conjugate of the amphospinor is extended into the amphometric space  as
\begin{align}
\left[\bm\xi^{\dagger_{\hspace{-.1em}\epsilon}}(\theta)\right]^A:=\left[\bm\xi(\theta)\right]^*_A\left[\bm\epsilon(\theta)\right]^{AB}\implies
\<\bpsi(\theta),\bm\zeta(\theta)\>=\bm\xi^{\dagger_{\hspace{-.1em}\epsilon}}(-\theta){\bm\zeta}(\theta)=
\bm\xi^t(\theta)\cdot\bm\epsilon(\theta)\cdot{\bm\zeta}(\theta).
\end{align}

Consequently, the Higgs spinor and dotted and undotted Weyl spinors have been unified to the amphospinor with the causal group under the following diagram:
\begin{align}
\begin{array}{rccccc}
&\theta=-1&{\longleftrightarrow}&\theta=0&{\longleftrightarrow}&\theta=1\\
&\downarrow&~&\downarrow&~&\downarrow\\
\bm\xi(\theta)=&\bm\xi(-1)&{\longleftrightarrow}&\bm\xi(0)&{\longleftrightarrow}&\bm\xi(1)\\
&\rotatebox[origin=c]{90}{=}&~&\rotatebox[origin=c]{90}{=}&~&\rotatebox[origin=c]{90}{=}\\
&\dot{\bm\xi}&{\longleftrightarrow}&
\bm\phi^{~}_\textsc{h}&{\longleftrightarrow}&
\bm\xi\label{AmphoSpinorDiagram}
\end{array}
\end{align}
The following diagram shows a schematic view of relations among Lie algebras $\sss\uuu(2)$, ${\sss\lll(2,\C)}$ and $\overline{\sss\lll(2,\C)}$ in the amphometric space:
\begin{align}
\begin{array}{ccccccccc}
\theta=-1&{\longleftrightarrow}&-1<\theta<0&{\longleftrightarrow}&\theta=0&{\longleftrightarrow}&0<\theta<1&{\longleftrightarrow}&\theta=1\\
\downarrow&~&\downarrow&~&\downarrow&~&\downarrow&~&\downarrow\\
\overline{\sss\lll(2.\C)}&\longleftrightarrow&\LCa(2)&\longleftrightarrow&\sss\uuu(2)&\longleftrightarrow&\LCa(2)&\longleftrightarrow&\sss\lll(2.\C)\\
\rotatebox[origin=c]{90}{$\in$}&&\rotatebox[origin=c]{90}{$\in$}&&\rotatebox[origin=c]{90}{$\in$}&&\rotatebox[origin=c]{90}{$\in$}&&\rotatebox[origin=c]{90}{$\in$}\\
\gBa{r}(\varphi;-1)&\longleftrightarrow&\gBa{r}(\varphi;\theta)&\longleftrightarrow&\gBa{r}(\varphi;0)&\longleftrightarrow&\gBa{r}(\varphi;\theta)&\longleftrightarrow&\gBa{r}(\varphi;1)\\
\rotatebox[origin=c]{90}{=}&&&&\rotatebox[origin=c]{90}{=}&&&&\rotatebox[origin=c]{90}{=}\\
{\gBa{r}(\varphi;1)}^*&&&&\gBa{r}(\varphi)&&&&\pij{\hlf}(\gB{r})\\
\rotatebox[origin=c]{90}{=}\\\pij{\overline\hlf}(\gB{r})
\end{array}\!.
\end{align}

%% file: Sec3-4.tex
%
%
\subsection{Hodge-dual operator and chiral bases}\label{CaAM} 
We extend the Hodge-dual operator into the amphometric space modifying \textbf{Definition \ref{ctm}} for $\aaa^{~}_\theta,\bbb^{~}_\theta\in\Omega(\TsMQ)$ as
\begin{subequations}
\begin{align}
\text{(\ref{HodgeD})}\rightarrow
\aaa^{~}_\theta\wedge\HDQ(\bbb^{~}_\theta)&:=
\text{det}[\betathe]^\hlf\hspace{.1em}\<\aaa^{~}_\theta|\bbb^{~}_\theta\>\vvv^{~}_\theta,\label{HodgeDQ}
\intertext{where we utilise a convention as}
\text{det}[\betathe]^\hlf&=e^{3i\pi\theta/2}\!.\label{DetEtaHlf}
\end{align}
Thus, a real part of $\text{det}[\betathe]^\hlf$ can be negative.
\end{subequations}
The $\Ca(2)$-invariant bilinear form and pseudo-norm are also extended accordingly into the amphometric space as
\begin{align*}
\<\aaa^{~}_{\hspace{-.1em}\bullet}|\bbb^{~}_{\hspace{-.1em}\bullet}\>
\xrightarrow{\betaB\rightarrow\betathe}
\<\aaa^{~}_\theta|\bbb^{~}_\theta\>
~~\text{and}~~
\|\aaa^{~}_{\hspace{-.1em}\bullet}\|\vvv^{~}_\bullet
\xrightarrow{\betaB\rightarrow\betathe}
\|\aaa^{~}_\theta\|\hspace{.2em}\vvv^{~}_\theta,~~\bullet\in\{\textsc{e},\textsc{l}\},
\end{align*}
with
\begin{align*}
\vvv^{~}_\theta:=\frac{1}{4!}\epsilon^{~}_{\bcdots\bcdots}
\eeeQ^\bcdot\wedge\eeeQ^\bcdot\wedge\eeeQ^\bcdot\wedge\eeeQ^\bcdot
~~\text{and}~~
\aaa^{~}_\theta=\frac{1}{p!}{a}_{i_1\cdots i_p}\eee^{i_1}_\theta\wedge\cdots\wedge\eee^{i_p}_\theta,
\end{align*}
where $\eeeQ^a{\in}\Omega^1(\TsMQ)$ $(a\in\{0,1,2,3\}$) are the standard bases in $\TsMQ$  yielding
\begin{align*}
\eeeQ^a\big|^{~}_{\theta=0}=\eee^a_\textsc{e}~~\text{and}~~
\eeeQ^a\big|^{~}_{\theta=\pm1}=\eee^a_\textsc{l}.
\end{align*}
We introduce the ampho-base-vectors as 
\begin{align}
\eee_\theta^a&=\(\eee_\theta^0,\eee_\theta^1,\eee_\theta^2,\eee_\theta^3\):=
\(\eee^0,i\kappa(\theta)\eee^1,i\kappa(\theta)\eee^2,i\kappa(\theta)\eee^3\)\!,\label{eeeQ}
\end{align}
which can be understood analogically as analytic continuation owing to the replacement $t\rightarrow it$ utilised in the quantum field theory.
The pseudo-norm (\ref{pnormsq}) has complex values at other than $\theta=0,\pm1$; thus, it is not the norm and is called the amphonorm. 
Co-derivative and Laplace--Beltrami operators defined in (\ref{codiff}) and (\ref{LBop}) are extended into the amphometric space accordingly.
E.g., the Laplace--Beltrami operator in the amphometric space is
\begin{align*}
\Delta^{~}_\theta&=d^{~}_\theta\hspace{.1em}{\cod}^{~}_\theta+{\cod}^{~}_\theta\hspace{.1em}d^{~}_\theta=
\partial^{~}_0+e^{i\pi\theta}(\partial^{~}_1+\partial^{~}_2+\partial^{~}_3).
\end{align*}

%
%
The Hodge-dual operator acts on two-form objects in the amphometric space as
\begin{subequations}
\begin{align}
&\HDQ({\eee_\theta^0}\wedge{\eee_\theta^1})={\eee_\theta^2}\wedge{\eee_\theta^3},~~~
\HDQ({\eee_\theta^0}\wedge{\eee_\theta^2})={\eee_\theta^3}\wedge{\eee_\theta^1},~~~
\HDQ({\eee_\theta^0}\wedge{\eee_\theta^3})={\eee_\theta^1}\wedge{\eee_\theta^2},\notag\\
&\HDQ({\eee_\theta^1}\wedge{\eee_\theta^2})={\eee_\theta^0}\wedge{\eee_\theta^3},~~~
\HDQ({\eee_\theta^1}\wedge{\eee_\theta^3})={\eee_\theta^2}\wedge{\eee_\theta^0},~~~
\HDQ({\eee_\theta^2}\wedge{\eee_\theta^3})={\eee_\theta^0}\wedge{\eee_\theta^1},\notag
\intertext{and the $\Z_2$-grading two-form bases are provided, e.g., as}
&\SSS^{+\hspace{.1em}a}_\theta:=\left\{
{\eee_\theta^0}\wedge{\eee_\theta^1}+{\eee_\theta^2}\wedge{\eee_\theta^3},\hspace{.2em}
{\eee_\theta^0}\wedge{\eee_\theta^2}-{\eee_\theta^1}\wedge{\eee_\theta^3},\hspace{.2em}
{\eee_\theta^0}\wedge{\eee_\theta^3}+{\eee_\theta^1}\wedge{\eee_\theta^2}\right\},\label{Splu}\\
&\SSS^{-\hspace{.1em}a}_\theta:=\left\{
{\eee_\theta^0}\wedge{\eee_\theta^1}-{\eee_\theta^2}\wedge{\eee_\theta^3},\hspace{.2em}
{\eee_\theta^0}\wedge{\eee_\theta^2}+{\eee_\theta^1}\wedge{\eee_\theta^3},\hspace{.2em}
{\eee_\theta^0}\wedge{\eee_\theta^3}-{\eee_\theta^1}\wedge{\eee_\theta^2}\right\},\label{Smin}
\intertext{yielding}
&\PH{\pm}\SSS^{\pm\hspace{.1em}a}_\theta=\pm\SSS^{\pm\hspace{.1em}a}_\theta\label{PSpm}
\intertext{and}
&\SSS^{\pm\hspace{.1em}a}_\theta\wedge\SSS^{\pm\hspace{.1em}b}_\theta=
\begin{cases}
~~~0&a\hspace{.3em}{\neq}\hspace{.3em}b\\
\pm2\hspace{.1em}e^{3i\pi\theta/2}\hspace{.1em}\vvv^{~}_\theta&a=b
\end{cases},
~~\text{and}~~
\SSS^{\pm\hspace{.1em}a}_\theta\wedge\SSS^{\mp\hspace{.1em}b}_\theta=0.
\end{align}\label{SSv}
\end{subequations}
They provide $(\ref{DHES})$ by replacement ``$\textsc{e}$''$\mapsto$``$\theta$''.
One can easily confirm that
\begin{align}
\SSS^{\pm\hspace{.1em}a}_1:=\left.\SSS^{\pm\hspace{.1em}a}_\theta\right|^{~}_{\theta=1}=\SSS^{\pm\hspace{.1em}a}_\textsc{l},~~~
\SSS^{\pm\hspace{.1em}a}_{-1}:=\left.\SSS^{\pm\hspace{.1em}a}_\theta\right|^{~}_{\theta=-1}=\SSS^{\mp\hspace{.1em}a}_\textsc{l}~~\text{and}~~
\SSS^{\pm\hspace{.1em}a}_0:=\left.\SSS^{\pm\hspace{.1em}a}_\theta\right|^{~}_{\theta=0}=\SSS^{\pm\hspace{.1em}a}_\textsc{e}.\label{SpmQ}
\end{align}
Operators $\PH{\pm}$ with $\HDB\rightarrow\HDQ$ in (\ref{projHodge}) are still the projection operator in the amphometric space fulfiling $\PH{\pm}\bcdot\PH{\pm}=\PH{\pm}$ and $\PH{\pm}\bcdot\PH{\mp}=0$. 
Here, we do not put the subscript ``$\theta$'' from the projection operators as $\PH{\pm}$ for simplicity  since the metric is clear owing to the operand.
The projection operator in the amphometric space yields
\begin{align*}
(\ref{PSpm})\implies
P^{\pm}_{0}&:=
\PH{\pm}\big|^{~}_{\theta=0}=P^{\pm}_\textsc{e}
~~\text{and}~~
P^{\pm}_{1}:=\PH{\pm}\big|^{~}_{\theta=1}=P^{\pm}_\textsc{l}=P^{\mp}_{-1};
\end{align*}
thus, the middle formula of (\ref{SpmQ}) means that
\begin{align*}
P^{\pm}_{-1}\SSS^{\pm}_{-1}=P^{\mp}_{\textsc{l}}\SSS^{\mp}_{\textsc{l}}=\SSS^{\mp}_{\textsc{l}}.
\end{align*}
The $\Z_2$-grading space of curvature two-forms and their Hodge-dual  in the amphometric space are provided as
\begin{align}
\FFFQ^\pm:=\PH{\pm}\hspace{.1em}\FFFQ
~~\textrm{and}~~
\HDQ(\FFFQ^\pm)=\pm\FFFQ^\pm
\in \VH{\pm}\left(\OmegaQ^2(\TsMQ)\right){\otimes}Ad(\ggg),\label{FpmQ}
\end{align}
where $\FFFQ^\pm$ are, respectively,  spanned by $\SSS^\pm_\theta$.

The curvature is the two-form object from the structure equation concerning the one-form connection.
The Hodge-dual operator is not endomorphism for one-form objects; thus, their space does not have the $\Z_2$-grading structure.
Nevertheless, splitting the one-form object space into two subspaces concerning the $\Z_2$-grading structure of the curvature two-form is useful.
We discuss this subject in the amphometric space below.
We introduce a chiral basis for the one-form objects as
\begin{align}
\eee_{ch}^a&:=\(\eee^0-\kappa(\theta)\eee^3,\eee^0+\kappa(\theta)\eee^3,\eee^1+i\eee^2,\eee^1-i\eee^2\)\!.\label{chiraleee}
\end{align}
In the Lorentzian metric space, $\eee_{ch}^3$ ($\eee_{ch}^4$) corresponds to a chiral-right (-left) polarization vector for a massless vector-boson travelling along the third axis, respectively.
Though the one-form object cannot be an eigenvector concerning the Hodge-dual operator, we observe the chiral structure of this one-form basis, such as
\begin{align}
\begin{array}{lc}
\eee^1_{ch}{\wedge}\eee^3_{ch}=
[\SSS^+_\theta]^1{+}i\hspace{.1em}[\SSS^{+}_\theta]^2\in \VH{+}\left(\OmegaQ^2\right),&
\eee^1_{ch}{\wedge}\eee^4_{ch}=
[\SSS^-_\theta]^1{-}i\hspace{.1em}[\SSS^{-}_\theta]^2\in \VH{-}\left(\OmegaQ^2\right)\\
\eee^2_{ch}{\wedge}\eee^3_{ch}=
[\SSS^-_\theta]^1{+}i\hspace{.1em}[\SSS^{-}_\theta]^2\in \VH{-}\left(\OmegaQ^2\right),&
\eee^2_{ch}{\wedge}\eee^4_{ch}=
[\SSS^+_\theta]^1{-}i\hspace{.1em}[\SSS^{+}_\theta]^2\in \VH{+}\left(\OmegaQ^2\right)
\end{array}\!.\label{eeSS}
\end{align}
Owing to this chiral structure, we assign base vectors $\eee^1_{ch}$ as the chiral-right and $\eee^2_{ch}$ as the chiral-left.
A two-form object in $\VH{+}\left(\OmegaQ^2\right)$ ($\VH{-}\left(\OmegaQ^2\right)$) is, respectively, SD (ASD); thus, the chiral-right polarization corresponds to the SD two-form and the chiral-right to ASD.
We have exploited the chiral representation for the Clifford algebra, which corresponds to taking the third-axis $\eee^3$ as the spin quantisation axis.
Thus, the chiral-right (-left) spinor moving along the third axis has a positive (negative) helicity, respectively.
We show a relation between the chirality of the Weyl spinor and the two-form object schematically as
\begin{align}
&\hspace{-2em}\text{Weyl spinor with chiral representation}\longrightarrow\text{spin quantisation-axis}=\eee^3\!,\notag\\
&\longrightarrow\left\{
\begin{array}{lllcr}
\eee^3_{ch}\in\text{right polarization}&\xrightarrow{\text{(\ref{eeSS})}}&\{\eee^1_{ch},\eee^3_{ch}\}\in\text{chiral-right}&=&\text{SD}\\
\eee^4_{ch}\in\text{left polarization}&\xrightarrow{\text{(\ref{eeSS})}}&\{\eee^2_{ch},\eee^4_{ch}\}\in\text{chiral-left}&=&\text{ASD}
\end{array}\right.\!.
\end{align}

%% file: Sec3-3.tex
%
%
\subsection{Clifford algebra and Dirac operator}\label{CaAM}
A representation of Clifford algebra in $Spin(\M_\theta)$ with the fixed $\theta$-value, namely the \textit{ amphoClifford algebr}a is defined to fulfil
algebra 
\begin{align}
&\{\gamma_{\theta}^{\hspace{.3em}a},\gamma_{\theta}^{\hspace{.3em}b}\}=
2\hspace{.1em}\eta_\theta^{\hspace{.2em}ab}\hspace{.1em}\bm{1}_\Sp
\implies\gamma^{\hspace{.3em}0}_{\theta}:=\gamma^{\hspace{.3em}0}_{\textsc{e}}
~~\textrm{and}~~\gamma^{\hspace{.3em}a}_{\theta}:=e^{i\pi\theta/2}\hspace{.1em}\gamma^{\hspace{.3em}a}_{\textsc{e}}
~~\textrm{for}~~a\in\{1,2,3\}.\label{gamma-theta}
\end{align}
The definition (\ref{gamma-theta}) fulfils the amphoClifford algebra at any values of $\theta\in[-1,1]$.
The amphoClifford algebra provides Euclidean and Lorentzian ones as
\begin{align}
\bm\gamma^{~}_\theta=
\begin{cases}
\bm\gamma^{~}_\textsc{l},&\theta=+1\\
\bm\gamma^{~}_\textsc{e},&\theta=\hspace{.8em}0\\
\bar{\bm\gamma}^{~}_\textsc{l},&\theta=-1\\
\end{cases}.
\end{align}
According to introducing the amphoClifford algebra, we obtain a generator of $Ampho(2)$ as
\begin{align}
S^{ab}_{\theta}:=
i\hspace{.1em}\left[\frac{\gamma^{\hspace{.3em}a}_{\theta}}{2},\frac{\gamma^{\hspace{.3em}b}_{\theta}}{2}\right].\label{Asp}
\end{align}

For the chiral operator in the amphometric space, namely the \textit{amphochiral operator}, we exploit the definition (\ref{Gamma}) with $\bm\eta=\bm\eta^{~}_\theta$ and obtain it as
\begin{align}
\Gamma^{~}_{\!\theta}&=\sqrt{\frac{1}{\text{det}[\betaQ]}}
\gamma_{\theta}^{\hspace{.3em}0}\gamma_{\theta}^{\hspace{.3em}1}\gamma_{\theta}^{\hspace{.3em}2}\gamma_{\theta}^{\hspace{.3em}3}=
f_\Gamma(\theta)
\gamma_{\textsc{e}}^{\hspace{.2em}0}\gamma_{\textsc{e}}^{\hspace{.2em}1}\gamma_{\textsc{e}}^{\hspace{.2em}2}\gamma_{\textsc{e}}^{\hspace{.2em}3}
\label{GammaQ1}
\quad\text{with}\quad
f_\Gamma(\theta):=\frac{e^{3i\pi\theta/2}}{\sqrt{e^{3i\pi\theta}}}.
\end{align}
Function $f_\Gamma(\theta)$ provides a factor at boundaries of the amphometric space as 
\begin{align}
f_\Gamma(\theta)=
\begin{cases}
+1,&\theta\rightarrow1+0,\quad\theta\rightarrow-1-0\\
-1,&\theta\rightarrow1-0,\quad\theta\rightarrow-1+0
\end{cases},
\quad\text{owing to the standard convention }\text{Re}[\sqrt{\bullet}]\geq0.
\end{align}
Thus, in the amphometric space with $\theta\in[-1,1]$, we obtain $f_\Gamma(1)=-f_\Gamma(-1)=1$, and it has discontinuity at $\theta=\pm1/3$.
To avoid this discontinuity, we exploit the convention 
\begin{align}
\sqrt{e^{3i\pi\theta}}=e^{3i\pi\theta/2}\implies
f_\Gamma(\theta)=1
\end{align}
for the amphochiral operator (\ref{GammaQ1}).
As a result, the factor in the chiral operator compensates a $\theta$-dependence in $\gamma^{~}_\theta$, and we have a $\theta$-independent amphochiral operator as
\begin{align}
\text{(\ref{GammaQ1})}&=
\gamma_{\textsc{e}}^{\hspace{.2em}0}\gamma_{\textsc{e}}^{\hspace{.2em}1}\gamma_{\textsc{e}}^{\hspace{.2em}2}\gamma_{\textsc{e}}^{\hspace{.2em}3}=
\begin{cases}
\(\!
\begin{array}{rc}
 -\bm{1}_2 & \bm{0}_2\\
  \bm{0}_2 & \bm{1}_2
\end{array}
\!\),&\text{dim-4}\\
\hspace{.5em}\(\!
\begin{array}{rc}
 -1 & 0\\
  0 & 1
\end{array}
\!\),&\text{dim-2}
\end{cases},
\label{GammaQ}
\intertext{yielding the $\theta$-independent projection operator as}
P^\pm_{\!\theta}&:=\frac{1}{2}\(\bm{1}_\Sp\pm\Gamma_{\!\theta}\)=\text{(\ref{Ppm})}\implies
P^\pm_{\!\theta}{\bcdot}P^\pm_{\!\theta}=P^\pm_{\!\theta},~
P^\pm_{\!\theta}P^\mp_{\!\theta}=0.\label{proj1Q}
\end{align}
The chirality of the amphospinor is consistently defined owing to the projection operators (\ref{proj1Q}) at any $\theta\in[-1,1]$; thus, the amphospinor space $\VV^\theta_{\!\hlf}$ splits into a disjoint pair as
\begin{subequations}
\begin{align}
\Gamma^{~}_{\!\theta}\hspace{.2em}\bm\xi^{~}_\textsc{u}(\theta)&=-\bm\xi^{~}_\textsc{u}(\theta)\implies 
P^-_{\!\theta}\bm\xi(\theta)=\bm\xi^{~}_\textsc{u}(\theta)
\in\VV^{\theta\textsc{u}}_{\!\hlf},\label{VpmthetaR}\\
\Gamma^{~}_{\!\theta}\hspace{.2em}\bm\xi^{~}_\textsc{d}(\theta)&=+\bm\xi^{~}_\textsc{d}(\theta)\implies
P^+_{\!\theta}\bm\xi(\theta)=\bm\xi^{~}_\textsc{d}(\theta)
\in\VV^{\theta\textsc{d}}_{\!\hlf}.\label{VpmthetaL}
\end{align}
\end{subequations}
We extend amphospinors, shown in the diagram (\ref{AmphoSpinorDiagram}), into the chiral amphospinors such that
\begin{align}
\begin{array}{rccccc}
&\theta=-1&{\longleftrightarrow}&\theta=0&{\longleftrightarrow}&\theta=1\\
&\downarrow&~&\downarrow&~&\downarrow\\
\bm\xi^{~}_\textsc{u}(\theta)=&\bm\xi^{~}_\textsc{u}(-1)=\dot{\bm\xi}^{~}_\textsc{u}&{\longleftrightarrow}
                                       &\bm\xi^{~}_\textsc{u}(0)=\bm\phi^{~}_\textsc{u}&{\longleftrightarrow}
                                       &\bm\xi^{~}_\textsc{u}(1)=\bm\xi^{~}_\textsc{u}\\
\\
\bm\xi^{~}_\textsc{d}(\theta)=&\bm\xi^{~}_\textsc{d}(-1)=\dot{\bm\xi}^{~}_\textsc{d}&{\longleftrightarrow}
                                       &\bm\xi^{~}_\textsc{d}(0)=\bm\phi^{~}_\textsc{d}&{\longleftrightarrow}
                                       &\bm\xi^{~}_\textsc{d}(1)=\bm\xi^{~}_\textsc{d}\\
&\downarrow&~&\downarrow&~&\downarrow\\
\text{(\ref{VpmthetaR})}=&\text{(\ref{phiLp2})}&{\longleftrightarrow}&\text{(\ref{phiEm})}&{\longleftrightarrow}&\text{(\ref{phiLm1})}\\ \\
\text{(\ref{VpmthetaL})}=&\text{(\ref{phiLm2})}&{\longleftrightarrow}&\text{(\ref{phiEp})}&{\longleftrightarrow}&\text{(\ref{phiLp1})}
\label{AmphoSpinorDiagram2}
\end{array}.
\end{align}

The Dirac operator in the amphometric space is defined as
\begin{align}
\ds_{\theta}&:=\iota_\gamma{d_{\theta}}=\slash{\partial}_{\theta}=
\gamma^a_{\theta}\hspace{.2em}\partial_a^{~}\implies
\slash{\partial}^{~}_{\theta} P^\pm_{\!\theta}=P^\mp_{\!\theta}\hspace{-.2em}\slash{\partial}^{~}_{\theta}.\label{DiracQ}
\end{align}
The operator flips the chirality; thus, it is the Dirac operator.
The generalised Lapracian is provided as
\begin{align}
\(\slash{\partial}_{\theta}\)^2&=\Delta^{~}_{\theta}\hspace{.1em}\bm{1}_\Sp
~~\text{where}~~
\Delta^{~}_{\theta}:=(\partial^{~}_0)^2+e^{i\pi\theta}\((\partial^{~}_1)^2+(\partial^{~}_2)^2+(\partial^{~}_3)^2\)\!.\label{DeltaQ}
\end{align}
We can confirm that
\begin{align*}
\slash{\partial}_{\textsc{v}\theta}&\xrightarrow{\theta=0}\slash{\partial}_{\textsc{ve}}\Rightarrow\text{(\ref{diracdE})},\hspace{.9em}
\slash{\partial}_{\textsc{v}\theta}\xrightarrow{\theta=1}\slash{\partial}_{\textsc{vl}}\Rightarrow\text{(\ref{diracd})},\\
\slash{\partial}_{\textsc{s}\theta}&\xrightarrow{\theta=0}\slash{\partial}_{\textsc{se}}\Rightarrow\text{(\ref{gammadelE})},~
\slash{\partial}_{\textsc{s}\theta}\xrightarrow{\theta=1}\slash{\partial}_{\textsc{sl}}\Rightarrow\text{(\ref{gammadelL})},~~
\slash{\partial}_{\textsc{s}\theta}\xrightarrow{\theta=-1}\overline{\slash{\partial}}_{\textsc{sl}}\Rightarrow\text{(\ref{gammadelL})}.
\end{align*}

%% file: Sec4.tex
\subsection{\YMU theory with amphometric}\label{CFT}
\textbf{Section \ref{YMUtheory}} formulated the classical \YMU theory in the local inertial manifold with internal $\SUW$ and local $SO(1,3)$ groups.
This section extends the theory into the amphometric space.
$SO(4)$ and $SO(1,3)$ groups compose the ampshometric space as the one-parameter subgroup of $\SL2C$. 
As a result, we showed symmetries of the Euclidean space are homotopic to those of the Minkowski space through the amphometric, which allows us to import some kinds of mathematical achievements of the \YMU theory in the Euclidean space into the Minkowski space\cite{Kurihara:2022sso}.

%
%
\input{Sec4-2}

%
%
\input{Sec4-3}

%
%
\input{Sec4-Higgs}

%
%
\input{Sec4-Fmass}

%% file: Sec4-2.tex
%
%
\subsubsection{Gauge bosons}
%
%
\paragraph{a) Yang--Mills gauge boson:}\label{YMinThetaB}
We extend the Yang--Mills gauge boson Lagrangian into the amphometric space.
In the spinor-gauge bundle, we introduced connection $\AAA^{~}_\SU$ and curvature $\FFF^{~}_\SU$ in \textbf{section \ref{YMgauegbundle}}. 
We replace the metric tensor in connections and curvatures as discussed in \textbf{section \ref{CaAM}} and denote them as $\AAA^{~}_\SU\rightarrow\AAA^{~}_{\theta}$ and $\FFF^{~}_\SU\rightarrow\FFF^{~}_\theta$, where we omit subscript ``$\SU$'' for simplicity.
The gauge boson Lagrangian is defined in the amphometric space as
\begin{subequations}
\begin{align}
\text{(\ref{YMLL})}&\implies
\LLL_{\textsc{y\hspace{-.1em}m}}(\AAA^{~}_\theta):=\text{det}[\betathe]^\hlf\hspace{.2em}
\Tr\left[\|\FFF^{~}_\theta(\AAA^{~}_\theta)\|^2\right]\vvv^{~}_\theta=
\Tr\left[\FFF^{~}_\theta(\AAA^{~}_\theta)\wedge\hat{\FFF}^{~}_\theta(\hat\AAA^{~}_\theta)\right],\label{LLLGB}
\end{align}
where we treat square-root function det$[\betathe]^\hlf$ through (\ref {DetEtaHlf}).
We utilise Carten's \textit{rep\'{e}re mobile} formalism and represent Lagrangians the coordinate-free way like (\ref{LLLGB}).
Although this representation can show the theory in a frame-independent way, the role of the metric tensor is not apparent.
Here, we show the Lagrangian form using a component representation as,
\begin{align}
\text{(\ref{LLLGB})}&=
-\frac{1}{4}\etaQ^{\bcdot\star}\etaQ^{\bcdot\star}
\f^{~}_{\hspace{-.1em}\theta\hspace{.1em}\bcdots}\hspace{.2em}
\f^{~}_{\hspace{-.1em}\theta\hspace{.1em}\stars}\hspace{.2em}\vvv^{~}_\theta
~~\text{and}~~\f^{~}_{\hspace{-.1em}\theta\hspace{.1em}ab}:=\f^{I}_{\hspace{-.1em}\theta\hspace{.1em}ab}\hspace{.2em}\tau_I,\label{LLLGB2}
\end{align}
\end{subequations}
and clarify the role of metric tensors.
In the amphometric space, we obtain a component representation of the equation of motion for the $\SUW$ Yang--Mills gauge boson as
\begin{align}
\hat{d}_{\AAAQ}\FFFQ&=
\hat{d}\FFFQ+\frac{c^{~}_\SU}{2}\left[\hat\AAA^{~}_\theta,\FFFQ\right]^{~}_{\hspace{-.1em}\wedge}=
{\etaQ}^\bcdots\(
\partial_\bcdot\f^I_{\hspace{-.1em}\theta\hspace{.1em}\bcdot\star}+{c^{~}_\SU}\hspace{.1em}f^I_{~JK}\hspace{.1em}\Aa^J_{\theta\hspace{.1em}\bcdot}\hspace{.1em}\f^K_{\hspace{-.1em}\theta\hspace{.1em}\bcdot\star}
\){\VVV_\theta}^\star=\dot{0},\label{ELEoMpm}
\end{align}
corresponding to (\ref{YMGBEoM3}) in the Lorentzian metric space, where $\Aa^I_{\theta\hspace{.1em}a}$ is a coefficient function of $\AAA^{~}_\theta$.
The structure constant $f^I_{~JK}$ is due to the internal $\SUW$ and has nothing to do with the metric tensor of the space-time manifold.

A space of the curvature two-form has the $\Z_2$-grading structure owing to the projection operator concerning the Hodge-dual as discussed in \textbf{section \ref{HDO}}.
We can decompose the curvature two-form and its Hodge-dual, respectively, as
\begin{align}
\FFF^{~}_\theta&=\FFF^{+}_\theta+\FFF^{-}_\theta
~~\text{and}~~
\hat\FFF^{~}_\theta=\FFF^{+}_\theta-\FFF^{-}_\theta,
\end{align}
where $\FFF^{\pm}_\theta$ is defined by (\ref{FpmQ}).
Consequently, we obtain the Lagrangian form using $\FFF^{\pm}_\theta$ as
\begin{align}
\text{(\ref{LLLGB})}&=\(\|\FFF^{+}_\theta\|^2+\|\FFF^{-}_\theta\|^2\)\vvv^{~}_\theta.
\end{align}
The extremal of the bosonic action integral concerning a amphoconnection $\AAA^{~}_\theta$ is given by the SD curvature ($\FFF^{+}_\theta$) or ASD curvature ($\FFF^{-}_\theta$) in the \textbf{Euclidean space} as known well.
We prove this remark for investigating the difference between Euclidean and Minkowski spaces.
\begin{remark}\label{SDorASD}
The Yang--Mills action in the closed and oriented Euclidean metric space has extremal at the SD- or ASD-curvature.
\end{remark}
\begin{proof}
The Yang--Mills action has the representation using SD and ASD curvatures as
\begin{align}
\I_{\hspace{-.2em}\textsc{y\hspace{-.1em}m}}^{~}&=\text{det}[\betathe]^{\hlf}\hspace{.2em}\int_\TsMQ\|\FFF^{~}_\theta\|^2\vvv^{~}_\theta=
\text{det}[\betathe]^{\hlf}\hspace{.2em}\int_\TsMQ\(\|\FFF^{+}_\theta\|^2+\|\FFF^{-}_\theta\|^2\)\vvv^{~}_\theta,\label{IymQ22}
\end{align}
where we assume $\TsMQ$ is the closed and oriented manifold.
On the other hand, the Chern index in the amphometric space is provided as
\begin{align}
c^{~}_2\left(\FFF^{~}_\theta\right)&=\frac{1}{8\pi^2}
\int_\TsMQ\left(\Tr\left[
\FFF^{~}_\theta\wedge\FFF^{~}_\theta\right]
-\Tr\left[\FFF^{~}_\theta\right]^2
\right)\vvv^{~}_\theta=
\frac{\text{det}[\betathe]^\hlf\hspace{.2em}}{8\pi^2}
\int_\TsMQ\left(
\|\FFF^{+}_\theta\|^2-\|\FFF^{-}_\theta\|^2\right)\vvv^{~}_\theta,\label{ChernClass}
\end{align}
where we use $\Tr\left[\FFF^{~}_\theta\right]=0$.
When $\theta=0$, equivalently, when the space-time manifold has the Euclidean metric, $\|\FFF^{\pm}_\textsc{e}\|^2$ is positive definite and we obtain that
\begin{align*}
\I_{\hspace{-.2em}\textsc{y\hspace{-.1em}m}}^{~}& \geq 8\pi^2\left|{c}_2\left(\FFF^{~}_\textsc{e}\right)\right|.
\end{align*}
Therefore, the extremal of the action integral is given by $\|\FFF^{+}_\textsc{e}\|=0$ or $\|\FFF^{-}_\textsc{e}\|=0$. 

At the general point other than $\theta=\pm1,0$, the amphonorm $\|\FFF^{\pm}_\theta\|^2$ is complex valued in general; thus, it is not the ordinal number, and the magnitude relationship cannot be defined.
In the Lorentzian metric space, the pseudo-norm $\|\bullet\|^{~}_\textsc{l}\in\R$ is neither positive definite nor non-degenerate.
Therefore, \textbf{Remark \ref{SDorASD}} is maintained only in the Euclidean metric space.
We note that when the Lagrangian density is given by the curvature norm squared in the local inertial space-time, \textbf{Remark \ref{SDorASD}} holds even with $\www\neq0$.
On the other hand, the whole \YMU action includes the $\SUW$ gauge connection, and the Lagrangian density is not simply given by the $\SUW$ curvature norm squared as (\ref{IymQ22}); thus, the \textbf{Remark \ref{SDorASD}} holds only in vacuum.
\end{proof}

Though \textbf{Remark \ref{SDorASD}} holds only in the Euclidean space, we consider extending it to the entire amphometric space.
Hereafter, we assume a solution exists and is SD.
We introduce a space
\begin{align*}
\W^+&:=\MQ\otimes\theta\in[0,1],~~~\W^-:=\MQ\otimes\theta\in[-1,0]~~\text{and}~~\W:=\W^+\cup\W^-=\MQ\otimes\theta\in[-1,1].
\end{align*}
Suppose that the equation of motion $(\ref{ELEoMpm})$ has solutions at any $\theta\in[0,1)$.
At $\theta=1$, the solution exists at least in a subspace of $\TsML$.
More precisely, at $\theta=1$, we assume that singularities of the curvature are at most countable number of poles of the type 
\begin{align}
\FFFQ^{~}&=\sum_i\alpha_i(\|\bm{r}^i_\theta\|^{2}+\lambda)^{-m}+\beta\label{PoleSol}
\end{align}
on a space-like Cauchy surface, where $\bm{r}^i_\theta$ is position vectors of an $i^\text{th}$ pole on the Cauchy surface at $\theta=0$, $m\in\N$ and $0<\lambda\in\R$.
$\{\alpha_i,\beta\}$ are regular functions in $\TMQ$.
In this case, the singularity does not exist in $\W^+{\hspace{-.1em}\setminus\hspace{-.1em}}\W^+_1$ since $\|\bullet^{~}_\textsc{e}\|\geq0$ and $\Im \|\bullet^{~}_{\theta\neq1}\|\neq0$.
At $\theta=0$, we assume the equation of motion gives the SD curvature $\FFFO^+$ as its solution. 
Suppose $\AAAO^{+}$ is the vacuum solution of the Yang--Mills equation providing the SD curvature through the structure equation at $\theta=0$; we obtain that
\begin{align}
\hat{d}\FFFO^+=0
~~\text{and}~~
\FFFO^+=d\AAAO^+-i\hspace{.1em}c^{~}_\SU\hspace{.2em}\AAAO^+\wedge\AAAO^+.\label{ASDEoM}
\end{align}
Owing to the above assumptions, this solution can be extended into $\W^{+}{\hspace{-.1em}\setminus\hspace{-.1em}}{\W^{+}_1}$, and into $\W^{+}_1$ except at poles above.
Consequently, we obtain a diagram for the SD curvature $\FFFQ^+$ as
\begin{align}
&
\begin{array}{cccccccclc}
&&\theta=0&{\longleftrightarrow}&0<\theta<1&{\longleftrightarrow}&\theta=1\\
&&\downarrow&~&\downarrow&~&\downarrow\\
\Omega^2_\textsc{sd}\(\TsME\)&\hookrightarrow&\Omega^2_\textsc{sd}\(T^*\W^+_0\)&\longleftrightarrow&\Omega^2_\textsc{sd}\(T^*\W^+_\theta\)&\longleftrightarrow&\Omega^2_\textsc{sd}\(T^*\W^+_1\)&\hookleftarrow&
\Omega^2_\textsc{sd}\(\overline{\TsML}\)\\
\rotatebox[origin=c]{90}{$\in$}&&\rotatebox[origin=c]{90}{$\in$}&&\rotatebox[origin=c]{90}{$\in$}&&\rotatebox[origin=c]{90}{$\in$}&&\hspace{1em}\rotatebox[origin=c]{90}{$\in$}\\
\FFFE^+&=&\FFFO^+&\longleftrightarrow&\FFFQ^+&\longleftrightarrow&\FFF^+_1&=&\hspace{.9em}\FFFL^+
\end{array}\label{diagp}
\intertext{where}  
&\hspace{5em}T^*\W_\theta^+:=T^*\hspace{-.1em}\M_\theta\otimes\theta\in[0,1],\quad
\overline{\TsML}:=\TsML{\hspace{-.1em}\setminus\hspace{-.1em}}\left\{\bm{r}^i_\textsc{l}(t)\big|\|\bm{r}^i_\textsc{l}(t)\|^{2}+\lambda=0\right\},\label{traj}
\intertext{and $\Omega^2_\textsc{sd}$ is a vector space $\Omega^2$ spanned by the SD basis as}  
&\hspace{5em}\aaa\in\Omega^2_\textsc{sd}\(T^*\W_\theta\)\implies\HDQ\(\aaa\)=+\aaa.
\end{align}
The connection and curvature forms with $\theta\in[0,1)$ also belong to the adjoint representation of the causal subgroup, and they at $\theta=1$ belong to $\SL2C$.
(See (\ref{SU2SU2SL2CSL2C})).
We set the Lie algebra assignment as
\begin{align}
\FFFL^+{\hspace{.1em}\in\hspace{.1em}}\Omega^2_\textsc{sd}\(\overline{\TsML}\)\otimes{Ad}\(\sss\lll(2,\C)\).\label{Lieassignp}
\end{align}
The Euclidean SD solution can be extended to $\W^-$, too.
The diagram for $\W^-$ is
\begin{align}
\begin{array}{cccccccclc}
&&\theta=-1&{\longleftrightarrow}&-1<\theta<0&{\longleftrightarrow}&\theta=0\\
&&\downarrow&~&\downarrow&~&\downarrow\\
\Omega^2_\textsc{sd}\(\overline{\TsML}\)&\hookrightarrow&\Omega^2_\textsc{sd}\(T^*\W^-_{-1}\)&\longleftrightarrow&\Omega^2_\textsc{sd}\(T^*\W^-_\theta\)&\longleftrightarrow&\Omega^2_\textsc{sd}\(T^*\W^-_0\)&\hookleftarrow&
\Omega^2_\textsc{sd}\(\TsME\)\\
\rotatebox[origin=c]{90}{$\in$}&&\rotatebox[origin=c]{90}{$\in$}&&\rotatebox[origin=c]{90}{$\in$}&&\rotatebox[origin=c]{90}{$\in$}&&\hspace{1em}\rotatebox[origin=c]{90}{$\in$}\\
\FFFL^+&=&\FFF_{-1}^+&\longleftrightarrow&\FFFQ^+&\longleftrightarrow&\FFF^+_0&=&\hspace{.9em}\FFFE^+.
\end{array}\label{diagm}
\end{align}

We define the Chern index in the Lorentzian metric space as the limit 
\begin{align}
\infty>\left|\lim_{\theta\rightarrow\pm1}\frac{\text{det}[\betathe]^\hlf\hspace{.2em}}{8\pi^2}\int_{\overline\TsMQ}\|\FFF^{+}_\theta\|^2\hspace{.1em}\vvv^{~}_\theta\right|&=:
\left|c^{~}_2\left(\FFFL\right)\right|=
\left|\I_{\hspace{-.2em}\textsc{y\hspace{-.1em}m}}^{\textsc{l}}\right|,\label{ChernClassL}
\end{align}
when it exists as, at least,  the one-sided limit of $\theta\rightarrow\pm1\mp\varepsilon$ with $0<\epsilon\in\R$.
This quantity is homotopy equivalence to the Chern index in the Euclidean space; thus, the Chern index has the same number as that in the Euclidean space.  
The Lorentzian SD curvature $\FFFL^+$ is provided from Lorentzian gauge-connection $\AAAL$, which fulfils the structure equation and equation of motion in the Lorentzian metric space.
Consequently, when the Euclidean Yang--Mills gauge curvature is SD and fulfils (\ref{PoleSol}), the Lorentzian gauge curvature is also SD providing the Chern index as above.
We discuss it in detail from the cobordism point of view in \textbf{Appndix \ref{IAMS}} for the instanton solution.

We showed that SD solutions in the Lorentzian metric space correspond to those in the Euclidean metric space above.
We can also show the Lorentzian Yang--Mills equation has no solutions other than them.
Suppose the Lorentzian Yang--Mills equation has solutions of a type
\begin{align}
\FFFL&=\alpha\hspace{.1em}\FFFL^++\beta\hspace{.1em}\FFFL^- ~~\text{with}~~\alpha,\beta\in\C,\label{PoleSolLorentz}
\end{align}
in a domain $\overline\TsML$ and singularity of solution $\FFFL$ is a type  (\ref{PoleSol}).
Solutions $\FFFL^\pm$ in (\ref{PoleSolLorentz}) are, respectively, spanned by basis $\SSS^{\pm\hspace{.1em}a}_\textsc{l}$ given by (\ref{SpL}) and (\ref{SmL}).
We introduce $\overline{T^*\W^+_1}$ as submersion of $\overline\TsML$ by means of a differentiable map
\begin{align} 
\SSS^{\pm\hspace{.1em}a}_\textsc{l}\rightarrow\SSS^{\pm\hspace{.1em}a}_\theta
~~\text{and}~~
\bm\eta^{~}_\textsc{l}\rightarrow\bm\eta^{~}_\theta ~~\text{in}~~ \theta\in [-1,1].
\end{align}
The same replacement applying to the Lorentzian Yang--Mills equation provides  the amphometric Yang--Mills equation having the solution $\FFFQ$ in $T^*\W^+_{\theta}$, which is
closer of $\overline{T^*\W^+_{\theta}}$ at $\theta\neq1$.
In this case, $\FFFQ$ is a solution of the Yang--Mills equation and the structure equation in $T^*\W^+_{\theta}$ at $\theta\in[0,1)$.
The Chern index has the same number in the Lorentzian and Euclidean metric spaces owing to the cobordism; thus, $\|\FFFE^-\|=0\implies\|\FFFL^-\|=0$.
Consequently, we have concluded that either the SD or ASD solution gives the solution of the Lorentzian Yang--Mills equation.

%
%
\paragraph{b) gravitational gauge boson:}
The gravitational Lagrangian form, replacing the metric tensor from Lorentzian to amphometric, provides the Lagrangian as
\begin{align}
(\ref{EHLL})&\implies
\LLL_{\textsc{g}\hspace{-.05em}\textsc{r}}(\wwwQ^{~},\eeeQ^{~}):=-\frac{\text{det}[\betathe]^\hlf}{\hbar\kappa^{~}_\textsc{E}}
\Tr^{~}_\cP\hspace{-.2em}
\left[\|\FFF^{\theta}_\cP\|^2_{\theta}\right]\vvv^{~}_\theta=
-\frac{1}{\hbar\kappa^{~}_\textsc{E}}\Tr^{~}_\cP\hspace{-.2em}\left[\FFF^\theta_\cP\wedge\hat\FFF^\theta_\cP\right],\label{EHLformula}
%
%
\end{align}
where co-Poincar\'{e} connection and curvature are given by (\ref{cPLieA1}) and (\ref{cPLieA2}) with the Lorentzian metric.
$\FFF^\theta_\cP$ is a Co-Poincar\'{e} curvature written by $\wwwQ$ and $\eeeQ$.
A space of the two-form objects is split into two subspaces owing to the Hodge-dual operator as (\ref{superspace}).
According to this splitting, we obtain the splitting  
\begin{align*}
\Tr_\SO^{~}[\RRR_\theta^{~}(\wwwQ^{~})\wedge\SSS_\theta^{~}(\eeeQ^{~})]&=\Tr_\SO^{~}[\RRR^+_\theta(\wwwQ^{~})\wedge\SSS^+_\theta(\eeeQ^{~})]+
\Tr_\SO^{~}[\RRR^-_\theta(\wwwQ^{~})\wedge\SSS^-_\theta(\eeeQ^{~})],
\intertext{where}
\RRR^\pm_\theta(\wwwQ^{~})&:=\PH{\pm}\RRR_\theta^{~}(\wwwQ^{~})~~\text{and}~~\SSS^\pm_\theta(\eeeQ^{~}):=\PH{\pm}\SSS(\eeeQ^{~}).
\end{align*}
We can write them using the Rich and scalar curvature as
\begin{align*}
&-\Tr_\SO^{~}[\RRR_\theta^{~}(\wwwQ^{~})\wedge\SSS_\theta^{~}(\eeeQ^{~})]=R(\wwwQ^{~})\hspace{.2em}\vvv^{~}_\theta
~~\text{and}~~
-\Tr_\SO^{~}[\RRR^\pm_\theta(\wwwQ^{~})\wedge\SSS^\pm_\theta(\eeeQ^{~})]=
\frac{1}{2}\left(R(\wwwQ)\pm(R^0+
R_0
)
\right)\vvv^{~}_\theta,
\intertext{where}
&\hspace{1em}R^0:=i\frac{\kappa(-\theta)}{2}
\epsilon^{\hspace{1.3em}\stars}_{\theta\hspace{.1em}0\bcdot}\hspace{.2em}R^{0\bcdot}_{\hspace{.7em}\stars}(\wwwQ^{~})=
i\kappa(-\theta)\(R^{01}_{\hspace{.7em}23}-R^{02}_{\hspace{.7em}13}+R^{03}_{\hspace{.7em}12}\)(\wwwQ^{~})
\intertext{and}
&\hspace{1em}R_0:=i\frac{\kappa(\theta)}{2}
\epsilon^{\hspace{1.3em}0\star}_{\theta \bcdots}\hspace{.2em}R^{\bcdots}_{\hspace{.7em}0\star}(\wwwQ^{~})=
i\kappa(\theta)\(R^{12}_{\hspace{.7em}03}-R^{02}_{\hspace{.7em}13}+R^{23}_{\hspace{.7em}01}\)(\wwwQ^{~}).
\end{align*}
%
%
%
%
When the torsion is null,  space-time curvature has additional symmetry.
Owing to the definition (\ref{torsionFM}), the torsion-less condition yields
\begin{align*}
0&={\wwwQ^{~}}^{a}_{\hspace{.3em}\bcdot}\wedge\TTT^\bcdot_\theta=
{\wwwQ^{~}}^{a}_{\hspace{.3em}\bcdot}{\wedge}d\eeeQ^\bcdot+
{\wwwQ^{~}}^{a}_{\hspace{.3em}\bcdot}\wedge{\wwwQ^{~}}^{\bcdot}_{\hspace{.3em}\star}\wedge{\eeeQ^{~}}^{\star}=
{\etaQ}^{~}_\bcdots\hspace{.1em}\RRR_{~}^{a\bcdot}(\wwwQ^{~})\wedge\eee^\bcdot-d({\wwwQ^{~}}^{a}_{\hspace{.3em}\bcdot}{\wedge}\eeeQ^\bcdot),
\end{align*}
where ${\wwwQ^{~}}^{a}_{\hspace{.3em}b}:={\wwwQ^{~}}^{a\bcdot}{\etaQ}^{~}_{b\bcdot}$.
A total-derivative term does not contribute the equation of motion; thus, we obtain 
\begin{align}
\TTT^a(\eeeQ^{~})=\dot{0}&\implies
{\etaQ}^{~}_\bcdots\hspace{.1em}\RRR_{~}^{a\bcdot}(\wwwQ^{~})\wedge\eee^\bcdot=\dot{0},\label{TrlessR}
\end{align}
yielding
\begin{align}
\eta^{~}_{\theta a\bcdot}\eta^{~}_{\theta b\bcdot}\hspace{.1em}R^{\bcdots}_{~~cd}(\wwwQ^{~})&=
\eta^{~}_{\theta c\bcdot}\eta^{~}_{\theta d\bcdot}\hspace{.1em}R^{\bcdots}_{~~ab}(\wwwQ^{~}),\label{RabcdeqRcdab}
\end{align}
as the solution to the algebraic equation (\ref{TrlessR}) for any values of $\theta$.
We denote (\ref{RabcdeqRcdab}) shortly as
\begin{align}
R^{~}_{ab\hspace{.2em}cd}&=R^{~}_{cd\hspace{.2em}ab}.\label{RabcdeqRcdab2}
\end{align}
Symmetry (\ref{RabcdeqRcdab2}) induces the relation
\begin{align*}
R^0=-R_0.
\end{align*}
Consequently, one of $\RRR^\pm$ can yield the scalar curvature, like
\begin{align}
\Tr_\SO^{~}[\RRR^+_\theta(\wwwQ^{~})\wedge\SSS^+_\theta(\eeeQ^{~})]&=
\Tr_\SO^{~}[\RRR^-_\theta(\wwwQ^{~})\wedge\SSS^-_\theta(\eeeQ^{~})]=
-\frac{1}{2}
R(\wwwQ^{~})\hspace{.1em}\vvv^{~}_\theta;\label{RpmS}
\end{align}
thus, the Einstein--Hilbert Lagrangian can be written using one of the subspaces $\VV^\pm(\Omega^2)$.
When we choose SD curvature $\RRR^+_\theta(\wwwQ^{~})$ to construct the Lagrangian, the gravitational Lagrangian form is represented as
\begin{align*}
\LLL_{\textsc{g}\hspace{-.05em}\textsc{r}}(\wwwQ^{~})&=-2
\frac{\text{det}[\betathe]^\hlf}{\hbar\kappa}\Tr_\SO^{~}[\RRR^+_\theta(\wwwQ^{~})\wedge\SSS^+_\theta(\eeeQ^{~})],
~~\textrm{with}~~
\TTT^a(\eeeQ)=0.
\end{align*}
The volume form can also be written using one of $\SSS^\pm_\theta(\eeeQ^{~})$ such that: 
\begin{align*}
\frac{1}{3!}\vvv&=-f(\theta)\hspace{.2em}\Tr_\SO^{~}\left[\SSS^+_\theta(\eeeQ^{~})\wedge\SSS^+_\theta(\eeeQ^{~})\right]=
f(\theta)\hspace{.2em}\Tr_\SO^{~}\left[\SSS^-_\theta(\eeeQ^{~})\wedge\SSS^-_\theta(\eeeQ^{~})\right],
\intertext{where}
f(\theta)&:=\frac{i}{4}\(
3\kappa(-\theta)+\kappa(\theta)
\)=\left\{
\begin{array}{crr}
1,&\theta=&0\\
-i/2,&\theta=&\pm1
\end{array}
\right. .
\end{align*}

%% file: Sec4-3.tex
%
%
\subsubsection{Fermions}\label{SPrep}
The fermion Lagrangian in the Lorentzian metric space is provided by (\ref{WspinLagR}) and (\ref{WspinLagL}).
We can extend them to the amphometric space by replacing the Lorentzian metric with the amphometric. 
Correspondingly, a weak pair of the dual-spinor is extended to that of the dual-amphospinor as
\begin{align}
\bm\Xi^{~}_{\textsc{w}}(\theta_\nu,\theta_e)&:=
\bm\xi^\nu_{~}(\theta_\nu)\otimes\hbphiUw+\bm\xi^e_{~}(\theta_e)\otimes\hbphiDw
\longrightarrow
\begin{cases}
{\bm\xi}^\nu_{~}\otimes\hbphiUw+{\bm\xi}^e_{~}\otimes\hbphiDw
=\text{(\ref{VVXIenu1})},&{\theta_\nu=\theta_e=+1}\\
\dot{\bm\xi}^\nu_{~}\otimes\hbphiUw+\dot{\bm\xi}^e_{~}\otimes\hbphiDw
=\text{(\ref{VVXIenu2})},&{\theta_\nu=\theta_e=-1}
\end{cases}
;\label{VVXIenu2Ampho}
\end{align}
thus, dotted and undotted weak dual-spinor spinors are unified.
We introduce the chiral-left fermion Lagrangian in the amphometric space as
\begin{subequations}
\begin{align}
&\LL^{\theta\!L}_{\textsc{f}_{\hspace{-.1em}\textsc{m}}}
(\bm\Xi^{~}_{\textsc{w}},\bm\Xi^{\dagger_{\hspace{-.1em}\epsilon}}_{\textsc{w}}):=-
\hspace{.1em}\la{\bm\Xi}^{*}_{\textsc{w}}(\theta_\nu,\theta_e),
i\ds^{\hspace{.2em}\SG}_{\Theta}\hspace{.1em}{\bm\Xi}^{~}_{\textsc{w}}(\theta_\nu,\theta_e)\ra\!,\label{LfmenuQ2}
\intertext{where}
&i{\ds}^{\hspace{.2em}\SG}_{\Theta}:=
i{\ds}^{\hspace{.2em}\SG}_{\theta_\nu}\hbphiUw+i{\ds}^{\hspace{.2em}\SG}_{\theta_e}\hbphiDw,~~\text{with}~~
i{\ds}^{\hspace{.2em}\SG}_{\theta}:={\gamma}^\bcdot_{\theta}
\(\bm1_\SU\otimes\(i\hspace{.1em}\partial_\bcdot
-\frac{1}{2}\cG\hspace{.2em}\omega^{\hspace{.6em}\stars}_{\theta\bcdot}\hspace{.1em}{S}^{~}_{\theta\stars}\)
+{c^{~}_\SU}\hspace{.2em}\hspace{.1em}\Aa^J_{\theta\hspace{.1em}\bcdot}\)\!.\label{LfmenuQ3}
\end{align}
\end{subequations}
From (\ref{LfmenuQ2}), the chiral-left fermion-Lagrangian is provided  as
\begin{align}
\text{(\ref{LfmenuQ2})}\xrightarrow{\theta_\nu=\theta_e=-1}
\LL^{L}_{\textsc{f}_{\hspace{-.1em}\textsc{m}}}&=\text{(\ref{WspinEoML})}.
\end{align}
For the equation of motion (\ref{YMGBEoMDWS}) with the source term in the Lorentzian metric space, we implement the source term in the equation of motion (\ref{ELEoMpm}) as
\begin{align}
&\hat{d}_{\AAAQ}\FFFQ=c^{~}_\SU\hspace{.2em}{\bm{\Xi}}^{\dagger_{\hspace{-.1em}\epsilon}}(\theta_\nu,\theta_e)\hspace{.1em}
\gamma^\bcdot_{\textsc{s}\theta}\hspace{.1em}{\bm{\Xi}}(\theta_\nu,\theta_e)\hspace{.1em}
\VVV_{\theta\hspace{.1em}\bcdot}\xrightarrow{\theta_\nu=\theta_e=-1}\text{(\ref{YMGBEoMDWS})}.
\end{align}
The chiral-right fermion-Lagrangian in the amphhometric space consists of the undotted weyl spinor as
\begin{align}
&\LL^{\theta\!R}_{\textsc{f}_{\hspace{-.1em}\textsc{m}}}(\bm\xi^e_{~},\bpsi_{~}^{e\dagger_{\hspace{-.1em}\epsilon}})=
\<\bpsi^{e*}_{~}(\theta),i{\ds^{\hspace{.1em}\Sp}_{\theta}}\hspace{.1em}\bpsi^e_{~}(\theta)\>
\xrightarrow{\theta=1}
\LL^{R}_{\textsc{f}_{\hspace{-.1em}\textsc{m}}}=\text{(\ref{WspinLagR})},\label{LfmenuQ}~\text{with}~
i{\ds}^{\hspace{.2em}\Sp}_{\theta}:={\gamma}^\bcdot_{\theta}
\(
i\hspace{.1em}\partial_\bcdot
-\frac{1}{2}\cG\hspace{.2em}\omega^{\hspace{.6em}\stars}_{\theta\bcdot}\hspace{.1em}{S}^{~}_{\theta\stars}
\)\!.
\end{align}

We draw a structure of the dual-spinors in the whole $\W$ space schematically as
\begin{align}
\begin{array}{ccrclcc}
\theta=-1&{\longleftrightarrow}&&\theta=0&&{\longleftrightarrow}&\theta=+1\hspace{.6em}\\
\downarrow&&\hspace{2.5em}\rotatebox[origin=d]{-60}{$\downarrow$}&&\rotatebox[origin=d]{60}{$\downarrow$}\hspace{2.5em}&&\downarrow\hspace{.6em}\\
(\dot{\bm\xi}^\nu_{~}&\otimes&\hbphiUw)&&\CRed{(\hat{\bm\phi}_\textsc{w}^{'\textsc{u}}}&\CRed\otimes&{{\bm\xi}^\nu_{~})}\\
&+&&&&\CRed+\\
(\dot{\bm\xi}^e_{~}&\otimes&\hbphiDw)&&(\CRed{\hat{\bm\phi}_\textsc{w}^{'\textsc{d}}}&\CRed\otimes&\bm\xi^e_{~})\\
\rotatebox[origin=c]{90}{$\in$}&\rotatebox[origin=c]{90}{$=$}&\rotatebox[origin=c]{90}{$\in$}\hspace{1.5em}&&\hspace{.5em}\CRed{\rotatebox[origin=c]{90}{$\in$}}&\rotatebox[origin=c]{90}{\CRed{=}}&\rotatebox[origin=c]{90}{$\in$}\hspace{.6em}\\
\overline{\sss\lll(2,\C)}&\hspace{.4em}\dot{\bm\Xi}^{~}_\textsc{w}&(\sss\uuu_L^{~}(2)&\oplus&\sss\uuu_R^{~}(2))&\hspace{.4em}\CRed{{\bm\Xi}^{~}_\textsc{w}}&{\sss\lll(2,\C)}\\
&&&\rotatebox[origin=c]{90}{$=$}\\
&&&\sss\ooo(4)\\
\end{array}.
\label{lassign}
\end{align}
The table draws the right-handed Higgs spinor and chiral-right spinors as symmetric to chiral-left ones. 
In reality, the right-handed Higgs spinor shown in red is not observed.
Though the right-handed neutrino is not directory observed, a neutrino oscillation phenomenon indirectly proves its existence; thus, we draw it in black.
In this understanding, the electron--spinor and the neutrino-spinor themselves are not $\SUW$ symmetric; the $\SUW$-group acts on them though the Higgs-spinor in the dual-spinor as shown in (\ref{tauXi}). 

Suppose a single source spinor exists in the Lorentzian metric space as a classical current of the $SU(2)$ charge. 
Singular points of the connection form given in (\ref {traj}) configure the trajectory of the classical particle in the Lorentzian metric space as $\bm{j}(t):=c^{~}_\SU\hspace{.2em}\bm{r}(t)$, which is nothing other than the classical weak current. 
Therefore, the equation of motion  (\ref{YMGBEoMDWS}) with the source term is reduced to the vacuum equation (\ref{ASDEoM}) in $\overline{\TsML}$.
Consequently, the vacuum solutions of the Yang--Mills gauge boson in the Lorentzian metric space are homotopically equivalent to the SD curvature through the amphometric. However, the solution is not SD at $\theta=1$. 
The current author precisely discuss the existence of the solution of the Poisson equation in $\W^+{\hspace{-.1em}\setminus\hspace{-.1em}}\W^+_1$ in Ref.\cite{Kurihara:2023ftr}.


%% file: Sec4-Higgs.tex
%
%
\subsubsection{Higgs spinor}\label{HiggsinSwA}
The Higgs spinor Lagrangian in the Lorentzian metric space given as (\ref{LHiggs1})=(\ref{HiggsEoM2}) is cast into the amphometric one by replacing metric tensors and spinors as
\begin{subequations}
\begin{align}
\text{(\ref{LHiggs1})}\rightarrow
\LLL^{~}_{\textsc{h}}({\bm{\xi}}(\theta),{\bm{\xi}}^{\dagger_{\hspace{-.1em}\theta}}_{~}(\theta))&:=
\LLL^{kin}_{\textsc{h}}({\bm{\xi}}(\theta),{\bm{\xi}}^{\dagger_{\hspace{-.1em}\theta}}_{~}(\theta))-V\({\bm{\xi}}(\theta),{\bm{\xi}}^{\dagger_{\hspace{-.1em}\theta}}_{~}(\theta)\)\vvv^{~}_\theta\label{LHiggsQ1}
\intertext{with}
\LLL^{kin}_{\textsc{h}}({\bm{\xi}}(\theta),{\bm{\xi}}^{\dagger_{\hspace{-.1em}\theta}}_{~}(\theta)):=&
\frac{1}{4}(i{\ds}_{\theta}){\bcdot}(i{\ds}_{\theta})
\<{\bm{\xi}}(\theta),{\bm{\xi}}(\theta)\>\vvv^{~}_\theta.\label{LHiggsQ2}
\end{align}
\end{subequations}
When we set the amphometric parameter at $\theta=0$, the Higgs Lagrangian (\ref{LHiggsQ1}) reproduces (\ref{LHiggs1}).
In \textbf{section \ref{Higgsspinor}}, we discussed operator $\widehat{\sslash{\FFF}}^{~}_\SU$ based on Clifford algebra (\ref{ClQ}), and the results are independent of their component representations.
Therefore, amphoClifford algebra (\ref{gamma-theta}) ensures that a simple replacement of the metric tensor provides the Higgs Lagrangian in the amphometric space.  
According to this change, we obtain the equation of motion as
\begin{align}
&\frac{\delta\I_{\textsc{h}}({\bm{\xi}}(\theta),{\bm{\xi}}^{\dagger_{\hspace{-.1em}\theta}}_{~}(\theta))}{\delta{\bm{\xi}}^{\dagger_{\hspace{-.1em}\theta}}_{~}(\theta)}=0\implies
\(\Delta^{~}_\theta+i{c^{~}_\SU}\sslash{\FFF}^{~}_\theta+\frac{\delta V({\bm{\xi}}(\theta),{\bm{\xi}}^{\dagger_{\hspace{-.1em}\theta}}_{~}(\theta))}{\delta{\bm{\xi}}^{\dagger_{\hspace{-.1em}\theta}}_{~}(\theta)}\){\bm{\xi}}(\theta)=0,
~~\text{where}~~
\sslash{\FFF}^{~}_\theta:=\sslash{\FFF}^{~}_\SU\big|^{~}_{\gamma^{~}_\Sl\rightarrow\gamma^{~}_\theta}.\label{HiggsEoMQ}
\end{align}

As mentioned in \textbf{section \ref{sgb}}, the Higgs spinor does not interact directly with the spin connection since the $\SUW$ group is not the space-time symmetry but the internal symmetry.
When we identify the Higgs spinor with the amphospinor ${\bm{\xi}}(\theta)$ at $\theta=0$, it belongs to the $\SU(2)$ group owing to the space-time $\SO(4)$ symmetry as in  (\ref{lassign}) and may directly couple with the spin connection.   
The Higgs Lagrangian is provided through the $Spin(4)\otimes\SU(2)$ invariant Dirac operator in Euclidean space-time.
In this case, the Lichnerowicz formula in curved space-time is provided as
\begin{align}
&\text{(\ref{Lf0})}\rightarrow
-\(i{\ds}_{\theta}^{~}\)^2=
\Delta_\theta+\widehat{\sslash{\FFF}}^{~}_\theta,
~~\text{where}~~
\widehat{\sslash{\FFF}}^{~}_\theta\hspace{.1em}:=\iota_{\gamma_\theta}\iota_{\gamma_\theta}
(d^{~}_\SG{\wedge}\hspace{.1em}d^{~}_\SG).\label{LichnerowiczQ}
\end{align}
The twisting curvature of the Clifford module is provided owing to  (\ref{codiffSp1}) as
\begin{align}
\widehat{\sslash{\FFF}}^{~}_\theta&=
\iota_{\gamma_\theta}\iota_{\gamma_\theta}
\(d-i\hspace{.1em}{c^{~}_\SU}({\AAA}_{\SU}\otimes\bm{1}_\Sp)-i\hspace{.1em}\cG(\bm{1}_\SU\otimes{{\AAA}}_\Sp)\)^{2\wedge},\notag\\&=
-i\hspace{.2em}\iota_{\gamma_\theta}\iota_{\gamma_\theta}\(
  {c^{~}_\SU}(d{\AAA}_{\SU}-i\hspace{.1em}{c^{~}_\SU}{\AAA}_{\SU}\wedge{\AAA}_{\SU})
+{\cG}(d{\AAA}_{\Sp}-i\hspace{.1em}{\cG}{\AAA}_{\Sp}\wedge{\AAA}_{\Sp})
\)
\end{align}
The connection in the spinor bundle in the amphometric space is provided as
\begin{align*}
\text{(\ref{connectSp})}\rightarrow\slash{\AAA}_{\Sp}&
=\frac{i}{2}
(\iota_{\gamma_\theta}\www_\theta^{\hspace{.3em}\bcdots}){S}_{\Sl\hspace{.1em}\bcdots}\big|^{~}_{\gamma_\Sl\rightarrow\gamma_\theta}=
-\frac{1}{8}\omega_\diamond^{\hspace{.3em}\bcdot\star}\gamma_{\theta}^\diamond\hspace{.1em}
\etaQ_\bcdots\etaQ_\stars\left[\gamma^\bcdot_{\theta},\gamma^\star_{\theta}\right]=-\frac{1}{4}\omega_\diamond^{\hspace{.3em}\bcdot\star}\gamma_{\theta}^\diamond\hspace{.1em}
\etaQ_\bcdots\etaQ_\stars\gamma^\bcdot_{\theta}\gamma^\star_{\theta}.
\end{align*}
As a result, we obtain the twisting spin-curvature of the Clifford algebra as
\begin{align}
\widehat{\sslash{\RRR}}&=-{\cG}\iota_{\gamma_\theta}\iota_{\gamma_\theta}\sslash{\RRR}=
-\frac{\cG}{8}R^{\hspace{.3em}\bcdots}_{\theta\hspace{.7em}\stars}\hspace{.2em}
\gamma_{\theta\bcdot}\hspace{.1em}\gamma_{\theta\bcdot}(\eta_\theta^{\hspace{.3em}\star\diamond}\gamma_{\theta\diamond})
(\eta_\theta^{\hspace{.3em}\star\diamond}\gamma_{\theta\diamond}).\label{RssQ}
\end{align}
Hereafter, we omit subscript $\theta$ on Riemann curvature $R_{\hspace{.3em}\cdot\cdot}^{\cdot\cdot}$ to avoid too many subscripts.
We first calculate the twisting spin-curvature of the Clifford algebra in the Lorentzian metric space.
The Bianchi identity (\ref{RBianchi}) leads to an identity among space-time curvature components as a=b.
\begin{align}
&R^{a}_{\hspace{.5em}b\hspace{.2em}cd}+R^{a}_{\hspace{.5em}c\hspace{.2em}db}+R^{a}_{\hspace{.5em}d\hspace{.2em}bc}=0
{\quad\text{for}\quad}R^{a}_{\hspace{.5em}b\hspace{.2em}cd}:=
\etaL_{\hspace{.1em}b\bcdot}R^{a\bcdot}_{\hspace{.8em}cd},
\intertext{yielding}
&\sum_{c\neq b\neq d} R^{a}_{\hspace{.5em}b\hspace{.2em}cd}\hspace{.2em}\gamma_\Sl^{b}\hspace{.1em}\gamma_\Sl^{c}\hspace{.1em}\gamma_\Sl^{d}
=0.\label{RBianchicoe}
\end{align}
Thus, we obtain the twisting spin-curvature of the Clifford algebra as
\begin{align}
R^{~}_{ab\hspace{.2em}cd}\hspace{.2em}
\gamma_\Sl^{a}\hspace{.1em}\gamma_\Sl^{b}\hspace{.1em}\gamma_\Sl^{c}\hspace{.1em}\gamma_\Sl^{d}
&\overset{\text{(\ref{RBianchicoe})}}{=}
\etaL^\bcdots
\(-R^{~}_{\bcdot b\hspace{.2em}a\bcdot}
  +R^{~}_{\bcdot b\hspace{.2em}\bcdot a}\)\gamma_\Sl^{a}\hspace{.1em}\gamma_\Sl^{b}=
2\etaL^\bcdots\hspace{.1em}
R^{~}_{b\bcdot\hspace{.2em}a\bcdot}\hspace{.1em}\gamma_\Sl^{a}\hspace{.1em}\gamma_\Sl^{b},\notag\\
&\overset{\text{(\ref{RabcdeqRcdab2})}}{=}2
R^{\bcdot\star}_{\hspace{.7em}\bcdot\star}\implies
\text{(\ref{RssQ})}\xrightarrow{\theta=1}-\cG\frac{R^{~}_\theta}{4}.\label{RssL}
\end{align}
In the above calculations, we have utilised the Clifford algebra (\ref{ClQ}) independent from its representation and a metric tensor.
The final result (\ref{RssL}) is given by the scalar curvature after contractions between upper and lower indices in their intrinsic position without metric tensors; thus, (\ref{RssL}) is the metric independent. 
Consequently, we obtain the Lichnerowicz formula in the amphometric space as
\begin{align}
\text{(\ref{LichnerowiczQ})}\rightarrow-\(i{\ds}_{\theta}^{~}\)^2&=\Delta(\theta)
-i\hspace{.1em}{c^{~}_\SU}\sslash{\FFF}^{~}_\theta-{\cG}\frac{R^{~}_\theta}{4}.
\end{align}
The equation of motion for the Higgs spinor in the amphometric space is provided as
\begin{align}
\text{(\ref{HiggsEoM2})}&\implies
\(\Delta^{~}_\theta+i{c^{~}_\SU}\sslash{\FFF}^{~}_\theta+{\cG}\frac{R^{~}_\theta}{4}+\frac{\delta V({\bm{\xi}}(\theta),{\bm{\xi}}^{\dagger_{\hspace{-.1em}\theta}}_{~}(\theta))}{\delta{\bm{\xi}}^{\dagger_{\hspace{-.1em}\theta}}_{~}(\theta)}\){\bm{\xi}}(\theta)=0.
\end{align}

%% file: Sec4-Fmass.tex
%
%
\subsubsection{Fermion mass term}\label{intrHWQ}
The fermion mass term in the Lorentzian metric space is provided owing to the interaction between the weak dual-spinor and the Higgs spinor with the SSB as shown in \textbf{section \ref{intrHW}}.
It utilised the $\SL2C$-invariant bilinear form between dotted and undotted Weyl spinors.
Since the dual-spinor unifies dotted and undotted spinors in the amphometric space,  the fermion mass term consists of only the amphospinor, setting appropriate $\theta$-values.

We replace a dual-spinor in a Higgs-fermion interaction term (\ref{me2}) by the dual-amphospinor (\ref{VVXIenu2Ampho}) as
\begin{align}
\{\text{(\ref{XwPhiw})}\leftarrow\text{(\ref{VVXIenu2Ampho})}\}&\implies
\<{\bm\Xi}^{~}_{\textsc{w}}(\theta_\nu,\theta_e),\bpsi^\textsc{h}_{~}(\theta_\textsc{h}=0)\>,\label{dual-amphospinor-bilinear}
%
\intertext{where we note}
\bpsi^\textsc{h}_{~}(0)=\bphiw.&\notag
\end{align}
We obtain the fermion mass term using  (\ref{dual-amphospinor-bilinear}) as
\begin{align}
\LL^{\theta m_e}_{\textsc{f}_{\hspace{-.1em}\textsc{m}}}=
\lambda\<\<{\bm\Xi}^{*}_{\textsc{w}}(-\theta_\nu,-\theta_e),\bphiw\>,{\bm\Xi}^{~}_{\textsc{w}}(\theta_\nu,\theta_e)\>
=&-\lambda\hspace{.1em}\phi^{~}_\textsc{d}
\(\<\bpsi^{e*\hspace{.1em}}_\textsc{u}(-\theta_e\hspace{.1em}),\bpsi^e_\textsc{d}(\theta_e\hspace{.1em})\>-
   \<\bpsi^{e*\hspace{.1em}}_\textsc{d}(-\theta_e\hspace{.1em}),\bpsi^e_\textsc{u}(\theta_e\hspace{.1em})\>\)\notag\\
&-\lambda\hspace{.1em}\phi^{~}_\textsc{u}
\(\<\bpsi^{\nu*}_\textsc{u}(-\theta_\nu),\bpsi^\nu_\textsc{d}(\theta_\nu)\>-
   \<\bpsi^{\nu*}_\textsc{d}(-\theta_\nu),\bpsi^\nu_\textsc{u}(\theta_\nu)\>\)\!,\label{meQ2}
\end{align}
We obtain the fermion mass term (\ref{me2}) in the Lorentzian metric space by setting $\theta$-values as
\begin{align}
\text{(\ref{meQ2})}\big|^{~}_{ \theta^\nu=\theta^e=1}\implies\text{(\ref{me2})}.
\end{align}

%% file: Sec5.tex
\section{Graviweak correspondence}\label{GWc}
In this report, we have constructed the Yang--Mills theory including the gravitational force as the local $SO(1,3)$ gauge interaction, namely the Yang--Mills--Utiyama theory, in which the gauge theory of the space-time symmetry is treated similarly to that of the internal gauge symmetry.
We have extended the \YMU theory in the Lorentzian metric space further into the amphometric space, which smoothly connects the Lorentzian metric to the Euclidean one.
The amphometric method has connected two metric spaces and allowed us to import mathematical achievements established in Euclidean space into Lorentzian one.
This report showed two examples: a role of the SD (ASD) curvature and an instanton solution in the Lorentzian metric space.

Furthermore, the amphometric has connected two symmetries $SU\!(2)$ and $\SL2C$.
We have schematically drawn several gauge and spinor structures in the amphometric space, revealing the novel relation between weak and gravitational interactions.
This section discusses the relationship between internal and space-time symmetries and examines their correspondence.

\subsection{Gauge bosons}
First, we overview a chiral structure of the bosonic part of the \YMU theory.
A space of two-form objects in a four-dimensional space has the $\Z_2$-grading structure owing to the Hodge-dual operator.
The eigenvectors of the Hodge-dual operator are named SD (eigenvalue$=+1$) and ASD (eigenvlue$=-1$).
We assigned the chiral-right to the SD and the chiral-left to the ASD curvature.
Both the Yang--Mills gauge-boson Lagrangian and the second Chern index consist of a squared curvature.
The SD or the ASD curvature gives the solutions of the equation of motion in the Euclidean metric space; this is proven by comparing it with the second Chern index.
Whether the curvature is SD or ASD depends on the orientation of the manifold.
This is not true in the Lorentzian metric space.
Even though the solutions in the Lorentzian metric space are not SD nor ASD, they are homotopical equivalent to SD or ASD curvature in the Euclidean metric space in the amphometric space.

On the other hand, the Einstein--Hilbert gravitational Lagrangian, $\LLL_{\hspace{-.1em}E\hspace{-.1em}H}=\RRR\wedge\SSS$, consists of the curvature two-form and the surface two-form.
Thus, the extremal of the Lagrangian is not bounded by the second Chern index, consisting of the curvature squared.
However, when we construct the gravitational Lagrangian in the co-Poincar\'{e} bundle, the Lagrangian consists of the co-Poincar\'{e} curvature squared; the Lagrangian is bounded owing to the second Chern-index as a functional of the co-Poincar\'{e} curvature.
In reality, we can construct the Einstein--Hilbert gravitational  Lagrangian in half the space-time degree of freedom, the SD or the ASD two-form objects.
The chirality is shared between Yang--Mills and space-time curvatures; in this study, we fixed it as SD (chiral-right).

The structure groups of the gauge and spinor bundles are, respectively, $\SUW$ and $SO(1,3)$; they have the Lie-algebra homomorphism as (\ref{ismorSO4}) and  (\ref{ismorSO13}).
The space-time curvature belongs to an adjoint representation of $\sss\lll(2,\C)$, which is a half of $\sss\ooo(1,3)$ owing to (\ref{ismorSO13}).
We prepare the SD curvature with the causal subgroup $\FFF^+_\theta{\in}\Omega^2_\textsc{sd}(T^*\W^+_\theta){\oplus}Ad\!\(\Ca_\theta(2)\)$ in the amphometric space $\W^+_\theta$; we obtain the following diagram:
\begin{align}
\begin{array}{cclcl}
\RRR^+_\textsc{l}&{\in}&\Omega^2_\textsc{sd}(\TsML)&{\oplus}&Ad\!\(\LCa_{\theta=1}(2)=\sss\lll(2,\C)\)\\
\uparrow&&\theta\rightarrow1\uparrow&&\hspace{1.7em}\uparrow\theta\rightarrow1\\
\FFF^+_\theta&{\in}&\Omega^2_\textsc{sd}(T^*\W^+_\theta)&{\oplus}&Ad\!\(\LCa_\theta(2)\)\\
\downarrow&&\hspace{-.2em}\theta\rightarrow1\downarrow&&\hspace{1.7em}\downarrow\theta\rightarrow0\\
\FFF^+_\SU&{\in}&\Omega^2_\textsc{sd}(\TsML)&{\oplus}&Ad\!\(\LCa_{\theta=0}(2)=\sss\uuu^{~}_{\!R}(2)\)
\end{array},\label{RFF}
\end{align}
where we assume the existence of curvatures.
In the diagram, the Yang--Mills gauge curvature and the space-time curvature have a common origin in the amphometric space; However, both curvatures are in the Lorentzian metric space setting $\theta=1$,  the structure group split into $\SUW$ and $\SL2C$ setting $\theta=0$ and $\theta=1$, respectively. 
This is one consequence of the graviweak correspondence.

\subsection{Higgs and Weyl spinor interaction}
The fermion Lagrangian consists of the weak dual-spinor in the Lorentz metric space as (\ref{WspinEoML}), and it is extended to the weak dual-amphospinor in the amphometric space schematically shown in (\ref{lassign}).
When we allow mixing different $\theta$ values in the dual-amphospinor as (\ref{RFF}), the dual-amphospinor $\bm\Xi(\theta_1,\theta_2)$ has the diagram in the amphometric space under the graviweak point of view like
\begin{subequations}
\begin{align}
&
\begin{array}{ccccccccc}
\bm\Xi(\theta_1,\theta_2)&=&\bpsi(\theta_1)&\otimes&&\bpsi(\theta_2)\\
\downarrow&&\downarrow&&&\downarrow\\
\bm\Xi(-1,0)&=&\dot{\bpsi}&\otimes&(\bphiUh&\oplus&\bphiDh)\\
&&\rotatebox[origin=c]{90}{$\in$}&&\rotatebox[origin=c]{90}{$\in$}&&\rotatebox[origin=c]{90}{$\in$}\\
&&\Gamma^{~}_{\hspace{-.1em}\textsc{w}}\(\ML,\VV^{\textsc{l}}_{\overline{\hlf}}\)&\otimes
&\(\Gamma^{~}_{\hspace{-.1em}\textsc{w}}\(\ML,\VV^{\textsc{eu}}_{\hlf}\)\right.&\oplus
&\left.\Gamma^{~}_{\hspace{-.1em}\textsc{w}}\(\ML,\VV^{\textsc{ed}}_{\hlf}\)\)\\
\end{array}\!,\\
&\xrightarrow{\{P^-_\textsc{se},P^+_\textsc{se}\}}
\hspace{9em}\left\{\dot{\bm\Xi}^{\textsc{u}}_\textsc{w}=\dot{\bpsi}\oplus\bphiUh
,\dot{\bm\Xi}^{\textsc{d}}_\textsc{w}=\dot{\bpsi}\oplus\bphiDh\right\}.
\end{align}
\end{subequations}
It suggests the $\SUW$ symmetry of the Higgs spinor originates at the space-time $\SU\!(2)$ symmetry.

When we accept the graviweak correspondence, we can understand a fermion mass is induced by the Yukawa coupling of $\SL2C$ Weyl-spinor and $\SUW=\SU_{\hspace{-.15em}L}\hspace{-.1em}(2)$ Higgs spinor.
From the quantum field theoretic view, We reconsider an electron mass term here.
We respect the Higgs field as the space-time $SU(2)$ spinor and utilise the first differential formalism for a gravitational interaction term, the same as the Weyl spinor;
\begin{align}
(\ref{LfmenuQ3})\implies
-\frac{1}{2}\cG\hspace{.1em} \omega^{\hspace{.3em}\stars}_\bcdot\hspace{.1em}
({\bphi}_\textsc{h}^\dagger{\gamma}^\bcdot_{~}{S}^{~}_{\stars}\hspace{.1em}{\bphi}_\textsc{h}),\label{cWgS}
\end{align}
where we omit the subscript $\textsc{l}$ on the spin connection and the generator for simplicity.
We consider a configuration in which an electron induces a gravitational field of the Schwarzschild solution around it, and the Higgs spinor interacts gravitationally with the curved space-time in the electron centre-of-mass system.
An electron has an electric charge and intrinsic angular momentum (a spin); thus, the Kerr-- Newman metric may be a solution to the Einstein equation for an electron.
We ignore a contribution of the electric charge since it is suppressed by $(\cG/c^{~}_\SU)^2\approx1/137$ and also ignore a contribution from spin $\hbar/2$ since it gives a more minor contribution than the mass does.

The Schwarzschild solution with a polar-coordinate $x^a=(t,r,\vartheta,\varphi)$ is
\begin{align}
ds_\text{\hspace{-.1em}Schw}^2&=f^2_\text{\hspace{-.1em}Schw}(r)dt^2-f^{-2}_\text{\hspace{-.1em}Schw}(r)dr^2-r^2\left(d\vartheta^2+\sin^2{\varphi}~d\varphi^2\right),
\quad\text{where}\quad
f^2_\text{\hspace{-.1em}Schw}(r)=1-16\pi\kE m_e/r\label{lmschw}
\end{align}
and $m_e$ is the electron mass.
The vierbein form can be extracted from (\ref{lmschw}) as
\begin{align}
\eee^a_\text{\hspace{-.1em}Schw}&=\left(
f_\text{\hspace{-.1em}Schw}dt,
f_\text{\hspace{-.1em}Schw}^{-1}dr,
r~d\vartheta,r\sin{\vartheta}~d\varphi\right).\label{virschw}
\end{align}
From the above solutions and the torsion-less condition, a unique solution of the spin form can be obtained\cite{fre2012gravity,Kurihara:2016nap}:
\begin{align}
\www_\text{\hspace{-.1em}Schw}^{~}=&\left(
\begin{array}{cccc}
0&-8\pi\kE m_e/r^2~dt~& 0 & 0 \\
~& 0 &  f_\text{\hspace{-.1em}Schw}d\vartheta &
f_\text{\hspace{-.1em}Schw}\sin{\vartheta}~d\varphi \\
~&~&0&\cos{\vartheta}~d\varphi\\
~&~&~&0\\
\end{array}
\right).
\end{align}
In the region $r\ll1$, component $\omega_t^{\hspace{.3em}tr}=-m_e/r^2$ dominates over other components.
The gravitational coupling between the Higgs-spinor and an electron is provided as
\begin{align} 
\text{(\ref{cWgS})}&\approx-\frac{1}{2}\cG\sum_{p^{~}_i\in\{t,r,\vartheta,\varphi\}}\eta_{p^{~}_1p^{~}_2}^{~}\hspace{.1em}\eta_{p^{~}_3p^{~}_4}^{~}\hspace{.1em}
\omega_t^{\hspace{.3em}p^{~}_1p^{~}_3}\hspace{.1em}\gamma^{t}_{~}\hspace{.1em}S_{~}^{p^{~}_2p^{~}_4},\label{cWgS2}
\intertext{where $\bm\gamma_{~}^{p}$ is the Clifford algebra with the polar coordinate as}
\gamma_{~}^t&=\gamma_\VL^0,\notag\\
\gamma_{~}^r&=
\sin{\vartheta}\cos{\varphi}\hspace{.2em}\gamma_\VL^1\hspace{.2em}+
\sin{\vartheta}\sin{\varphi}\hspace{.2em}\gamma_\VL^2+
\cos{\vartheta}\hspace{.2em}\gamma_\VL^3,\notag\\
\gamma_{~}^\vartheta&=
\cos{\vartheta}\cos{\varphi}\hspace{.2em}\gamma_\VL^1+
\cos{\vartheta}\sin{\varphi}\hspace{.2em}\gamma_\VL^2+
\sin{\vartheta}\hspace{.2em}\gamma_\VL^3,\notag\\
\gamma_{~}^\varphi&=\hspace{8.4em}-\sin{\varphi}\hspace{.2em}\gamma_\VL^2+\cos{\varphi}\hspace{.2em}\gamma_\VL^3,\notag\\
\implies\quad
\eta^{~}_{tt}&=1,\quad\eta^{~}_{rr}=\eta^{~}_{\vartheta\vartheta}=\eta^{~}_{\varphi\varphi}=-1,\quad\text{otherwise}\quad \eta^{~}_{ab}=0,\notag
\intertext{yielding a Clifford algebra as}
\text{(\ref{ClQ})}\implies&\{\gamma_{~}^{p^{~}_1},\gamma_{~}^{p^{~}_2}\}=2\hspace{.1em}\eta_\textsc{l}^{p^{~}_1p^{~}_2}\bm{1}_\Sp
\quad\text{for}\quad p^{~}_i\in\{t,r,\vartheta,\varphi\}.\notag
\end{align}
The direct calculations provide the coupling as
\begin{align}
\text{(\ref{cWgS2})}=
4\pi{i}\hspace{.1em}\cG\hspace{.1em}\frac{\kE m_e}{r^2}\la{\bphi}_\textsc{h},\gamma^1_\VL{\bphi}_\textsc{h}\ra.
\end{align}
We note that $m_e$ is an electron's gravitational mass observed by the asymptotic observer appearing in the Schwarzschild solution.
Though we may obtain a quantum field theoretic coupling constant after appropriate averaging over $r$ and a renormalisation, we can expect that it is proportional to an electron's gravitational mass.
This coupling appears as the Yukawa coupling between an electron and the Higgs boson and provides a pole in the electron propagator, the QFTheoretic mass.
We can rephrase Einstein's equivalence principle in the QFT'c context as ``\textit{The pole in the fermion propagator is equivalent to the gravitational mass }''.
The graviweak correspondence exemplifies the equivalent principle realised in the QFT'c context.

%% file: Appendix.tex
\section{Induced representation}
\paragraph{Subgroups of $\bm{\sl(2)}$:}\label{inducedrep}
When a simple representation $\g$ of a group $G$ is given, we can construct new representations induced from the original one, namely the induced representation.
E.g., $G=S\hspace{-.1em}L\!\(2,\Ff\in\{\R,\C\}\)$ has following subgroups:
\begin{subequations}
\begin{align}
K&:=
\begin{cases}
SO(2)&(\Ff=\R)\\
SU(2)&(\Ff=\C)
\end{cases},\label{A1K}\\
T&:=\left\{\left.
t(a)=\(
\begin{array}{cl}
a & 0\\
0 &a^{-1}
\end{array}
\)\right|
a\in\Ff\hspace{-.1em}\setminus\hspace{-.2em}\{0\}
\right\},\label{A1T}\\
A&:=\left\{\left.
a(\varphi)=\(
\begin{array}{cc}
e^{\varphi} & 0\\
0 &e^{-\varphi} 
\end{array}
\)\right|
\varphi\in\R
\right\},\label{A1A}\\
N&:=\left\{\left.
\(
\begin{array}{cc}
1 & b\\
0 &1
\end{array}
\)\right|
b\in\Ff
\right\},\label{A1N}\\
B:=TN^t&=\left\{\left.
\(
\begin{array}{cl}
a & 0\\
b &a^{-1}
\end{array}
\)\right|
a\in\Ff\hspace{-.1em}\setminus\hspace{-.2em}\{0\},b\in\Ff
\right\},\label{A1B}\\
M:=K{\cap}T&=
\begin{cases}
\hspace{2em}\pm\bm{1}_2&(\Ff=\R)\\
\(
\begin{array}{cc}
e^{i\varphi} & 0\\
0 &e^{-i\varphi} 
\end{array}
\)&(\Ff=\C)
\end{cases}\label{A1M}.
\end{align}
\end{subequations}
They yield decomposition
\begin{subequations}
\begin{align}
T=M\hspace{-.1em}A,~~G=KTN=K\hspace{-.1em}AN,~~\text{and}~~
G=KTN^t=K\hspace{-.1em}AN^t.
\end{align}
Especially, the decomposition
\begin{align}
G=K\hspace{-.1em}AN=K\hspace{-.1em}AN^t
\end{align}
\end{subequations}
is known as the \emph{Iwasawa decomposition}.

For the Euclidean vector $\hat{\bm v}^{~}_\textsc{e}$ in the spinor representation (\ref{SigmaEv}), two subgroups $K=SU(2)$ in (\ref{A1K}) and $M$ in (\ref{A1M}) are introduced as
\begin{subequations}
\begin{align}
K&:=\left\{\left.
k(\vartheta)=\(
\begin{array}{rr}
\cos{\(\vartheta/2\)}&i\sin{\(\vartheta/2\)}\\
i\sin{\(\vartheta/2\)}&\cos{\(\vartheta/2\)}
\end{array}
\)\right|
0\leq\vartheta<4\pi
\right\},\\
M&:=\left\{\left.
m(\varphi)=\(
\begin{array}{rr}
e^{i\varphi/2}&0\\
0&e^{-i\varphi/2}
\end{array}
\)\right|
0\leq\varphi<4\pi
\right\}.
\end{align}
\end{subequations}
From these subgroups, we obtain an induced (unitary) representation as
\begin{subequations}
\begin{align}
&m(\varphi)k(\vartheta)m(\psi)=
\(
\begin{array}{ll}
\hspace{.5em}\cos{\(\vartheta/2\)}e^{i(\varphi+\psi)/2}&i\sin{\(\vartheta/2\)}e^{i(\varphi-\psi)/2}\\
i\sin{\(\vartheta/2\)}e^{-i(\varphi-\psi)/2}&\hspace{.4em}\cos{\(\vartheta/2\)}e^{-i(\varphi+\psi)/2}
\end{array}\),\label{generalrepSU2}
\intertext{with}
&0\leq\psi<4\pi,~~~
0\leq\vartheta<\pi,~~\text{and}~~
0\leq\varphi<2\pi,
\end{align}
\end{subequations}
where $(\psi,\vartheta,\varphi)$ is called the Euler angle.

(\ref{generalrepSU2}) provides the representation of dimension-2 for the unit Euclidean vector $\hat{\bm v}^{~}_\SE$, yielding the representation in dimension-4 as
\begin{align}
&(\ref{generalrepSU2})=\hat{\bm v}^{~}_\SE\implies
\hat{\bm v}^{~}_\VE=\Sigma^{-1}_\textsc{e}\hat{\bm v}^{~}_\SE=
\left(
     \begin{array}{r}
	\cos{\(\vartheta/2\)}\cos{\(\(\varphi+\psi\)/2\)}\\
	\sin{\(\vartheta/2\)}\cos{\(\(\varphi-\psi\)/2\)}\\
	-\sin{\(\vartheta/2\)}\sin{\(\(\varphi-\psi\)/2\)}\\
	\cos{\(\vartheta/2\)}\sin{\(\(\varphi+\psi\)/2\)}
      \end{array}
\right),
\intertext{yielding a normalized ($\sss\uuu(2)\hspace{-.2em}=\hspace{-.2em}\sss\ooo(4)$)-invariant measure as}
&\overline{\VVV}^{0}_\textsc{e}=\frac{1}{16\pi^2}\sin{\hspace{-.1em}\vartheta}\hspace{.1em}\sin{\varphi}\hspace{.2em}d{\vartheta}\hspace{.1em}d{\varphi}\hspace{.1em}d\psi.
\end{align}
We have another representation of $\hat{\bm v}^{~}_\textsc{e}$ using the four-dimensional polar coordinate as
\begin{align}
&\hat{\bm v}^{~}_\VE=
\left(
     \begin{array}{l}
	\cos{\vartheta}\\
	\sin{\vartheta}\cos{\varphi}\\
	\sin{\vartheta}\sin{\varphi}\cos{\psi}\\
	\sin{\vartheta}\sin{\varphi}\sin{\psi}
      \end{array}
\right)
~~\text{with}~~
0\leq\psi\leq2\pi,~~
0\leq\vartheta<\pi,~~\text{and}~~
0\leq\varphi<\pi,
\intertext{yielding the normalized  invariant measure as}
&\overline{\VVV}^{0}_\textsc{e}=\frac{1}{2\pi^2}\sin^{2}{\hspace{-.2em}\vartheta}\hspace{.1em}\sin{\varphi}\hspace{.1em}d{\vartheta}\hspace{.1em}d{\varphi}\hspace{.1em}d\psi.
\end{align}

%% file: amphoYMB.tex
\section{Instanton in amphometric space}\label{IAMS}
An instanton is a solution to the classical equation of motion for the Yang--Mills gauge field in Euclidean space-time, given by the SD- or the ASD-curvature.  
We extend it into the amphometric space and discuss topological aspects of the instanton solution in the Lorentzian metric space.
We utilise the index theorem for the Dirac operator, and the cobordism.
This appendix summarises the main conclusion of Ref.\cite{Kurihara:2022sso}.
\subsection{Cobordism}
Ren\'{e} Tom has proposed a method to categorise compact and orientable $C^\infty$-manifolds into the equivalence classes named ``\emph{cobordant}\cite{tom1,tom2}'' in 1953.
When a disjoint union of two $n$-dimensional manifolds is the boundary of a compact $(n\hspace{-.1em}+\hspace{-.2em}1)$-dimensional manifold, they are cobordant to each other; the cobordant is an equivalence relation and quotient space
\begin{align*}
\Omega_C^n:=\{n\textrm{-dimensional~closed~manifolds}\}/\hspace{-.2em}\sim 
\end{align*}
forms an abelian group, where $\sim$ is an equivalent relation concerning cobordism.
We present a definition and elementary remarks on cobordism in this section.
For details of the cobordism theory, see textbooks, e.g., Chapter 7 in Ref.\cite{hirsch1976differential}.
\begin{definition}[cobordant]\label{cob}
Suppose $M$ and $N$ are compact and oriented $n$-dimensional smooth manifolds, and $\omega_\bullet$ denotes an orientation of oriented manifold $\bullet$.
$(M,\omega_M)$ and $(N,\omega_N)$ are called ``cobordant'' to each other, if there exists a compact and oriented $(n\hspace{-.1em}+\hspace{-.2em}1)$-dimensional smooth manifold $(W,\omega_W)$ such that
\begin{align*}
(\partial{W},\hspace{.2em}\partial{\omega_W})&=(M,-\omega_M)\amalg(N,\omega_N),
\end{align*}
and $W$ is called a cobordism from $M$ to $N$.
When $M$ and $N$ are cobordant to each other owing to $W$, we denote them as $M\sim_WN$ in this report.
Cobordant induces an equivalent class among compact and oriented $n$-dimensional smooth manifolds, namely, a cobordant class denoted by $\Omega_{C}^n$.
\end{definition}

We utilise the following remarks and lemma to discuss the topological properties of the instanton solutions in the amphometric space.
\begin{remark}
Cobordant class $\Omega_{C}^n$ forms the abelian group concerning the disjoint-union operation.
\end{remark}
\begin{proof}
When map $M_0\rightarrow M_1$ is diffeomorphic, $M_0$ and $M
_1$ are cobordant; thus, $M_0$ and $M_1$ are cobordant with respect to closed period $\theta=[0,1]$; $M_0\sim_{M_0\otimes[0,1]}M_1$ such that
$
\partial{(M_\theta\otimes[0,1])}=(M_0\otimes0)\amalg(M_1\otimes1)
$
with appropriate orientations.
It fulfils the associative and commutative laws trivially.
Any compact manifold $V=\partial{W}$ is the zero element owing to $(M\amalg V)\sim_{W}M$.
The inverse of $(M,\omega_M)$ is given by $(M,-\omega_M)$.
Therefore, the remark is maintained.
\end{proof}

\begin{remark}[Pontrjagin]\label{prg}~\\
Characteristic numbers are cobordism invariant.
\end{remark}
\begin{proof}
A proof is provided for the Pontrjagin number; it is sufficient to show that the Pontrjagin numbers on $M$ vanish when $M=\partial W$.
There exists one-dimensional trivial bundle $\xi$ such that  $TW\bigl|_M=TM\oplus\xi$; thus, for any inclusion map $\iota:M\hookrightarrow{W}$, relation $p_i(M)=\iota^\#(p_i(W))$ is fulfilled.
Therefore, for any invariant polynomial $f\in\WWW(M)$ to define characteristic numbers, it is provided such that:
\begin{align*}
f\left(p_1,p_2,\cdots\right)[M]=\int_Mf\left(p\left(M\right)\right)=
\int_{\partial M}\iota^\#\left(f\left(p\left(W\right)\right)\right)
=\int_Wdf\left(p\left(W\right)\right)=0,
\end{align*}
owing to the Stokes theorem; thus, the lemma is maintained.
Here, $[M]$ denotes a cobordism class of manifold $M$, and $\iota^\#(\bullet)$ shows the pull of $\bullet$.
\end{proof}

\begin{remark}\label{remA4}
A transformation between the Euclidean space and the Minkowski space owing to the amphometric is a homeomorphism.
\end{remark}
\begin{proof}
Suppose $\M_\textsc{e}$ and $\M_\textsc{l}$ are $n$-dimensional (pseud-)Riemannian manifolds with, respectively, Euclidean and Lorentzian metrics, which are topological spaces with the $\R^n$ Euclidean topology.
We introduce the $(n\hspace{-.1em}+\hspace{-.1em}1)$-dimensional manifold $\W_\theta:=\R^n\otimes(\theta\in[0,1])$ which has the submersion of
\begin{align*}
\M_\textsc{e}\hookrightarrow\W_\theta:\M_\textsc{e}=\W_0~~&\text{and}~~~
\M_\textsc{l}\hookrightarrow\W_\theta:\M_\textsc{l}=\W_1.
\end{align*}
The rotation-invariant (isotropic) bilinear form $\langle \bullet|\bullet\rangle_{\theta}$ of two vectors, $\bm{v}=(v^1,\cdots,v^n,\theta=\text{const.})$ and $\bm{u}=(u^1,\cdots,u^n,\theta=\text{const.})$, in the foliation $\W_{\theta}$  is defined as
\begin{align*}
\langle v|u\rangle^{~}_{\theta}:=\sum_{a,b=1,n}\left[\bm{\eta}^{~}_{\theta}\right]_{ab}v^au^b\in\C,
~~&\text{where}~~
\eta_{\theta}^{~}:=\text{diag}
\left[1,e^{i\pi\theta},e^{i\pi\theta},\cdots,e^{i\pi\theta}\right].
\end{align*}
Tensor $\bm\eta^{~}_{\theta}$ provides the metric tensor at $\theta=0$ and $\theta=1$.
A map 
\begin{align*}
\pi_\theta:\M_\textsc{e}\rightarrow\M_\textsc{l}:\(x^1,\cdots,x^n,0\)\in\W_0\mapsto\(x^1,\cdots,x^n,1\)\in\W_1
\end{align*}
is a bijection, continuous, and the inverse $\pi^{-1}_\theta$ is continuous.
Moreover, $\mathrm{det}[\bm\eta^{~}_{\theta}]\neq0$ at any point in $\theta\in[0,1]$; thus, the tensor is invertible.
Therefore, $\M_\textsc{e}$ and $\M_\textsc{m}$ are homeomorphism as the $n$-dimensional manifold equipping the isotropic bilinear form. 
\end{proof}
\noindent

\subsection{Instanton}
The four-dimensional Euclidean space-time has the SD- or ASD curvature as a solution of the equation of motion obtained as the stationary point of the Yang--Mills action as shown in \textbf{Remark \ref{SDorASD}}.
The instanton solution\cite{ATIYAH1978185}, is an example of the (A)SD solution, and the instanton number is a topological invariant in the four-dimensional Euclidean space-time.
This section treats the instanton solution in a flat space-time ($\www=0$) with the amphometric and the instanton number in Euclidean and Lorentzian metric spaces.
We set coupling constant ${c^{~}_\SU}$ to unity for simplicity.

\paragraph{a) unphysical solution:}
The instanton connection at $\bm\rho^{t}=(t,x,y,z)$ in the amphometric space is provided as
\begin{subequations}
\begin{align}
\Aa_\theta(\bm\rho)&
=i\kappa(\theta)\frac{\bm\rho^2}{\bm\rho^2+\lambda^2}\bm{g}(\bm\rho)\hspace{.1em}d\bm{g}^{-1}(\bm\rho))~~\text{with}~~
0<\lambda\in\R,\label{instantonsol}
\intertext{where we set}
\bm{g}(\bm\rho)&:=\frac{1}{\sqrt{\bm\rho^{2}}}\hspace{.2em}\bm\sigma^{~}_{\hspace{-.1em}\theta}\cdot\bm\rho~~\textrm{with}~~
\bm\rho^2:=\eta^{~}_{\theta\hspace{.1em}\bcdots}\hspace{.1em}
\rho^\bcdot\hspace{.1em}\rho^\bcdot.\label{instantonsol2}
\end{align}
\end{subequations}
Direct calculations  of (\ref{instantonsol}) with (\ref{instantonsol2}) provide instanton connection as  
\begin{align*}
\AAA_\theta^I&=\left[\Aa_\theta^I\right]_\bcdot \eee^\bcdot_\theta=
\frac{1}{\bm\rho^2+\lambda^2}\times\left\{
\begin{array}{cl}
x^{~}_\theta\hspace{.1em}dt-t^{~}_{~}dx^{~}_\theta
+z^{~}_\theta\hspace{.1em}dy^{~}_\theta-y^{~}_\theta\hspace{.1em}dz^{~}_\theta,&(I=1)\\
y^{~}_\theta\hspace{.1em}dt-z^{~}_\theta\hspace{.1em}dx^{~}_\theta
-t^{~}_{~}\hspace{.1em}dy^{~}_\theta+x^{~}_\theta\hspace{.1em}dz^{~}_\theta,&(I=2)\\
z^{~}_\theta\hspace{.1em}dt+y^{~}_\theta\hspace{.1em}dx^{~}_\theta
-x^{~}_\theta\hspace{.1em}dy^{~}_\theta-t^{~}_{~}\hspace{.1em}dz^{~}_\theta,&(I=3)
\end{array}\!,
\right.
\end{align*}
where $(x^{~}_\theta,y^{~}_\theta,z^{~}_\theta):=i\kappa(\theta)(x,y,z)$ and $\eee^a_\theta=(\ref{eeeQ})$.
The instanton connections in Euclidean and Lorentzian metric spaces are provided by setting $\Aa_\textsc{e}:=\Aa_\theta\big|_{\theta\rightarrow0}$ and $\Aa_\textsc{l}:=\Aa_\theta\big|_{\theta\rightarrow\pm1}$, respectively.
The structure equation provides the instanton curvature as
\begin{align}
&\FFF_\theta^I=\frac{1}{2}\left[\f_\theta^I\right]_\bcdots \eee^\bcdot_\theta\wedge\eee^\bcdot_\theta=
-\frac{\lambda^2}{\bm\rho^2+\lambda^2}\times\left\{
\begin{array}{cl}
dt{\wedge}dx_\theta+dy_\theta{\wedge}dz_\theta=\SSS_\theta^{+\hspace{.1em}1}&(I=1)\\
dt{\wedge}dy_\theta-dx_\theta{\wedge}dz_\theta=\SSS_\theta^{+\hspace{.1em}2}&(I=2)\\
dt{\wedge}dz_\theta+dx_\theta{\wedge}dy_\theta=\SSS_\theta^{+\hspace{.1em}3}&(I=3)
\end{array}\!.
\right.\label{SDInsF}
\end{align} 
Owing to the last expression of (\ref{SDInsF}) with (\ref{PSpm}), this is the SD curvature at any values of $\theta\in[0,1)$.
The amphometric instanton-curvature provides Euclidean and Lorentzian curvatures as 
\begin{align*}
\FFF^{~}_\theta\big|_{\theta=0}=\FFF^{+}_\textsc{e},~~
\FFF^{~}_\theta\big|_{\theta=1}=\FFF^{+}_\textsc{l}.
\end{align*}
Owing to the \textbf{Remark \ref{SDorASD}}, the SD curvature is a solution of the Yang--Mills equation in the Euclidean space.
In reality, direct calculations show the instanton curvature (\ref{SDInsF}) is the solution of the Yang--Mills equation (\ref{ELEoMpm}) in the amphometric space.
We consider to extent the \textbf{Remark \ref{SDorASD}} to the amphometric space here.

Compact manifold $\overline\ME$ is constructed by the following compactification of $\ME$:
First, we introduce a four-dimensional polar coordinate in $\ME$ as
\begin{align*}
\bm\rho^{t}_\textsc{e}&:=\(
r_\textsc{e}\hspace{.1em}\cos{\vartheta_1},
r_\textsc{e}\hspace{.1em}\sin{\vartheta_1}\cos{\vartheta_2},
r_\textsc{e}\hspace{.1em}\sin{\vartheta_1}\sin{\vartheta_2}\cos{\vartheta_3},
r_\textsc{e}\hspace{.1em}\sin{\vartheta_1}\sin{\vartheta_2}\sin{\vartheta_3}
\),
\intertext{yielding}
\int\vvv^{~}_\textsc{e}&=
\int_0^{\infty}dr_\textsc{e}
\int_0^{\pi}d\vartheta_1
\int_0^{\pi}d\vartheta_2
\int_0^{2\pi}\hspace{-.1em}d\vartheta_3\hspace{.2em}
r^3_\textsc{e}\hspace{.1em}\sin^2{\hspace{-.2em}\vartheta_1}\hspace{.1em}\sin{\hspace{-.1em}\vartheta_2}.
\end{align*}
We note that $\<\bm\rho^{~}_\textsc{e},\bm\rho^{~}_\textsc{e}\>=r^{2}_\textsc{e}>0$.
We apply variable transformations 
\begin{align}
r^{~}_\textsc{e}\mapsto r'_\textsc{e}=
\begin{cases}
r^{~}_\textsc{e}/\lambda,&0\leq r^{~}_\textsc{e}\leq\lambda\\
\lambda/r^{~}_\textsc{e},&\lambda<r^{~}_\textsc{e}<\infty  
\end{cases}.\label{rechange}
\end{align}
When the one-point compactification introduces the infinite point equating with the origin, we obtain two compact manifolds $D^4$ as $D^4_\textrm{in}~(r^{~}_\textsc{e}\leq\lambda)$ and $D^4_\textrm{out}~(r^{~}_\textsc{e}\geq\lambda)$.
Then, surfaces of $D^4_\textrm{in}$ and $D^4_\textrm{out}$ are equated to each other with keeping three angular variables; thus, $\M_E$ is compacitified into union of two compact manifolds such that 
\begin{align*}
\overline\ME&=D_\textrm{in}^4\cup D_\textrm{out}^4
~~\text{with}~~
D^4_\textrm{in}\cap D^4_\textrm{out}=\partial{D^4_\textrm{in}}=-\partial{D^4_\textrm{out}}= S^3.
\end{align*}
Consequently, $\overline\ME$ is boundary-less: $\partial(\overline\ME)=\emptyset$.

The second Chern class defined as (\ref{ChernClass}) in the amphometric space is calculated from (\ref{SDInsF}) as
\begin{align*}
c^{~}_2\(\FFFQ\)&=-\frac{1}{8\pi^2}\int\Tr_\SU\left[\FFFQ\wedge\FFFQ\right]=
e^{3i\pi\theta/2}\frac{6}{\pi^2}\int\left(\frac{\lambda}{\bm\rho^{2}_\theta+\lambda^2}\right)^4\vvv^{~}_\theta.
\end{align*}
After the variable transformations (\ref{rechange}) we obtain the Chern class in Euclidean disks $D_\textrm{in}^4$ and  $D_\textrm{out}^4$ as
\begin{align*}
\int_{D^4_\textrm{in}}c^{~}_2(\FFFE)&=
\int_{D^4_\textrm{out}}c^{~}_2(\FFFE)=
\frac{6}{\pi^2}\int_0^1
\frac{r'^{3}_\textsc{e}}{(r'^{2}_\textsc{e}+1)^4}
dr'_\textsc{e}\int_{S^3}d\Omega_4=\frac{1}{2},
\end{align*}
where $d\Omega_4$ is an integration measure for an angular integration on $S^3$, which gives $\int_{S^3}d\Omega_4=2\pi^2$.
The Chern index is obtained as an integration of  them as
\begin{align*}
ch(\overline\ME)=
\int_{D^4_\textrm{in}}c^{~}_2(\FFFE)+
\int_{D^4_\textrm{out}}c^{~}_2(\FFFE)=1.
\end{align*}
Consequently, we obtain the first Pontrjagin index (equal to the second Chern index) of the instanton curvature in the compact Euclidean manifold as unity. 
We referred to this integer as the instanton number.

On intersection $D^4_\textrm{in}\cap D^4_\textrm{out}=S^3$, angular variables coincide as
\begin{align*}
\partial D_\textrm{in}\ni\vartheta_i^\textrm{in}=\vartheta_i^\textrm{out}\in-\partial D_\textrm{out} ,
\end{align*}
for $i\in\{1,2,3\}$ with keeping an orientation on their surfaces.
When $\vartheta_3^\textrm{in}$ maps to $\vartheta_3^\textrm{out}$ as $n\times\hspace{.2em}\vartheta_3^\textrm{in}=\vartheta_3^\textrm{out}$ with $0<n\in\Z$, the instanton number is given as $c^{~}_2(\overline\ME)=n$ due to the $\vartheta_3$ integration from $0$ to $2n\pi$, which corresponds to the winding number around infinity.

The first Pontrjagin index is defined in the entire space of $\W_\theta$.
After the same compactification as in a Euclidean space in $\W_\theta$, a four-dimensional polar coordinate is introduced in the amphometric space as $\bm\rho^{~}_\theta:=\bm\rho^{~}_\textsc{e}\big|^{~}_{r_\textsc{e}{\rightarrow}r_\theta}$.
The replacement (\ref{rechange}) provides the Chern index as
\begin{align*}
ch(\overline{\M_\theta})&=
e^{i\pi\theta/2}\frac{6\cdot2^4}{\pi^2}
\int_0^1{dr^{~}_\theta}\int_0^\pi{d\vartheta_1}\int_0^\pi{d\vartheta_2}\int_0^{2\pi}{d\vartheta_3}
\hspace{.2em}r^{3}_\theta\hspace{.1em}\sin^2{\hspace{-.2em}\vartheta_1}\sin{\vartheta_2}\\ 
&\times\left[
\(2+r^2_\theta\(1+\cos{2\vartheta_1}+2e^{i\pi\theta}\sin^2{\hspace{-.2em}\vartheta_1}\)\)^{-4}
+
\(1+2r^2_\theta+\cos{2\vartheta_1}+2e^{i\pi\theta}\sin^2{\hspace{-.2em}\vartheta_1}\)^{-4}
\right],\\&=
e^{i\pi\theta/2}\frac{2^3}{\pi}
\int_0^\pi{d\vartheta_1}
\frac{\sin^2{\hspace{-.2em}\vartheta_1}}{\(1+\cos{2\vartheta_1}+2e^{i\pi\theta}\sin^2{\hspace{-.2em}\vartheta_1}\)^2}
\\&=
\begin{cases}
+1&4n-1<\theta<4n+1\\
-1&4n+1<\theta<4n+3\\
\text{indefinite}&\theta=2n+1
\end{cases},
~~~~~~~~~~~n\in\Z.
\end{align*}
At $\theta=2n+1$ with $n\in\Z$, we obtain $e^{i\pi\theta}=-1$; thus, the integrand has a pole on the $\vartheta_1$ integration contour yielding indefinite.
With $\theta$ at other than an odd integer, the $\vartheta_1$ integration provides $\pm1$ depending on the sign of the imaginary part of the integrand.
Consequently, the Chern index with the Lorentzian metric is not well-defined.
A domain of the instanton solution in the Lorentzian metric space is not the entire space of $\ML$ but is $\ML\hspace{-.3em}\setminus\hspace{-.3em}\M_\lambda$, where $\M_\lambda:=\{\rho^{~}_\textsc{l}\in\ML\mid{\etaL_\bcdots\rho^{\bcdot}_\textsc{l}\rho^{\bcdot}_\textsc{l}+\lambda^2=0}\}$; thus, manifold $\ML$ is not closed as a principal bundle with the instanton connection.
Taking closure of manifold $\ML$ concerning $\theta$ from the left, $\ML$ has the Chern index as
\begin{align*}
ch(\ML):=ch(\M^{~}_{\theta=1-\epsilon})\big|^{~}_{\epsilon\rightarrow+0}=1.
\end{align*}
The instanton solution in the Minkowski space is an example of \textbf{Remark \ref{prg}}: manifolds $\M_0=\ME$ and $\M_1=\ML$ are cobordant to each other, and the instanton solutions in the Euclidean and Lorentzian metric are homotopical equivalent in $\W^{~}_\theta$; thus, the Chern index of $\ML$ is consistently yielding the same number as that of  $\ME$.

The instanton curvature (\ref{SDInsF}) with the Lorentzian metric is
\begin{align}
&\FFFL^I=-\frac{\lambda^2}{\bm\rho^2+\lambda^2}\SSS_\textsc{l}^{+\hspace{.1em}a=I}.\label{SDCL}
\end{align} 
In four-dimensional space, the SD two-forms have three bases as the same number as the $SU(2)$ generators; thus, we can express the instanton solution as in (\ref{SDCL}).
After the quantisation, it squared may provide a probability to find the instanton in the Lorentzian space-time.
For an observer at the origin of the coordinate, the probability peaks around $t^2-|\bm{r}|^2+\lambda^2=0$, which is in the space-like region of the observer.
In other words, the instanton has a tachyon trajectory and is unphysical.

\paragraph{b) physical solution:}
We consider the Euclidean instanton solution which provides a physical particle in the Lorentzian metric space. 
We modify connection (\ref{instantonsol}) as
\begin{align}
\Aa_\theta(\bm\rho)&=(\ref{instantonsol})\big|_{\lambda\rightarrow\lambda_\theta},
\quad\text{where}\quad
\lambda_\theta:=i\kappa(\theta)\lambda,\label{physInstA}
\intertext{yielding}
(\ref{physInstA})&\implies\FFFQ^I=
\begin{cases}
-{\lambda^2}/{(\bm\rho^2+\lambda^2)}\hspace{.2em}\SSS_\textsc{e}^{+\hspace{.1em}a=I},&\theta=0\\
+{\lambda^2}/{(\bm\rho^2-\lambda^2)}\hspace{.2em}\SSS_\textsc{l}^{+\hspace{.1em}a=I},&\theta=\pm1
\end{cases}.\label{physInst}
\end{align} 
Physical instanton (\ref{physInst}) has a pole on the Cauchy surface only at $\theta=\pm1$ in $\W^{~}_\theta$.
This curvature fulfils the Yang--Mills equation in vacuum with $\theta\in(-1,1)$.
Direct calculations with the same compactification as above show the Chern index of this curvature as
\begin{align}
1=ch(\MQ)=ch(\ME)=ch(\M^{~}_{\theta=1-\epsilon})\big|^{~}_{\epsilon\rightarrow+0}=:ch(\ML).
\end{align} 
After the quantisation, this solution squared may again provide a probability of finding the instanton in the Lorentzian space-time.
For an observer at the origin of the coordinate at $t=0$, the probability peaks around $t^2-|\bm{r}|^2=\lambda^2$, which is in the time-like region of the observer.
In this case, the instanton is physical.
In conclusion, we obtained the exact solution of the $SU(2)$ Yang--Mills equation in Lorentzian vacuum as follows: 
\begin{align}
\AAAL^I&=\left[\Aa_\textsc{l}^I\right]_\bcdot \eee^\bcdot_\textsc{l}=
\frac{1}{\bm\rho^2-\lambda^2}\times\left\{
\begin{array}{cl}
x\hspace{.1em}d\tt-\tt\hspace{.1em}dx-z\hspace{.1em}dy+y\hspace{.1em}dz,&(I=1)\\
y\hspace{.1em}d\tt+z\hspace{.1em}dx-\tt\hspace{.1em}dy-x\hspace{.1em}dz,&(I=2)\\
z\hspace{.1em}d\tt-y\hspace{.1em}dx+x\hspace{.1em}dy-\tt\hspace{.1em}dz,&(I=3)
\end{array}
\right.\hspace{-.5em},~~\text{with}~~
\bm\rho^2=t^2-x^2-y^2-z^2,
\end{align}
where $\tt:=it$.